\documentclass[aps,twocolumn,superscriptaddress,amsmath,showpacs,tightenlines]{revtex4}
\usepackage{epsfig,graphicx,times}
\usepackage{amstext}
\usepackage{amsmath}
\usepackage{amssymb}
\usepackage{graphicx}
\usepackage{latexsym}
\usepackage{bm}

\newcommand{\Emu}{48}
\newcommand{\Ea}{0.7}
\newcommand{\Ejv}{144}

\newcommand{\woA}{0.0257} 
\newcommand{\woC}{0.1319} 

\newcommand{\wvA}{3.69}   
\newcommand{\wvC}{18.99}  

\newcommand{\ioA}{0.5617} 
\newcommand{\ioC}{0.2916} 

\newcommand{\fB}{0.525}
\newcommand{\WoA}{0.162}  
\newcommand{\WoC}{0.0663} 

\newcommand{\WvA}{23.33}  
\newcommand{\WvC}{9.54}   

\newcommand{\IoA}{0.1353}   
\newcommand{\IoB}{0.0596}   
\newcommand{\IoC}{0.4495}   

\newcommand{\IoD}{0.5883}    
\newcommand{\IoE}{-0.1768}   
\newcommand{\IoF}{-0.0467}   

\begin{document}

\title{Electromagnetically induced transparency and Autler-Townes splitting in superconducting flux quantum circuits}

\author{Hui-Chen Sun}
\affiliation{Institute of Microelectronics, Tsinghua University, Beijing 100084, China}
\affiliation{CEMS, RIKEN, Saitama 351-0198, Japan}

\author{Yu-xi Liu}
\email{yuxiliu@mail.tsinghua.edu.cn}
\affiliation{Institute of Microelectronics, Tsinghua University, Beijing 100084, China}
\affiliation{Tsinghua National Laboratory for Information Science and Technology (TNList), Beijing 100084, China}
\affiliation{CEMS, RIKEN, Saitama 351-0198, Japan}

\author{J. Q. You}
\affiliation{Beijing Computational Science Research Center, Beijing 100084, China}
\affiliation{CEMS, RIKEN, Saitama 351-0198, Japan}

\author{E. Il'ichev}
\affiliation{Institute of Photonic Technology, D-07702 Jena, Germany}
\affiliation{Novosibirsk State Technical University, 20 Karl Marx Avenue, 630092 Novosibirsk, Russia}

\author{Franco Nori}
\affiliation{CEMS, RIKEN, Saitama 351-0198, Japan}
\affiliation{Physics Department, The University of Michigan, Ann Arbor, Michigan 48109-1040, USA}

\date{\today}

\begin{abstract}
We study the microwave absorption of a driven three-level quantum system,
which is realized by a superconducting flux quantum circuit (SFQC),
with a magnetic driving field applied to the two upper levels.
The interaction between the three-level system and its environment is studied within the Born-Markov approximation,
and we take into account the effects of the driving field on the damping rates of the three-level system.
We study the linear response of the driven three-level SFQC to a weak probe field.
The linear magnetic susceptibility of the SFQC can be changed by both the driving field and the bias magnetic flux.
When the bias magnetic flux is at the optimal point,
the transition from the ground state to the second excited state is forbidden
and the three-level SFQC has a ladder-type transition.
Thus, the SFQC responds to the probe field like natural atoms with ladder-type transitions.
However, when the bias magnetic flux deviates from the optimal point,
the three-level SFQC has a cyclic transition,
thus it responds to the probe field like a combination of natural atoms
with ladder-type transitions and natural atoms with $\Lambda$-type transitions.
In particular, we provide detailed discussions on the conditions for realizing
electromagnetically induced transparency and Autler-Townes splitting in three-level SFQCs.

\pacs{42.50.Gy, 42.50.Ct, 74.50.+r, 85.25.Cp}
\end{abstract}
\maketitle \pagenumbering{arabic}

\section{Introduction}

Superconducting quantum circuits (SQCs) with Josephson junctions have been
experimentally demonstrated to possess quantized energy levels
(e.g., see reviews~\cite{RMP,wendin,you-today,clarke,you-nature,Buluta,xiang}),
which are analogous to the quantized internal levels of natural atoms.
However, in contrast to natural atoms,
the quantized energy levels of SQCs can be tuned by externally controllable parameters.
These artificially fabricated SQCs have been extensively explored as qubits in quantum information processing.
They also provide us a controllable platform to test fundamental quantum phenomena at a macroscopic scale.
For example, quantum interference via Landau-Zener-St\"{u}ckelberg transitions~\cite{LZ1,LZ2,LZ3}
has been experimentally demonstrated in SQCs~\cite{oliver,mika,wilson,LZS-izma,lahaye,gzsun}.
Moreover, circuit quantum electrodynamics (circuit QED) of SQCs has been extensively explored
(e.g., see Refs.~\cite{you1,liue,blais,wallraff,circuitQED-R}).
Furthermore, the Sisyphus cooling of a harmonic oscillator via
a superconducting flux quantum circuit (SFQC)~\cite{orlando}
has also been studied~\cite{shnirman-cooling,miro-cooling,franco,ashhab-lasing} theoretically and experimentally.
Our theoretical prediction on the coexistence of one- and two-photon transitions~\cite{liu}
in three-level SFQCs have been experimentally demonstrated~\cite{deppe}.
This coexistence results from the controllable symmetry of the Hamiltonian for SFQCs~\cite{liu},
which are very different from natural atoms.

For three-level SQCs, quantum state control has been theoretically studied
in $\Lambda$-type transition configurations (e.g., Refs.~\cite{zhou,amin,kis,chuipingyang}).
The microwave-induced cooling of a superconducting qubit
via the third energy level has been experimentally demonstrated~\cite{valenzuela}.
This mechanism can be further used to cool the environment surrounding the qubit~\cite{you-liu}.
The inverse process of cooling~\cite{valenzuela} can be used for
single-photon production~\cite{you-liu1} and lasing~\cite{ashhab-lasing},
which has been experimentally demonstrated using superconducting charge quantum circuits~\cite{oa}.
SQCs also allow to experimentally explore atomic-physics phenomena~\cite{Buluta,you-nature} on microelectronic chips,
e.g., electromagnetically induced transparency (EIT)~\cite{EIT-T,harris,marangos,fleischhauer}
and Autler-Townes splitting (ATS)~\cite{ats}.
EIT and ATS both display a dip in the absorption spectrum of a three-level quantum system
to a weak resonant probe field when a strong driving field is appropriately applied.
However, EIT is due to Fano interference~\cite{fano},
while ATS is due to the driving-field-induced shift of the transition frequency which is probed.
The application of EIT in atomic systems to nonlinear optics~\cite{fleischhauer}
and quantum information theory~\cite{lukin} has been extensively studied.

\begin{figure}
\vspace{-0.5cm}
\includegraphics[bb=0 0 511 199, width=8.5 cm, clip]{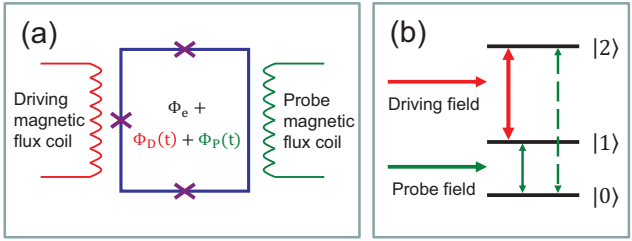}
\caption[]{(Color online)
(a) Schematic diagram of a SFQC with three Josephson junctions.
Here $\Phi_{e}$ is a bias magnetic flux (dc), while $\Phi_{\rm D}(t)=\Phi_{c}\cos(\omega_{0}t)$
is a strong driving magnetic flux (ac) provided by the left coil (in red color),
$\Phi_{\rm P}(t)$ is a weak probe magnetic flux (ac) provided by the right coil (in green color).
(b) Schematic diagram of the three-level SFQC.
The driving field is used to couple the two upper energy levels $|1\rangle$ and $|2\rangle$.
However, the probe field is used to couple the energy levels $|0\rangle$ and $|1\rangle$,
as well as the energy levels $|0\rangle$ and $|2\rangle$.}
\label{fig1}
\end{figure}

In this paper, we study the linear response of a three-level SFQC
to a weak probe field when the two upper levels are driven by a strong external microwave field.
Our motivation is quadfold:

\begin{enumerate}
\item
The microwave-induced transitions between different energy levels of SFQCs
can be adjusted by the bias magnetic flux~\cite{liu};
thus the linear response of SFQCs should depend on the magnetic bias.

\item EIT has been proposed as a promising method to probe the coherence of superconducting qubit states~\cite{murali,dutton}.
Tunable EIT has been studied in circuit QED systems by using dressed states~\cite{HouIan}.
Moreover, ATS~\cite{atsET,atsEP,atsEF,Bsanders,atsET-2,atsET-3} and coherent population trapping~\cite{CPT}
have been experimentally demonstrated in different types of SQCs with three energy levels.
ATS has been proposed as a basis for fast, high on/off ratio microwave routers~\cite{atsEP,atsET-2}.

\item One experiment~\cite{atsEF} in three-level SFQCs showed that the two peaks in the transmission spectrum
for ATS~\cite{Bsanders} have different heights, even when the driving field is resonantly applied,
and this phenomenon cannot be explained by using a simple Lindblad master equation.

\item Using weak continuous measurements~\cite{averin,korotkov,brink,pilgram,greenberg,clerk},
experimentalists studied magnetic susceptibilities to extract the information of SQCs~\cite{el,miro,izmalkov,shny}.
\end{enumerate}

This study mainly focuses on the following questions:
(i) how the linear response of SFQCs changes with the tunable bias magnetic flux;
(ii) what are the differences between the linear responses of SFQCs and natural atoms;
(iii) what are the conditions for realizing EIT and ATS in three-level SFQCs;
(iv) why the transmission spectrum in the ATS experiment~\cite{atsEF,Bsanders} is asymmetric.

Differences between our study for EIT with those in Refs.~\cite{murali,dutton} are:
(i) we consider the effect of the driving field on the dissipation of SFQCs
by using the method developed in Refs.~\cite{efremov,smirnov1,smirnov2,smirnov3}.
(ii) Refs.~\cite{murali,dutton} study EIT in the basis of the single-well states of SFQCs,
which have a $\Lambda$-type transition.
However our study is in the basis of the three lowest eigenstates of SFQCs,
which have a ladder-type transition (or cyclic transition)
when the bias magnetic flux is at (or deviates from) the optimal point~\cite{liu}.
(iii) In contrast to Refs.~\cite{murali,dutton},
the environmental temperature effects on the responses of the three-level SFQCs have also been studied here.
(iv) Moreover, we also provide detailed discussions on the relation between EIT and ATS for SFQCs.

Our paper is organized as follows.
In Sec.~\ref{TM}, we first briefly review the SFQC and write the Hamiltonian of the three-level SFQC,
which interacts with the strong driving field, the weak probe field, and the environment.
We also give the definition of the linear magnetic susceptibility of the three-level SFQC
to a weak probe magnetic field.
In Sec.~\ref{ES}, formal solutions of the operators of the three-level SFQC are given
by solving the Heisenberg-Langevin equations.
In Sec.~\ref{MS}, magnetic susceptibilities of the three-level SFQC are calculated
and the conditions for realizing EIT and ATS are derived.
In Sec.~\ref{NUM}, the numerical results for the magnetic susceptibilities are discussed
by using experimentally accessible parameters.
We finally give conclusions in Sec.~\ref{CS}.

\section{Theoretical Model}\label{TM}

\subsection{Hamiltonian of superconducting flux qubit circuits}

\begin{figure}
\vspace{-0.5cm}
\includegraphics[bb=40 210 520 600, width=8 cm, clip]{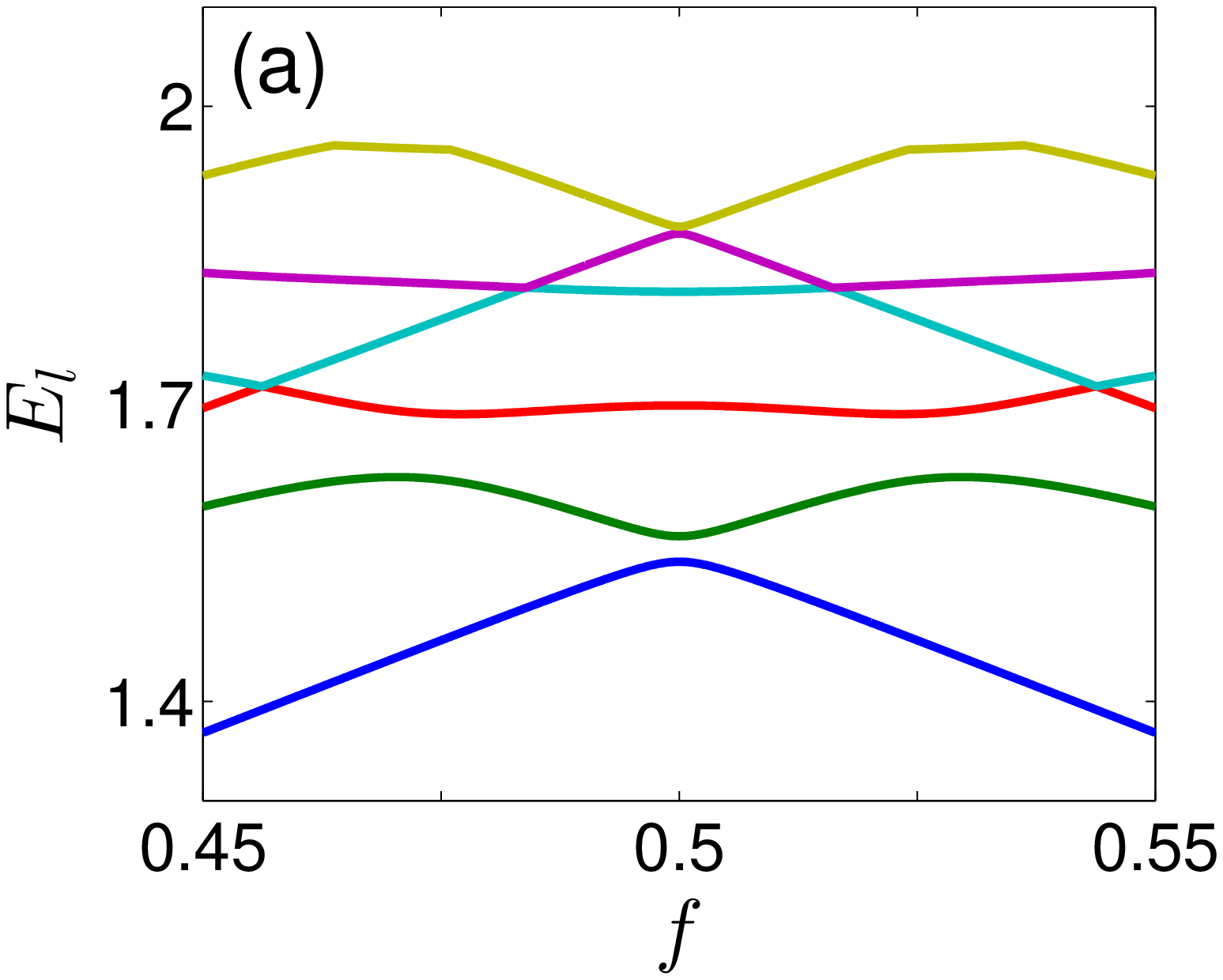}
\includegraphics[bb=30 210 510 600, width=4.2 cm, clip]{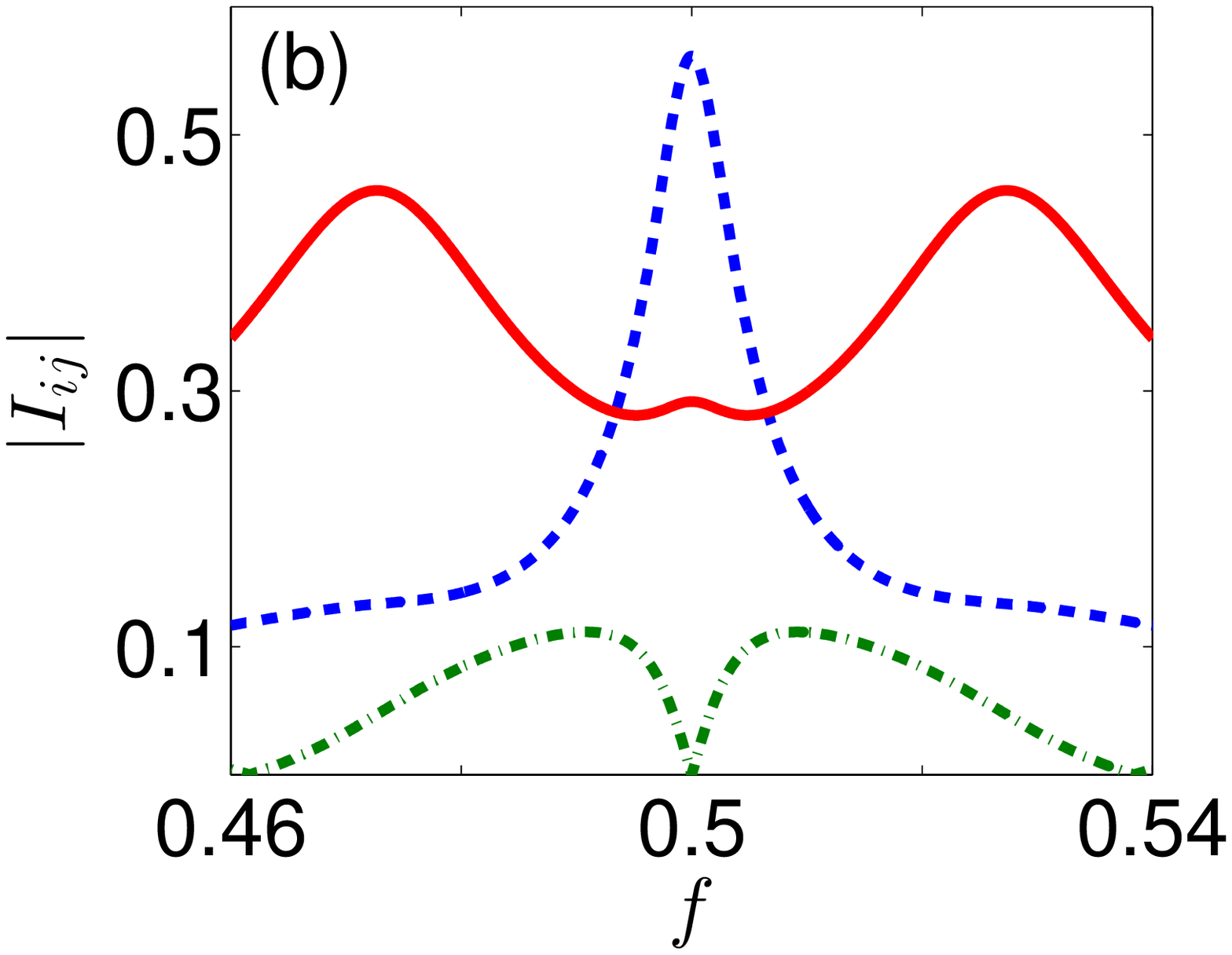} 
\includegraphics[bb=30 210 510 600, width=4.2 cm, clip]{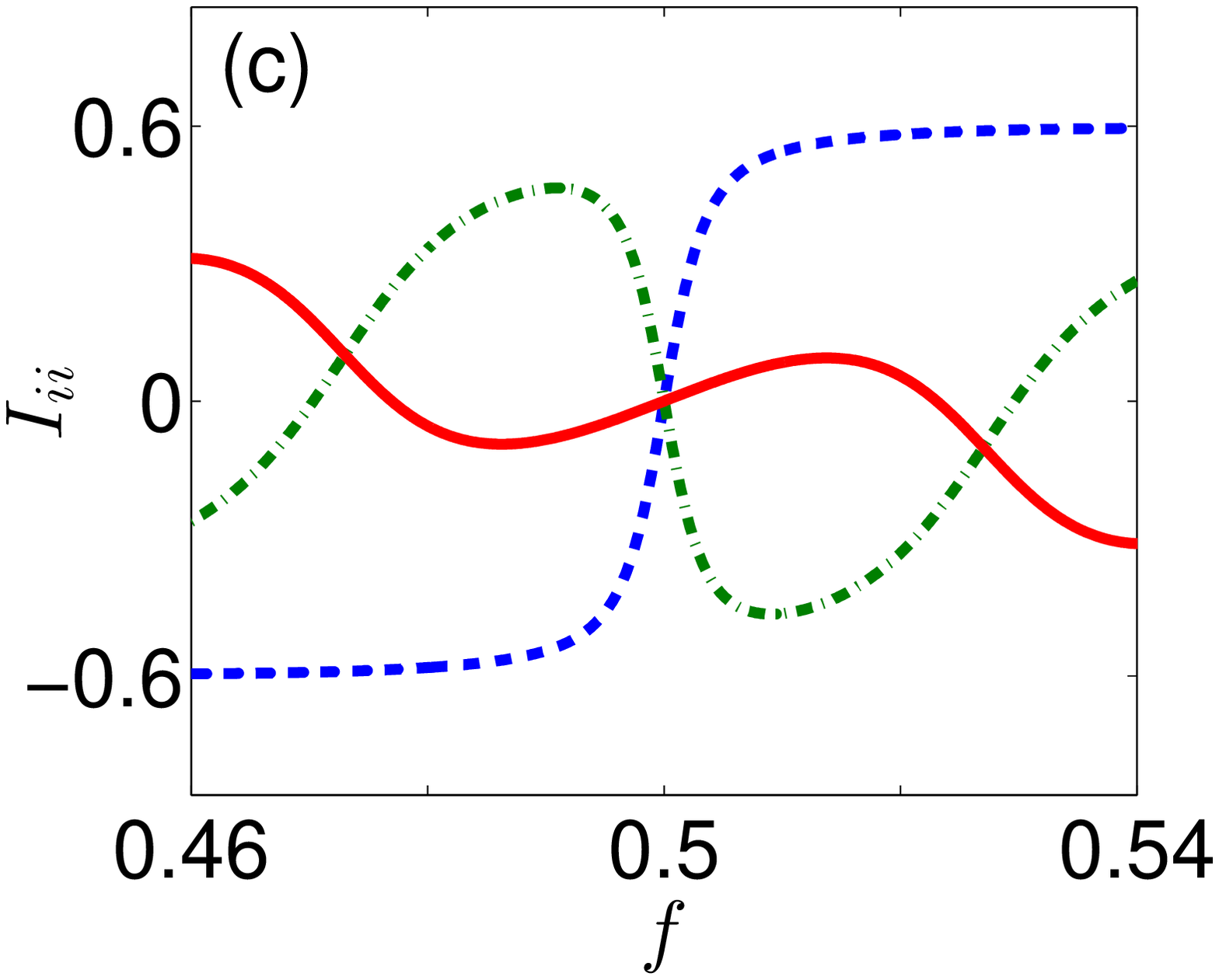}
\caption[]{(Color online)
(a) Eigenvalues $E_{l}$ of the SFQC versus the reduced magnetic flux $f$,
in units of $E_{\rm J}$, for the six lowest energy levels.
(b) Moduli of the loop current transition matrix elements $|I_{ij}|$ ($i<j$)
for the three lowest energy levels ($|0\rangle$, $|1\rangle$, and $|2\rangle$)
versus the reduced magnetic flux $f$ for $|I_{01}|$ (blue dashed curve),
$|I_{02}|$ (green dash-dotted curve), and $|I_{12}|$ (red solid curve).
(c) Loop current diagonal matrix elements $I_{ii}$
for the three lowest energy levels ($|0\rangle$, $|1\rangle$, and $|2\rangle$)
versus the reduced magnetic flux $f$ for $I_{00}$ (blue dashed curve),
$I_{11}$ (green dash-dotted curve), and $I_{22}$ (red solid curve).
In (b) and (c), the loop current matrix elements are in units of $I_{0}$.
Here we choose $\alpha=\Ea$ and $E_{\rm J}/E_{\rm c}=\Emu$.
}\label{fig2}
\end{figure}

We study a SFQC, as shown in Fig.~\ref{fig1}(a), consisting of three Josephson junctions
in a superconducting loop with negligible self-inductance.
Two junctions have equal size, each with Josephson energy $E_{\rm J}$ and capacitance $C_{\rm J}$.
The third one, which is smaller than the others,
has a Josephson energy $\alpha E_{\rm J}$ and capacitance $\alpha C_{\rm J}$, with $0.5<\alpha <1$.
The SFQC is threaded by a bias magnetic flux (dc) $\Phi_{e}$
and driven by a strong time-dependent magnetic flux (ac) $\Phi_{\rm D}(t)$.
A weak magnetic flux $\Phi_{\rm P}(t)$ as a probe field is also applied to the SFQC.
If the driving $\Phi_{\rm D}(t)$ and the probe $\Phi_{\rm P}(t)$ fields are not applied,
then the Hamiltonian of the SFQC with a dc bias $\Phi_{e}$ is written as (e.g., in Refs.~\cite{liu,orlando})
\begin{equation}\label{eq:1}
H_{0}=\frac{P_{p}^{2}}{2M_{p}} +\frac{P_{m}^{2}}{2M_{m}} +U(\varphi_{p},\varphi_{m}),
\end{equation}
with effective masses $M_{p}=2C_{\rm J}(\Phi_{0}/2\pi)^2$ and $M_{m}=M_{p}(1+2\alpha)$.
Here $\Phi_{0}$ is the flux quantum.
The quantum conjugate variables $\varphi_{p}$ and $\varphi_{m}$ of the effective momenta $P_{p}$ and $P_{m}$ are defined by
$\varphi_{p}=(\varphi_{1}+\varphi_{2})/2$ and $\varphi_{m}=(\varphi_{2}-\varphi_{1})/2$,
with the phase drops $\varphi_{1}$ and $\varphi _{2}$ across the two larger junctions.
The potential energy $U(\varphi_{p},\varphi_{m})$ is
\begin{align}\label{eq:2u}
U(\varphi_{p},\varphi_{m})
&= 2E_{\rm{J}}(1-\cos \varphi_{p}\cos \varphi_{m})
\nonumber\\
&+ \alpha E_{\rm J}\left[1-\cos \left(2\pi f+2\varphi_{m}\right)\right],
\end{align}
with the reduced magnetic flux $f=\Phi_{e}/\Phi_{0}$.
The bias magnetic flux $\Phi_{e}$ can be used to adjust
the shape of the potential energy between the symmetric double-well and the asymmetric double-well.
Thus, as shown in Fig.~\ref{fig2}(a), the eigenvalues $E_{l}$ of the SFQC can be
adjusted by $\Phi_{e}$ (or the reduced magnetic flux $f$).
In the basis of the eigenstates $|l\rangle$, corresponding to the $l$th eigenvalue $E_{l}$,
of the Hamiltonian $H_{0}$ in Eq.~(\ref{eq:1}), we can rewrite Eq.~(\ref{eq:1}) as
\begin{equation}\label{eq:2}
H_{0}=\sum_{l=0}^{N} E_{l}\,\sigma_{ll},
\end{equation}
with $\sigma_{ll}=|l\rangle\langle l|$.
As an example and for concreteness, in Fig.~\ref{fig2}(a),
the eigenvalues $E_{l}$ of the Hamiltonian in Eq.~(\ref{eq:1})
for the six lowest energy levels have been plotted as a function of
the reduced magnetic flux $f$ with $\alpha=\Ea$ and $E_{\rm J}/E_{\rm c}=\Emu$.
Here, the charging energy $E_{\rm c}=e^2/(2C_{\rm J})$.

\subsection{Hamiltonian of a driven and probed three-level system with its environment}

Let us now only consider the three lowest energy levels of the SFQC as in Refs.~\cite{liu,you-liu,you-liu1}.
That is, the free Hamiltonian of the SFQC can be given by Eq.~(\ref{eq:2}) with $l=0,\,1,\,2$.
As schematically shown in Fig.~\ref{fig1}(b), when the driving  $\Phi_{\rm D}(t)$
and the probe $\Phi_{\rm P}(t)$ fields are applied to the three-level SFQC,
the Hamiltonian of the three-level SFQC with its environment can be given by
\begin{align}\label{eq:4}
H_{\rm q}
&= \sum_{l=0}^{2} E_{l}\,\sigma_{ll} +\hbar\sum_{n}\omega^{\prime}_{n}b^{\dagger}_{n}b_{n} +H_{\rm I}.
\end{align}
The interaction Hamiltonian $H_{\rm I}$ is generally given by
\begin{align}\label{Hi}
H_{\rm I}
&= H_{\rm I, D} +H_{\rm I, P} +H_{\rm I, E}
\nonumber\\
&= -\Phi_{\rm D}(t)\hat{I} -\Phi_{\rm P}(t)\hat{I} -\hbar Q\hat{I},
\end{align}
where $\hbar Q$ represents the magnetic flux induced by the environment.
The average value of the operator $\hat{I}$ in Eq.~(\ref{Hi}) represents
the current through the SFQC loop, which is generally given as
\begin{equation}\label{Im}
\hat{I}=\sum_{i,j=0}^{2}I_{ij} \, \sigma_{ij},
\end{equation}
with $\sigma_{ij}=|i\rangle\langle j|$ in the basis of the three lowest eigenstates ($|0\rangle$, $|1\rangle$, and $|2\rangle$)
of the Hamiltonian in Eq.~(\ref{eq:1}).
Here, the matrix elements of the loop current operator $\hat{I}$ in Eq.~(\ref{Im}) are given as $I_{ij}=\langle i |\hat{I}_{g}|j\rangle$,
for the general definition~\cite{cxy} of the loop current operator
\begin{equation}\label{loopcurrent}
\hat{I}_{g}
=\frac{\alpha\,I_{0}}{1+2\alpha}
\left[ 2\cos\varphi_{p} \sin\varphi_{m} -\sin\left( 2\pi f +2\varphi_{m} \right) \right],
\end{equation}
with $I_{0}=2\pi E_{\rm J}/\Phi_{0}$.
For the completeness of the paper and further numerical discussions,
the moduli of the loop current transition matrix elements $|I_{ij}|$ ($i<j$) in Fig.~\ref{fig2}(b)
and the loop current diagonal matrix elements $I_{ii}$ in Fig.~\ref{fig2}(c) are plotted as a function of
the reduced magnetic flux $f$ for the three lowest energy levels,
with the same parameters as in Fig.~\ref{fig2}(a).
It is clear that these matrix elements can be controlled by the bias magnetic flux $\Phi_{e}$ (or saying $f$).

The environment is described by multimode harmonic oscillators,
each with the creation (annihilation) operator $b^{\dagger}_{n}$ ($b_{n}$) and frequency $\omega_{n}^{\prime}$.
The environmental variable is
\begin{equation}
Q=\sum_{n}\kappa_{n}(b^{\dagger}_{n}+b_{n}),
\end{equation}
where $\kappa_{n}$ characterizes the coupling between each mode of the environment and the three-level SFQC.

The driving field $\Phi_{\rm D}(t)$ is assumed to have frequency $\omega_{0}$ and amplitude $\Phi_{c}$,
resonantly or near resonantly applied to the two upper energy levels $|1\rangle$ and $|2\rangle$.
Thus, in the rotating-wave approximation (RWA), the driving-field-induced Hamiltonian
$H_{\rm I, D}$ in Eq.~(\ref{Hi}) can be given as
\begin{align}
H_{\rm I, D}=\hbar\Omega_{\rm D} \exp(i\omega_{0}t)\sigma_{12}
+\hbar\Omega_{\rm D}^{*} \exp(-i\omega_{0}t)\sigma_{21}.
\end{align}
The coupling constant $\Omega_{\rm D}$ between the driving field
$\Phi_{\rm D}(t)$ and the three-level SFQC is
\begin{equation}\label{eq:5}
\Omega_{\rm D}=-\frac{1}{2\hbar}\Phi_{c} \, I_{12},
\end{equation}
with $I_{12}$ given in Eq.~(\ref{Im}) for $i=1$ and $j=2$.
The Rabi frequency of the driving field is given by the modulus $|\Omega_{\rm D}|$ of the coupling constant.

In contrast to the driving field $\Phi_{\rm D}(t)$,
in this study, we assumed that the probe field $\Phi_{\rm P}(t)$
includes two components which can induce either the $|0\rangle\leftrightarrow|1\rangle$ transition
or the $|0\rangle\leftrightarrow|2\rangle$ transition.
Thus, under the RWA, the probe-field-induced Hamiltonian $H_{\rm I, P}$ in Eq.~(\ref{Hi}) can be given as
\begin{align}
H_{\rm I, P}= -\Phi_{\rm P}(t) \left(\sum_{i=1}^{2}I_{0i} \, \sigma_{0i} +{\rm h.c.}\right).
\end{align}

Based on the above discussions, the total Hamiltonian of the driven and probed three-level SFQC
with its environment can be given by
\begin{align}\label{Htotal}
H
&= \sum_{l=0}^{2} E_{l}\,\sigma_{ll} +\hbar\sum_{n}\omega^{\prime}_{n}b^{\dagger}_{n}b_{n}
\nonumber\\
&+ \hbar\Omega_{\rm D} \exp(i\omega_{0}t)\sigma_{12} +\hbar\Omega_{\rm D}^{*} \exp(-i\omega_{0}t)\sigma_{21}
\nonumber\\
&- \Phi_{\rm P}(t) \left(\sum_{i=1}^{2}I_{0i} \, \sigma_{0i} +{\rm h.c.}\right) -\hbar Q\hat{I},
\end{align}
with $\hat{I}$ given in Eq.~(\ref{Im}).

We emphasize that the Hamiltonian of the SFQC in Eq.~(\ref{eq:1}) has a well-defined parity
when the bias magnetic flux is at the optimal point, i.e., $f=0.5$.
In this case, the transition between the energy levels $|0\rangle$ and $|2\rangle$ is forbidden
and the selection rule in SFQCs is the same as in natural atoms~\cite{liu},
thus the three-level SFQC has a ladder-type transition,
and the probe field can only couple the energy levels $|0\rangle$ and $|1\rangle$.
However, when the bias magnetic flux deviates from the optimal point, i.e., $f\neq 0.5$,
the parity of the Hamiltonian in Eq.~(\ref{eq:1}) related to the variables $\varphi_{p}$ and $\varphi_{m}$ is broken
and any transition between two of the three lowest energy levels is possible.
In this case, the three-level SFQC has a cyclic transition~\cite{liu}
and the probe field can couple the energy levels $|0\rangle$ and $|1\rangle$,
as well as the energy levels $|0\rangle$ and $|2\rangle$.
This is a very important difference between three-level SFQCs and three-level natural atoms.

\subsection{Magnetic susceptibility of three-level SFQCs}

The linear response of the three-level SFQC to the probe field
can be characterized by the linear magnetic susceptibility.
By taking the same arguments and calculation method as in Ref.~\cite{smirnov3},
the linear magnetic susceptibility $\chi_{q}(\omega)$ of the three-level SFQC to the probe field
$\Phi_{\rm P}(t)$ can be obtained via the following Fourier transform
\begin{equation}\label{eq:14}
\left\langle \frac{\delta \hat{I}(t)}{\delta \Phi_{\rm P}(t_{1})}\right\rangle
= \int \frac{d\omega}{2\pi}\chi_{q}(\omega) \exp[-i\omega(t-t_{1})].
\end{equation}
Note that in this paper we use the integral sign $\int$
to denote the integration over the whole real axes, i.e., $\int \equiv \int_{-\infty}^{\infty}$.
The real and imaginary parts of the magnetic susceptibility $\chi_{q}(\omega)$
are used to characterize, respectively, the dispersion and the absorption of the probe field by the three-level SFQC.
The time-dependent loop current operator $\hat{I}(t)$ in Eq.~(\ref{eq:14})
can be expressed via Eq.~(\ref{Im}) as
\begin{equation}\label{eq:12}
\hat{I}(t)=\sum_{i,j =0}^{2}I_{ij} \, \sigma_{ij}(t).
\end{equation}
Note that the difference of the susceptibility between the three-level SFQCs in Eq.~(\ref{eq:14})
and the superconducting qubits in Ref.~\cite{smirnov3}
only involves replacing the loop current operator of
the superconducting qubits by that of the three-level SFQCs.
However, this simple replacement will result in many new results, as discussed below.

\section{Equations of motion and solutions of operators}\label{ES}

\subsection{Heisenberg equations and correlation functions}

To obtain the magnetic susceptibility $\chi_{q}(\omega)$ of the three-level SFQC,
we now have to obtain the solution of the loop current operator $\hat{I}(t)$
by solving the equations of motion for all operators $\sigma_{lm}(t)$.
In the rotating reference frame for the Hamiltonian
in Eq.~(\ref{Htotal}) with the unitary transform
\begin{equation}\label{eq:15}
U(t)=\exp\left( -i\omega_{1}\sigma_{11} \,t -i\omega^{\prime}\sigma_{22} \,t \right),
\end{equation}
the Heisenberg equation of motion for any operator $\sigma_{lm}(t)$ of the driven and probed three-level SFQC,
interacting with the environment, can be given by
\begin{align}\label{eq:16}
\frac{\partial \sigma_{lm}}{\partial t}
&= -i\Delta(\sigma_{l2}\delta_{m2}-\sigma_{2m}\delta_{2l})
\nonumber\\
&- i\Omega_{\rm D}(\sigma_{l2}\delta_{m1}-\sigma_{1m}\delta_{2l})
-i\Omega_{\rm D}^{*}(\sigma_{l1}\delta_{m2}-\sigma_{2m}\delta_{1l})
\nonumber\\
&+ \frac{i}{\hbar}(\sigma_{l0}\delta_{m1}-\sigma_{1m}\delta_{0l})
[\hbar Q+\Phi_{\rm P}(t)]I_{10} e^{i\omega_{1}t}
\nonumber\\
&+ \frac{i}{\hbar}(\sigma_{l1}\delta_{m0}-\sigma_{0m}\delta_{1l})
[\hbar Q+\Phi_{\rm P}(t)]I_{01} e^{-i\omega_{1}t}
\nonumber\\
&+ \frac{i}{\hbar}(\sigma_{l0}\delta_{m2}-\sigma_{2m}\delta_{0l})
[\hbar Q+\Phi_{\rm P}(t)]I_{20} e^{i\omega^{\prime}t}
\nonumber\\
&+ \frac{i}{\hbar}(\sigma_{l2}\delta_{m0}-\sigma_{0m}\delta_{2l})
[\hbar Q+\Phi_{\rm P}(t)]I_{02} e^{-i\omega^{\prime}t}
\nonumber\\
&+ \frac{i}{\hbar}(\sigma_{l1}\delta_{m2}-\sigma_{2m}\delta_{1l})
\hbar Q I_{21} e^{i\omega_{0}t}
\nonumber\\
&+ \frac{i}{\hbar}(\sigma_{l2}\delta_{m1}-\sigma_{1m}\delta_{2l})
\hbar Q I_{12} e^{-i\omega_{0}t}
\nonumber\\
&+ \frac{i}{\hbar}(\sigma_{l0}\delta_{m0}-\sigma_{0m}\delta_{0l}) \hbar Q I_{00}
\nonumber\\
&+ \frac{i}{\hbar}(\sigma_{l1}\delta_{m1}-\sigma_{1m}\delta_{1l}) \hbar Q I_{11}
\nonumber\\
&+ \frac{i}{\hbar}(\sigma_{l2}\delta_{m2}-\sigma_{2m}\delta_{2l}) \hbar Q I_{22},
\end{align}
with $\omega^{\prime}=\omega_{0}+\omega_{1}=\omega_{2}-\Delta$.
Here, $\Delta=\omega_{3}-\omega_{0}$ is the detuning between the frequency $\omega_{0}$ of the driving field
and the transition frequency $\omega_{3}$ from the energy level $|2\rangle$ to the energy level $|1\rangle$.
The transition frequencies $\omega_{i}$ ($i=1,\,2,\,3$) of the three-level SFQC are defined as
$\omega_{1}=(E_{1}-E_{0})/\hbar$, $\omega_{2}=(E_{2}-E_{0})/\hbar$, and $\omega_{3}=(E_{2}-E_{1})/\hbar$.
In the derivation of Eq.~(\ref{eq:16}), we have used the completeness relation
$|0\rangle\langle 0|+|1\rangle\langle 1|+|2\rangle\langle 2|=1$ for the three-level SFQC and neglected the constant term.
In Eq.~(\ref{eq:16}), all operators are Heisenberg operators, e.g.,
$\sigma_{lm}\equiv \sigma_{lm}(t)$ and also $Q\equiv Q(t)$.

The environment has an infinite number of degrees of freedom,
thus it is usually considered as a macroscopic system and assumed in a thermodynamic equilibrium state.
If the free environmental variable $Q^{(0)}(t)$
\begin{equation}\label{eq:15-1}
Q^{(0)}(t)=\sum_{n}\kappa_{n} \left[b^{\dagger}_{n}(0)e^{i\omega_{n}^{\prime}t} +b_{n}(0)e^{-i\omega_{n}^{\prime}t}\right]
\end{equation}
obeys Gaussian fluctuations or the interaction between the three-level SFQC and its environment is weak~\cite{efremov},
the environmental operator $Q(t)$ has the following solution
\begin{equation}
Q(t)=Q^{(0)}(t)+\int_{-\infty}^{t} \!\!\! dt^{\prime} \, \phi(t,t^{\prime}) \, \hat{I}_{r}(t^{\prime}),
\end{equation}
with the time-dependent loop current operator in the rotating reference frame
\begin{align}\label{eq:21}
\hat{I}_{r}(t^{\prime})
&= \left[ I_{01}\sigma_{01}(t^{\prime})e^{-i\omega_{1}t^{\prime}}
+I_{02}\sigma_{02}(t^{\prime})e^{-i\omega^{\prime}t^{\prime}} \right.
\nonumber\\
&+ \left. I_{12}\sigma_{12}(t^{\prime})e^{-i\omega_{0}t^{\prime}}+{\rm h.c.} \right]
+\sum_{i=0}^{2}I_{ii}\sigma_{ii}(t^{\prime}).
\end{align}
Note that $b^{\dagger}_{n}(0)$ and $b_{n}(0)$ in Eq.~(\ref{eq:15-1})
are the creation and annihilation operators of the $n$th environmental bosonic mode at the initial time.
The environmental linear response function $\phi(t,t^{\prime})$ is defined as the correlation
function of the free environmental variables averaged over the environmental equilibrium state
\begin{equation}\label{CrPhi}
\phi(t,t^{\prime})=\langle i[Q^{(0)}(t), Q^{(0)}(t^{\prime})]_{-} \rangle_{\rm E} \,\theta(t-t^{\prime}),
\end{equation}
where the Heaviside step function $\theta(t-t^{\prime})=1$ for $t-t^{\prime}>0$;
but $\theta(t-t^{\prime})=0$ for $t-t^{\prime}<0$.
The environmental susceptibility $\chi(\omega)$
is defined~\cite{efremov} by the Fourier transform of the response function $\phi(t,t^{\prime})$,
\begin{equation}\label{Xw}
\chi(\omega) =\chi^{\prime}(\omega)+i\chi^{\prime\prime}(\omega)
=\int d\tau \, \phi(\tau) \, \exp(i\omega\tau),
\end{equation}
with the time interval $\tau=t-t^{\prime}$ and
\begin{equation}\label{XwRI}
\chi^{\prime}(\omega)=-\frac{1}{\pi} \; \wp \!\! \int d\omega^{\prime} \, \frac{\chi^{\prime\prime}(\omega^{\prime})}{\omega-\omega^{\prime}}.
\end{equation}
Here $\wp$ stands for the Cauchy principal value.

Using the method developed in Refs.~\cite{efremov,smirnov1,smirnov2,smirnov3},
the product operators, e.g., $\sigma_{lm}(t)Q(t)$ in Eq.~(\ref{eq:16}), can be written as
\begin{align}\label{eq:22}
\sigma_{lm}(t)Q(t)
&= \frac{1}{2}\int dt^{\prime}\phi(t,t^{\prime})
\left[\sigma_{lm}(t), \hat{I}_{r}(t^{\prime}) \right]_{+} +\xi_{lm}(t)
\nonumber\\
&+ i\int dt^{\prime}\widetilde{M}(t,t^{\prime})
\left[\sigma_{lm}(t),\hat{I}_{r}(t^{\prime}) \right]_{-}.
\end{align}
Here, the upper bound of the integral in Eq.~(\ref{eq:22}) has been extended from the time $t$ to $\infty$
by using the step function $\theta(t-t^{\prime})=0$ for $t^{\prime} > t$.
The fluctuation force $\xi_{lm}(t)$ in Eq.~(\ref{eq:22}),
which has zero average value over the environmental equilibrium state, is expressed as
\begin{align}\label{eq:25}
\xi_{lm}(t)
&= \frac{1}{2}\left[\sigma_{lm}(t),Q^{(0)}(t)\right]_{+}
\nonumber\\
&- i\int dt^{\prime} \widetilde{M}(t,t^{\prime})
\left[ \sigma_{lm}(t), \hat{I}_{r}(t^{\prime}) \right]_{-},
\end{align}
with the causal correlation function $\widetilde{M}(t,t^{\prime})=M(t,t^{\prime})\theta(t-t^{\prime})$.
The symmetrized correlation function $M(t,t^{\prime})$
of the free environmental variable $Q^{(0)}(t)$ is defined by
\begin{equation}\label{CrM}
M(t,t^{\prime}) =\frac{1}{2}\langle [Q^{(0)}(t),Q^{(0)}(t^{\prime})]_{+}\rangle_{\rm E},
\end{equation}
with the average over the environmental equilibrium state.
Here, we need to mention that the Born approximation is made
when we take the average over the environmental equilibrium state in Eqs.~(\ref{CrPhi}) and (\ref{CrM}).
That is, the states of the three-level SFQC and the environment are factorized at any time
and the environment is always assumed in its equilibrium state.
The spectral density $S(\omega)$ is defined~\cite{efremov} by the Fourier transform of the correlation function $M(t,t^{\prime})$,
\begin{equation}\label{Sw}
S(\omega)=\int d\tau\, M(\tau) \, \exp(i\omega\tau).
\end{equation}
The Fourier transform $\widetilde{S}(\omega)$ of the correlation function $\widetilde{M}(\tau)$ can be given as~\cite{efremov}
\begin{equation}\label{SwT}
\widetilde{S}(\omega)=\frac{1}{2}\left[S(\omega)+iS^{\prime}(\omega)\right],
\end{equation}
with
\begin{equation}\label{SwRI}
S^{\prime}(\omega)=\frac{1}{\pi} \; \wp \!\! \int d\omega^{\prime}\,\frac{S(\omega^{\prime})}{\omega-\omega^{\prime}}.
\end{equation}

\subsection{Quantum Langevin equations}

Using Eqs.~(\ref{eq:16}), (\ref{eq:21}), and (\ref{eq:22}),
we can rewrite the Heisenberg equation in Eq.~(\ref{eq:16}) into a quantum Langevin equation.
For example, the operator $\sigma_{01}(t)$ obeys the following quantum Langevin equation
\begin{align}\label{eq:L-29}
\frac{\partial\sigma_{01}(t)}{\partial t}
& =i\frac{|I_{01}|^2}{2}\int dt^{\prime}
[S^{(+)}_{00,01}(t,t^{\prime})+iS^{(-)}_{00,01}(t,t^{\prime})]e^{i\omega_{1}\tau}
\nonumber\\
& -i\frac{|I_{01}|^2}{2}\int dt^{\prime}
[S^{(+)}_{11,01}(t,t^{\prime})+iS^{(-)}_{11,01}(t,t^{\prime})]e^{i\omega_{1}\tau}
\nonumber\\
& -i\frac{|I_{02}|^2}{2}\int dt^{\prime}
[S^{(+)}_{21,02}(t,t^{\prime})+iS^{(-)}_{21,02}(t,t^{\prime})]e^{i\omega^{\prime}\tau}
\nonumber\\
& +i\frac{|I_{12}|^2}{2}\int dt^{\prime}
[S^{(+)}_{02,21}(t,t^{\prime})+iS^{(-)}_{02,21}(t,t^{\prime})]e^{-i\omega_{0}\tau}
\nonumber\\
& -i\frac{(I_{00}-I_{11})I_{00}}{2}\int dt^{\prime}
[S^{(+)}_{01,00}(t,t^{\prime})+iS^{(-)}_{01,00}(t,t^{\prime})]
\nonumber\\
& -i\frac{(I_{00}-I_{11})I_{11}}{2}\int dt^{\prime}
[S^{(+)}_{01,11}(t,t^{\prime})+iS^{(-)}_{01,11}(t,t^{\prime})]
\nonumber\\
& -i\frac{(I_{00}-I_{11})I_{22}}{2}\int dt^{\prime}
[S^{(+)}_{01,22}(t,t^{\prime})+iS^{(-)}_{01,22}(t,t^{\prime})]
\nonumber\\
& -i\Omega_{\rm D}\sigma_{02}(t) +\frac{i}{\hbar}\Phi_{\rm P}(t)
I_{10}[\sigma_{00}(t)-\sigma_{11}(t)]e^{i\omega_{1}t}
\nonumber\\
& -\frac{i}{\hbar}\Phi_{\rm P}(t) I_{20}\sigma_{21}(t)e^{i\omega^{\prime}t}
+iI_{10}[\xi_{00}(t)-\xi_{11}(t)]e^{i\omega_{1}t}
\nonumber\\
& -iI_{20}\xi_{21}(t)e^{i\omega^{\prime}t} +iI_{12}\xi_{02}(t)e^{-i\omega_{0}t}
\nonumber\\
& -i(I_{00}-I_{11})\xi_{01}(t).
\end{align}
The commutators $S^{(+)}_{ij,lm}(t,t^{\prime})$ and
anti-commutators $S^{(-)}_{ij,lm}(t,t^{\prime})$ in Eq.~(\ref{eq:L-29}) are defined as
\begin{align}
S^{(+)}_{ij,lm}(t,t^{\prime})
& =\varphi(t,t^{\prime})[\sigma_{ij}(t),\,\sigma_{lm}(t^{\prime})]_{+},\label{cmtor}
\\
S^{(-)}_{ij,lm}(t,t^{\prime})
& =2\widetilde{M}(t,t^{\prime})[\sigma_{ij}(t),\,\sigma_{lm}(t^{\prime})]_{-}.\label{acmtor}
\end{align}
The fluctuation forces $\xi_{lm}(t)$ in Eq.~(\ref{eq:L-29}) can be given by Eq.~(\ref{eq:25}).
However, we are only interested in
the average dynamics of the three-level SFQC operators in our following discussions
and not interested in the correlation of the fluctuation forces;
thus, hereafter we average Eq.~(\ref{eq:L-29}) over the environmental equilibrium state,
and the fluctuation forces $\xi_{lm}(t)$ become zero.
Under the integral in Eq.~(\ref{eq:L-29}), we only keep the terms with exponential factors including the time interval $\tau$,
e.g., $\exp(-i\omega_{0} \tau)$, and the terms without exponential factors.
Other fast-oscillating terms with exponential factors, e.g., $\exp(i\omega_{1}t-i\omega_{0}t^{\prime})$,
have been neglected under the integral in Eq.~(\ref{eq:L-29})
because the contributions of these terms are negligibly small.

To obtain each integral in Eq.~(\ref{eq:L-29}), we have to first calculate
the commutators $S^{(+)}_{ij,lm}(t,t^{\prime})$ and anti-commutators $S^{(-)}_{ij,lm}(t,t^{\prime})$
with expressions $[\sigma_{ij}(t), \sigma_{lm}(t^{\prime})]_{\pm}$
in Eqs.~(\ref{cmtor}) and (\ref{acmtor}).
With the assumptions that the coupling between the three-level SFQC and its environment is weak,
and that the environmental correlation time $\tau_{c}$ is very small,
the relaxation of the three-level SFQC is negligible during the environmental correlation time $\tau_{c}$.
Thus the Bloch-Redfield approximation can be applied
and the time-dependent operators $\sigma_{lm}(t^{\prime})$ in Eqs.~(\ref{cmtor}) and (\ref{acmtor})
can be approximately obtained by neglecting the interaction between the three-level SFQC and its environment.
In this case, the operators $\sigma_{lm}(t^{\prime})$ can be easily expressed
in terms of the operators at the moment $t$ and the time interval $\tau$.
For example, the operator $\sigma_{01}(t^{\prime})$ can be expressed as
\begin{equation}\label{eq:35}
\sigma_{01}(t^{\prime})
=e^{i\frac{\Delta}{2}\tau} A^{*}(\tau) \sigma_{01}(t)
+ie^{i\frac{\Delta}{2}\tau} B(\tau) \sigma_{02}(t).
\end{equation}
Here
\begin{align}
A(\tau) &= \cos\left(\frac{\Omega\tau}{2}\right)
+i\sin\left(\frac{\Omega \tau}{2}\right)\cos2\theta,
\\
B(\tau) &= \nu \sin\left(\frac{\Omega \tau}{2}\right)\sin 2\theta,
\end{align}
with $\Omega=\sqrt{\Delta^2+4|\Omega_{\rm D}|^2}$ and $\nu=\Omega_{\rm D}/|\Omega_{\rm D}|$.
The detailed calculations on the operators $\sigma_{lm}(t^{\prime})$ are given in Appendix~\ref{ApCA}.
Using the relation of simultaneous operators
$\sigma_{ij}(t)\sigma_{lm}(t)=\sigma_{im}(t)\delta_{jl}$,
the commutators and anti-commutators in Eqs.~(\ref{cmtor}) and (\ref{acmtor}) can be
expressed by the operators at the moment $t$ and the time interval $\tau$.
Thus, substituting the operator relations (see Appendix~\ref{ApCA})
as in Eq.~(\ref{eq:35}) into Eq.~(\ref{eq:L-29}), and integrating over the time interval $\tau$,
the quantum Langevin equation Eq.~(\ref{eq:L-29}) can be simplified.

Based on above discussions, the quantum Langevin equations of other operators
can also be derived by using the same approach as for Eq.~(\ref{eq:L-29}).
All of the quantum Langevin equations related to
the operators $\sigma_{lm}(t)$ after averaging over the environmental equilibrium state are given below:
\begin{align}\label{SL01}
\frac{\partial \sigma_{01}(t)}{\partial t}
&= i\Gamma_{11}\sigma_{01}(t) +i(\Gamma_{12}-\Omega_{\rm D})\sigma_{02}(t)
\nonumber\\
& +\frac{i}{\hbar}\Phi_{\rm P}(t) I_{10}[\sigma_{00}(t)-\sigma_{11}(t)]e^{i\omega_{1}t}
\nonumber\\
& -\frac{i}{\hbar}\Phi_{\rm P}(t) I_{20}\sigma_{21}(t)e^{i\omega^{\prime}t},
\end{align}
\begin{align}\label{SL02}
\frac{\partial \sigma_{02}(t)}{\partial t}
&= i(\Gamma_{21}-\Omega_{\rm D}^{*})\sigma_{01}(t) +i(\Gamma_{22}-\Delta)\sigma_{02}(t)
\nonumber\\
& +\frac{i}{\hbar}\Phi_{\rm P}(t) I_{20}[\sigma_{00}(t)-\sigma_{22}(t)]e^{i\omega^{\prime}t}
\nonumber\\
& -\frac{i}{\hbar}\Phi_{\rm P}(t) I_{10}\sigma_{12}(t)e^{i\omega_{1}t},
\end{align}
\begin{align}\label{SL12}
\frac{\partial \sigma_{12}(t)}{\partial t}
&= i(\Gamma_{31}-\Delta)\sigma_{12}(t) +i\Gamma_{32}\sigma_{21}(t) +i\Gamma_{33}\sigma_{00}(t)
\nonumber\\
& +i(\Gamma_{34}-\Omega^{*}_{\rm D})\sigma_{11}(t) +i(\Gamma_{35}+\Omega^{*}_{\rm D})\sigma_{22}(t)
\nonumber\\
& +\frac{i}{\hbar}\Phi_{\rm P}(t)
[I_{20}\sigma_{10}(t)e^{i\omega^{\prime}t} -I_{01}\sigma_{02}(t)e^{-i\omega_{1}t}],
\end{align}
\begin{align}\label{SL00}
\frac{\partial \sigma_{00}(t)}{\partial t}
&= i\Gamma_{41}\sigma_{00}(t) +i\Gamma_{42}\sigma_{11}(t) +i\Gamma_{43}\sigma_{22}(t)
\nonumber\\
& +i\Gamma_{44}\sigma_{12}(t) +i\Gamma_{45}\sigma_{21}(t)
\nonumber\\
& +\frac{i}{\hbar}\Phi_{\rm P}(t) I_{01}\sigma_{01}(t)e^{-i\omega_{1}t}
-\frac{i}{\hbar}\Phi_{\rm P}(t) I_{10}\sigma_{10}(t)e^{i\omega_{1}t}
\nonumber\\
& +\frac{i}{\hbar}\Phi_{\rm P}(t) I_{02}\sigma_{02}(t)e^{-i\omega^{\prime}t}
-\frac{i}{\hbar}\Phi_{\rm P}(t) I_{20}\sigma_{20}(t)e^{i\omega^{\prime}t},
\end{align}
\begin{align}\label{SL11}
\frac{\partial \sigma_{11}(t)}{\partial t}
&= i\Gamma_{51}\sigma_{00}(t) +i\Gamma_{52}\sigma_{11}(t) +i\Gamma_{53}\sigma_{22}(t)
\nonumber\\
& +i(\Gamma_{54}-\Omega_{\rm D})\sigma_{12}(t) +i(\Gamma_{55}+\Omega^{*}_{\rm D})\sigma_{21}(t)
\nonumber\\
& -\frac{i}{\hbar}\Phi_{\rm P}(t) I_{01}\sigma_{01}(t)e^{-i\omega_{1}t}
+\frac{i}{\hbar}\Phi_{\rm P}(t) I_{10}\sigma_{10}(t)e^{i\omega_{1}t}
\nonumber\\
& +\frac{i}{\hbar}\Phi_{\rm P}(t) I_{12}\sigma_{12}(t)e^{-i\omega_{0}t}
-\frac{i}{\hbar}\Phi_{\rm P}(t) I_{21}\sigma_{21}(t)e^{i\omega_{0}t},
\end{align}
\begin{align}\label{SL22}
\frac{\partial \sigma_{22}(t)}{\partial t}
&= i\Gamma_{61}\sigma_{00}(t) +i\Gamma_{62}\sigma_{11}(t) +i\Gamma_{63}\sigma_{22}(t)
\nonumber\\
& +i(\Gamma_{64}+\Omega_{\rm D})\sigma_{12}(t) +i(\Gamma_{65}-\Omega^{*}_{\rm D})\sigma_{21}(t)
\nonumber\\
& -\frac{i}{\hbar}\Phi_{\rm P}(t) I_{02}\sigma_{02}(t)e^{-i\omega^{\prime}t}
+\frac{i}{\hbar}\Phi_{\rm P}(t) I_{20}\sigma_{20}(t)e^{i\omega^{\prime}t}
\nonumber\\
& -\frac{i}{\hbar}\Phi_{\rm P}(t) I_{12}\sigma_{12}(t)e^{-i\omega_{0}t}
+\frac{i}{\hbar}\Phi_{\rm P}(t) I_{21}\sigma_{21}(t)e^{i\omega_{0}t}.
\end{align}
The expressions for the complex coefficients $\Gamma_{lm}$ ($l,\,m=1,\,2$)
in Eqs.~(\ref{SL01}) and (\ref{SL02}) are given in Appendix~\ref{ApGM}.
The expressions for the complex coefficients $\Gamma_{lm}$
in Eqs.~(\ref{SL12})--(\ref{SL22}) are not given because they are not used in the following calculations.

\subsection{Steady-state values}

We are interested in the linear magnetic susceptibility of the driven three-level SFQC in the steady state;
therefore we need to obtain the steady-state values
and the probe-field-dependent average values of the operators $\sigma_{lm}$ of the three-level SFQC.
The steady-state values $\langle\sigma^{(s)}_{lm}\rangle$
can be obtained via Eqs.~(\ref{SL01})--(\ref{SL22}) by setting $\partial\sigma_{lm}(t)/\partial t=0$, $\Phi_{\rm P}(t)=0$,
and averaging over the initial state of the three-level SFQC.
From Eqs.~(\ref{SL01}) and (\ref{SL02}), we obtain that
the steady-state values $\langle\sigma^{(s)}_{01}\rangle$ and $\langle\sigma^{(s)}_{02}\rangle$ are zero, i.e.,
\begin{equation}\label{s0102}
\langle\sigma^{(s)}_{01}\rangle=\langle\sigma^{(s)}_{02}\rangle=0.
\end{equation}

The steady-state values $\langle\sigma^{(s)}_{12}\rangle$ and $\langle\sigma^{(s)}_{ll}\rangle$ ($l=0,\,1,\,2$)
can be obtained via Eqs.~(\ref{SL12})--(\ref{SL22}).
However, in the usual experiments with SFQCs,
the condition $k_{\rm B}T \ll |E_{i}-E_{j}|$ ($i,\,j=0,\,1,\,2,\,i\neq j$) is fulfilled.
Thus, the population of the energy levels $|1\rangle$ and $|2\rangle$ due to thermal excitations can be neglected,
and the steady-state values $\langle\sigma^{(s)}_{12}\rangle$ and $\langle\sigma^{(s)}_{ll}\rangle$ ($l=0,\,1,\,2$)
can be approximately given as
\begin{align}\label{note}
\langle\sigma^{(s)}_{12}\rangle \approx \langle\sigma^{(s)}_{11}\rangle
\approx \langle\sigma^{(s)}_{22}\rangle \approx0,\,\,\, {\rm and} \,\,\,\langle\sigma^{(s)}_{00}\rangle \approx 1.
\end{align}
By numerically solving Eqs.~(\ref{SL12})--(\ref{SL22}) (not shown in this paper),
we find that $|\langle\sigma^{(s)}_{12}\rangle|<10^{-2}$ and $\langle\sigma^{(s)}_{ll}\rangle<3 \times 10^{-2}$ ($l=1,\,2$)
with the parameters used in Sec.~\ref{NUM} for numerical calculations.
Thus the approximation in Eq.~(\ref{note}) is reasonable.

\subsection{Formal solution of quantum Langevin equations}

The time-dependent average values $\langle\sigma_{lm}(t)\rangle$
can be obtained by solving Eqs.~(\ref{SL01})--(\ref{SL22}) using the Fourier transform.
The solutions are only calculated to first order in $\Phi_{\rm P}(t)$.
Therefore, here first Eqs.~(\ref{SL01})--(\ref{SL22}) are
averaged over the initial state of the three-level SFQC,
and afterwards, in the terms including $\Phi_{\rm P}(t)$,
all the average values $\langle\sigma_{lm}(t)\rangle$
are replaced by their steady-state values $\langle\sigma_{lm}^{(s)}\rangle$.

As shown in Eq.~(\ref{s0102}), the steady-state values $\langle\sigma^{(s)}_{01}\rangle$
and $\langle\sigma^{(s)}_{02}\rangle$ are zero,
thus the time-dependent average values $\langle\sigma_{12}(t)\rangle$ and $\langle\sigma_{ll}(t)\rangle$ ($l=0,\,1,\,2$)
are independent of $\Phi_{\rm P}(t)$ when calculated to first order in $\Phi_{\rm P}(t)$,
and do not affect the linear magnetic susceptibility of the three-level SFQC.
Therefore, we only need to calculate the time-dependent average values $\langle\sigma_{01}(t)\rangle$
and $\langle\sigma_{02}(t)\rangle$.
The formal solutions of the average values $\langle\sigma_{01}(t)\rangle$
and $\langle\sigma_{02}(t)\rangle$ are expressed as
\begin{align}
\label{FS01}
\langle\sigma_{01}(t)\rangle
&= i\frac{I_{10}}{\hbar} \int dt^{\prime} G_{22}(\tau)
\Phi_{\rm P}(t^{\prime}) e^{i\omega_{1}t^{\prime}}
\nonumber\\
&- i\frac{I_{20}}{\hbar} \int dt^{\prime} G_{12}(\tau)
\Phi_{\rm P}(t^{\prime})e^{i\omega^{\prime} t^{\prime}}
\nonumber\\
&+ G_{22}(t)\langle\sigma_{01}(0)\rangle -G_{12}(t)\langle\sigma_{02}(0)\rangle,
\\
\label{FS02}
\langle\sigma_{02}(t)\rangle
&= i\frac{I_{20}}{\hbar}\int dt^{\prime} G_{11}(\tau)
\Phi_{\rm P}(t^{\prime})e^{i\omega^{\prime}t^{\prime}},
\nonumber\\
&- i\frac{I_{10}}{\hbar}\int dt^{\prime} G_{21}(\tau)
\Phi_{\rm P}(t^{\prime})e^{i\omega_{1}t^{\prime}}
\nonumber\\
&- G_{21}(t)\langle\sigma_{01}(0)\rangle +G_{22}(t)\langle\sigma_{02}(0)\rangle.
\end{align}

The Fourier transforms of the Green functions $G_{lm}(\tau)$ in Eqs.~(\ref{FS01}) and (\ref{FS02}),
are given by
\begin{align}
\label{G11}
G_{11}(\omega)
&= \int d\tau\, G_{11}(\tau)\, e^{i\omega\tau}
= \frac{i \left(\omega+\Gamma_{11}\right)}{D(\omega)},
\\
\label{G12}
G_{12}(\omega)
&= \int d\tau\, G_{12}(\tau)\, e^{i\omega\tau}
= \frac{i \left(\Gamma_{12}-\Omega_{\rm D}\right)}{D(\omega)},
\\
\label{G21}
G_{21}(\omega)
&= \int d\tau\, G_{21}(\tau)\, e^{i\omega\tau}
= \frac{i \left(\Gamma_{21}-\Omega_{\rm D}^{*}\right)}{D(\omega)},
\\
\label{G22}
G_{22}(\omega)
&= \int d\tau\, G_{22}(\tau)\, e^{i\omega\tau}
= \frac{i \left(\omega-\Delta+\Gamma_{22}\right)}{D(\omega)},
\end{align}
with the denominator
\begin{align}
D(\omega)
&= -\left(\omega+\Gamma_{11}\right) \left(\omega-\Delta+\Gamma_{22}\right)
\nonumber\\
& +\left(\Gamma_{12}-\Omega_{\rm D}\right) \left(\Gamma_{21}-\Omega_{\rm D}^{*}\right).
\end{align}

\subsection{Discussions}

The complex coefficients $\Gamma_{lm}$ in Eqs.~(\ref{SL01})--(\ref{SL22})
incorporate the effects of the environment on the three-level SFQC.
The real parts of $\Gamma_{lm}$ represent the Lamb frequency shifts of the three-level SFQC,
while the imaginary parts of $\Gamma_{lm}$ represent the damping rates of the three-level SFQC.

We can further simplify $\Gamma_{lm}$ via the fluctuation-dissipation theorem.
According to the fluctuation-dissipation theorem, the spectral density $S(\omega)$ in Eq.~(\ref{Sw})
and the imaginary part of the environmental susceptibility $\chi^{\prime\prime}(\omega)$ in Eq.~(\ref{Xw})
satisfy the following relation
\begin{equation}\label{FDT}
S(\omega)=\chi^{\prime\prime}(\omega)\coth\left(\frac{\hbar\omega}{2k_{\rm B}T}\right).
\end{equation}
Here, $T$ is the equilibrium temperature of the environment.
In our calculations, $\chi^{\prime\prime}(\omega)$ is approximately given
by an Ohmic spectrum with exponential cutoff~\cite{leggett1,leggett}
\begin{equation}\label{Xohmic}
\chi^{\prime\prime}(\omega)=\eta\,\omega \exp\left(-\frac{|\omega|}{\omega_{c}}\right).
\end{equation}
Here $\omega_{c}$ is the cutoff frequency typically assumed to be much larger than
all the other relevant frequency scales of the three-level SFQC.
The dimensionless constants $\eta |I_{ij}|^2/(2\pi)$ ($i,\,j= 0,\,1,\,2$)
represent the coupling strengths between the three-level SFQC and its environment.

Using Eqs.~(\ref{XwRI}), (\ref{SwRI}), (\ref{FDT}), and (\ref{Xohmic}),
we can simplify all formulae related to $\chi(\omega)$
and $\widetilde{S}(\omega)$ by $\chi^{\prime\prime}(\omega)$.
In the following calculations, we neglect the real parts of the complex coefficients $\Gamma_{lm}$,
which are responsible for the Lamb frequency shifts of the three-level SFQC.
In this way, the complex coefficients $\Gamma_{lm}$ in Eqs.~(\ref{SL01})--(\ref{SL22})
are replaced by $i\gamma_{lm}$, with $\gamma_{lm}={\rm Im}(\Gamma_{lm})$.

Now we make a comparison between the quantum Langevin equations in Eqs.~(\ref{SL01})--(\ref{SL22})
with the Lindblad master equation.
The method of deriving the quantum Langevin equations in Eqs.~(\ref{SL01})--(\ref{SL22})
is similar to the method in Ref.~\cite{open} of deriving the Lindblad master equation
with the Born-Markov approximation and the RWA.
And we can transform the quantum Langevin equations in Eqs.~(\ref{SL01})--(\ref{SL22})
into an equivalent Lindblad master equation~\cite{open}.
However, there are differences between the quantum Langevin equations in Eqs.~(\ref{SL01})--(\ref{SL22})
and the commonly-used type of Lindblad master equations.

(i) Usually, a commonly-used type of Lindblad master equation (denoted by cLME),
e.g., the Lindblad master equation used in Ref.~\cite{atsEF},
does not consider the effects of the driving field
on the coupling between the three-level SFQC and its environment.
That is to say, the frequency shifts and the damping rates induced by the environment in the cLME
are independent of the driving field.
However, in this paper we take these effects of the driving field into account as in Ref.~\cite{smirnov2}
and derive the driving-field-dependent complex coefficients $\Gamma_{lm}$ (see Appendix~\ref{ApGM}).

(ii) When the driving field $\Phi_{\rm D}$ is not applied,
the real (imaginary) parts of the complex coefficients $\Gamma_{lm}$ are equivalent to
the frequency shifts (the damping rates) in the cLME.
For example, $\gamma_{11}$ and $\gamma_{22}$, the imaginary parts of $\Gamma_{11}$ and $\Gamma_{22}$,
are equivalent to the damping rates of the off-diagonal matrix elements $\rho_{21}$ and $\rho_{31}$ in Ref.~\cite{atsEF}, respectively.
However, when the driving field $\Phi_{\rm D}$ is applied,
the complex coefficients $\Gamma_{lm}$ are modified by the Rabi frequency $|\Omega_{\rm D}|$
and the detuning $\Delta$ of the driving field $\Phi_{\rm D}$.
In addition, some of the complex coefficients $\Gamma_{lm}$,
e.g., $\Gamma_{12}$ and $\Gamma_{21}$, are not considered in the cLME
because these are nonzero only when the driving field $\Phi_{\rm D}$ is applied.
Thus, the phenomena induced by these complex coefficients $\Gamma_{lm}$,
e.g., $\Gamma_{12}$ and $\Gamma_{21}$, cannot be explained by using the cLME.
In Sec.~\ref{MS}, we will show that $\gamma_{12}$ and $\gamma_{21}$,
the imaginary parts of $\Gamma_{12}$ and $\Gamma_{21}$,
can make the two peaks in the absorption spectrum for ATS have different heights.

\section{Magnetic susceptibility}\label{MS}

To study the linear response of the three-level SFQC to the probe field
when the three-level SFQC is in the steady state,
we now calculate the linear magnetic susceptibilities.

\subsection{Susceptibility of the three-level SFQC}\label{s2}

The magnetic susceptibility $\chi_{q}(\omega)$ can be given by the Fourier transform of Eq.~(\ref{eq:14}) as
\begin{align}\label{XqF}
\chi_{q}(\omega)
&= \int d\tau_1 \left\langle
\frac{\delta\hat{I}(t)}{\delta \Phi_{\rm P}(t_{1})} \right\rangle \exp(i\omega\tau_1),
\end{align}
with the time interval $\tau_1=t-t_{1}$.
In the rotating reference frame,
the loop current operator $\hat{I}(t)$ in Eq.~(\ref{XqF}) takes the form as shown in Eq.~(\ref{eq:21}),
\begin{align}
\hat{I}(t)
&= \left[ I_{01}\sigma_{01}(t)e^{-i\omega_{1}t}
+I_{02}\sigma_{02}(t)e^{-i\omega^{\prime}t} \right.\nonumber
\\
&+ \left. I_{12}\sigma_{12}(t)e^{-i\omega_{0}t}+{\rm h.c.} \right]
+\sum_{i=0}^{2}I_{ii}\sigma_{ii}(t).
\end{align}

As discussed in Subsec.~D of Sec.~III, the average values $\langle\sigma_{12}(t)\rangle$
and $\langle\sigma_{ll}(t)\rangle$ ($l=0,\,1,\,2$) are independent of $\Phi_{\rm P}(t)$,
and do not affect the magnetic susceptibility $\chi_{q}(\omega)$.
Thus, the susceptibility $\chi_{q}(\omega)$ defined in Eq.~(\ref{XqF}) can be calculated via
the functional derivatives of the average values $\langle\sigma_{01}(t)\rangle$, $\langle\sigma_{02}(t)\rangle$,
$\langle\sigma_{10}(t)\rangle$, and $\langle\sigma_{20}(t)\rangle$ over the probe field $\Phi_{\rm P}(t_1)$.
Using Eqs.~(\ref{FS01}) and (\ref{FS02}), the functional derivatives of the average values
$\langle\sigma_{01}(t)\rangle$ and $\langle\sigma_{02}(t)\rangle$
over the probe field $\Phi_{\rm P}(t_1)$ can be given as
\begin{align}
\label{Dp01}
\frac{\delta\langle\sigma_{01}(t)\rangle}{\delta \Phi_{\rm P}(t_{1})}
&= i\frac{I_{10}}{\hbar} G_{22}(\tau_1) e^{i\omega_{1}t_{1}}
-i\frac{I_{20}}{\hbar} G_{12}(\tau_1) e^{i\omega^{\prime}t_{1}},
\\
\label{Dp02}
\frac{\delta \langle\sigma_{02}(t)\rangle}{\delta \Phi_{\rm P}(t_{1})}
&= i\frac{I_{20}}{\hbar} G_{11}(\tau_1) e^{i\omega^{\prime}t_{1}}
-i\frac{I_{10}}{\hbar} G_{21}(\tau_1) e^{i\omega_{1}t_{1}}.
\end{align}
The functional derivatives of the average values $\langle\sigma_{10}(t)\rangle$
and $\langle\sigma_{20}(t)\rangle$ over the probe field $\Phi_{\rm P}(t_1)$
can be obtained by taking the conjugates of Eqs.~(\ref{Dp01}) and (\ref{Dp02}).
Therefore, we can straightforwardly obtain
\begin{equation}\label{Xq}
\chi_{q}(\omega) =\chi_{01}(\omega) +\chi_{02}(\omega).
\end{equation}
Here, $\chi_{01}(\omega)$ and $\chi_{02}(\omega)$ is determined by
\begin{align}\label{X01F}
\chi_{01}(\omega)
&= \int d\tau_1 \left\langle
\frac{\delta[I_{01}\sigma_{01}(t)e^{-i\omega_{1}t}+{\rm h.c.}]}
{\delta \Phi_{\rm P}(t_{1})} \right\rangle e^{i\omega\tau_1},
\end{align}
\begin{align}\label{X02F}
\chi_{02}(\omega)
&= \int d\tau_{1} \left\langle
\frac{\delta [I_{02}\sigma_{02}(t)e^{-i\omega^{\prime}t}+{\rm h.c.}]}{\delta
\Phi_{\rm P}(t_{1})} \right\rangle e^{i\omega\tau_{1}}.
\end{align}
According to Eqs.~(\ref{X01F}) and (\ref{X02F}),
using Eqs.~(\ref{G11})--(\ref{G22}), (\ref{Dp01}), and (\ref{Dp02}),
we can obtain
\begin{align}
\label{X01}
\chi_{01}(\omega)
&= \frac{|I_{01}|^2 (\delta_{1}-\Delta+i\gamma_{22})}
{\hbar D_{1}(\delta_{1})},
\\
\label{X02}
\chi_{02}(\omega)
&= \frac{|I_{02}|^2 (\delta_{2}+i\gamma_{11})}
{\hbar D_{1}(\delta_{2})}.
\end{align}
Here
$\delta_{1}=\omega-\omega_{1}$, $\delta_{2}=\omega-\omega^{\prime}$,
and
\begin{align}\label{Dfunctions}
D_{1}(\omega)
&= -\left(\omega +i\gamma_{11}\right) \left(\omega-\Delta +i\gamma_{22}\right)
\nonumber\\
& +\left(i\gamma_{12}-\Omega_{\rm D}\right) \left(i\gamma_{21}-\Omega_{\rm D}^{*}\right).
\end{align}

We can prove from Eqs.~(\ref{X01F}) and (\ref{X02F}) that the magnetic susceptibility $\chi_{01}(\omega)$
and $\chi_{02}(\omega)$ are odd functions of the frequency $\omega$,
thus we only study the behavior of these magnetic susceptibilities in the regime $\omega>0$.
In this case, the functional derivatives of the average values $\langle\sigma_{10}(t)\rangle$
and $\langle\sigma_{20}(t)\rangle$ over the probe field $\Phi_{\rm P}(t_1)$
are neglected in obtaining Eqs.~(\ref{X01}) and (\ref{X02}),
because these functional derivatives are responsible for small anti-rotating wave terms when $\omega>0$.
We have also neglected all fast-oscillating terms, e.g., the second term in Eq.~(\ref{Dp01}),
in obtaining Eqs.~(\ref{X01}) and (\ref{X02}).
These fast-oscillating terms account for the three-wave mixing phenomenon of the three-level SFQC,
which is not in the scope of this paper.

The magnetic susceptibility $\chi_q(\omega)$ in Eq.~(\ref{Xq})
consists of two terms, $\chi_{01}(\omega)$ and $\chi_{02}(\omega)$.
$\chi_{01}(\omega)$ results from the $|0\rangle\leftrightarrow|1\rangle$ transition induced by the probe field,
while $\chi_{02}(\omega)$ results from the $|0\rangle\leftrightarrow|2\rangle$ transition induced by the probe field.
From Eqs.~(\ref{X01}) and (\ref{X02}), we can find that $\chi_{01}(\omega)$ and $\chi_{02}(\omega)$
are similar to the susceptibilities of three-level natural atoms~\cite{marangos,Abi},
driven at the $|1\rangle\leftrightarrow|2\rangle$ transition by a strong field, and probed by a weak field.
Here, $\chi_{01}(\omega)$ (or $\chi_{02}(\omega)$) is similar to the susceptibility of three-level natural atoms
with ladder-type (or $\Lambda$-type) transitions,
because in such natural atoms the weak probe field
can induce the $|0\rangle\leftrightarrow|1\rangle$ ($|0\rangle\leftrightarrow|2\rangle$) transition,
but cannot induce the $|0\rangle\leftrightarrow|2\rangle$ ($|0\rangle\leftrightarrow|1\rangle$) transition.

In this study, we assume that the Rabi frequency $|\Omega_{\rm D}|$,
the driving-field detuning $\Delta$, and the damping rates $\gamma_{ij}$ ($i,\,j=1,\,2$)
are all much smaller than the transition frequencies $\omega_{i}$ ($i=1,\,2,\,3$).
With these assumptions, we can find that $\chi_{01}(\omega)$ plays a major role
in the magnetic susceptibility $\chi_q(\omega)$ when $\omega$ is near resonant to
the $|0\rangle\leftrightarrow|1\rangle$ transition,
i.e., $\chi_q(\omega) \approx \chi_{01}(\omega)$ when $\omega \approx \omega_1$.
In contrast, $\chi_{02}(\omega)$ plays a major role
in the magnetic susceptibility $\chi_q(\omega)$ when $\omega$ is near resonant to
the $|0\rangle\leftrightarrow|2\rangle$ transition,
i.e., $\chi_q(\omega) \approx \chi_{02}(\omega)$ when $\omega \approx \omega_2$.
Hereafter, we denote the frequency range, in which $\omega$ is near resonant to
the $|0\rangle\leftrightarrow|1\rangle$ (or $|0\rangle\leftrightarrow|2\rangle$) transition,
as the $|0\rangle\leftrightarrow|1\rangle$ (or $|0\rangle\leftrightarrow|2\rangle$) frequency range.

As shown in Fig.~\ref{fig2}(b), the moduli of the loop current transition matrix elements
$|I_{01}|$ and $|I_{02}|$ are dependent on the bias magnetic flux.
Thus, we can find from Eqs.~(\ref{Xq}), (\ref{X01}), and (\ref{X02}) that
the magnetic susceptibility $\chi_q(\omega)$ can be tuned by the bias magnetic flux through $|I_{01}|$ and $|I_{02}|$.
According to the symmetric analysis~\cite{liu},
when the bias magnetic flux is at the optimal point,
$|I_{02}|=0$ and the probe field $\Phi_{\rm P}(t)$ cannot induce
the $|0\rangle\leftrightarrow|2\rangle$ transition of the three-level SFQC.
At this special point, $\chi_{q}(\omega)$ has the same function as $\chi_{01}(\omega)$,
thus the three-level SFQC can respond to the probe field in the $|0\rangle\leftrightarrow|1\rangle$ frequency range,
like natural atoms with ladder-type transitions. However, when the bias magnetic flux deviates from the optimal point,
the probe field $\Phi_{\rm P}(t)$ can induce both the $|0\rangle\leftrightarrow|1\rangle$
and $|0\rangle\leftrightarrow|2\rangle$ transitions via interactions with the loop current of the SFQC.
In this case, $\chi_{q}(\omega)$ is the summation of $\chi_{01}(\omega)$ and $\chi_{02}(\omega)$.
Thus the three-level SFQC can respond to the probe field in
both the $|0\rangle\leftrightarrow|1\rangle$ and $|0\rangle\leftrightarrow|2\rangle$ frequency ranges.
In other words, the three-level SFQC acts like a combination of natural atoms with ladder-type transitions
and natural atoms with $\Lambda$-type transitions.
This is an obvious difference between the linear responses of three-level SFQCs and three-level natural atoms.

\begin{figure}
\includegraphics[bb=60 70 660 470, width=8.5 cm, clip]{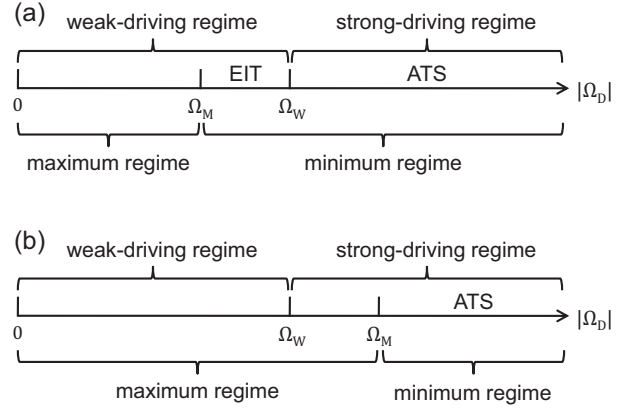}
\caption[]{
(a) Schematic diagram of the weak-driving regime, the strong-driving regime, the maximum regime,
and the minimum regime when $\Omega_{\rm M}<\Omega_{\rm W}$.
(b) Schematic diagram of the weak-driving regime, the strong-driving regime, the maximum regime,
and the minimum regime when $\Omega_{\rm M} \geq \Omega_{\rm W}$.
}\label{fig3}
\end{figure}

\subsection{Conditions for realizing EIT and ATS}

We know that the linear response of the driven three-level SFQC to the probe field can be used to
characterize EIT and ATS. Let us now study the conditions for realizing EIT and ATS in the three-level SFQC.
As discussed in Subsec.~A of Sec.~\ref{MS},
the three-level SFQC may respond to the probe field in two different frequency ranges,
the $|0\rangle\leftrightarrow|1\rangle$ frequency range and the $|0\rangle\leftrightarrow|2\rangle$ frequency range.
Thus EIT (ATS) may occur in these two frequency ranges.
In the $|0\rangle\leftrightarrow|1\rangle$ frequency range,
the EIT (ATS) spectrum is characterized by $\chi_{01}(\omega)$,
while in the $|0\rangle\leftrightarrow|2\rangle$ frequency range, the EIT (ATS) spectrum is characterized by $\chi_{02}(\omega)$.
We first introduce the method provided in Refs.~\cite{Bsanders,Ani,Abi} of distinguishing EIT from ATS.
Then we combine this method with other necessary conditions
to derive the conditions for realizing EIT and ATS.

References~\cite{Bsanders,Ani,Abi} provided a method of distinguishing EIT from ATS
by decomposing the linear response of a three-level quantum system into two resonances.
For example, the magnetic susceptibilities $\chi_{01}(\omega)$
and $\chi_{02}(\omega)$ can be decomposed as
\begin{align}
\label{X01D}
\chi_{01}(\omega)
&= R_{+}^{(01)}(\delta_1) +R_{-}^{(01)}(\delta_1)
\nonumber\\
&= \frac{\alpha_{01}}{\delta_{-}-\delta_{+}} \frac{\delta_{+} -\Delta+i\gamma_{22}}{\delta_1-\delta_{+}}
\nonumber\\
& +\frac{\alpha_{01}}{\delta_{+}-\delta_{-}} \frac{\delta_{-} -\Delta+i\gamma_{22}}{\delta_1-\delta_{-}},
\\
\label{X02D}
\chi_{02}(\omega)
&= R_{+}^{(02)}(\delta_2) +R_{-}^{(02)}(\delta_2)
\nonumber\\
&= \frac{\alpha_{02}}{\delta_{-}-\delta_{+}} \frac{\delta_{+} +i\gamma_{11}}{\delta_2-\delta_{+}}
\nonumber\\
& +\frac{\alpha_{02}}{\delta_{+}-\delta_{-}} \frac{\delta_{-} +i\gamma_{11}}{\delta_2-\delta_{-}},
\end{align}
where $\alpha_{0i} = |I_{0i}|^2/\hbar$ ($i=1,\,2$)
and $\delta_{\pm}$ are the complex roots of the equation $D_{1}(\delta)=0$.
If the driving-field detuning $\Delta$ is assumed to be zero and the damping rates $\gamma_{12}$ and $\gamma_{21}$ are neglected,
then $\delta_{\pm}$ can be given by
\begin{align}
\delta_{\pm}= \frac{1}{2} \left(-i\gamma_{11}-i\gamma_{22}
\pm \Omega_{\rm T}\right),
\end{align}
with $\Omega_{\rm T}=\sqrt{4|\Omega_{\rm D}|^2-(\gamma_{11}-\gamma_{22})^2}$.
For simplicity, the following discussions for EIT and ATS will first be limited to the case that $\Delta=0$ and $\gamma_{12}=\gamma_{21}=0$.
The effects of nonzero $\Delta$, $\gamma_{12}$, and $\gamma_{21}$ will be discussed in Subsec.~C of Sec.~\ref{MS}.

We first study the realization of EIT (ATS) in the $|0\rangle\leftrightarrow|1\rangle$ frequency range,
where the EIT (ATS) spectrum is characterized by $\chi_{01}(\omega)$.
According to Refs.~\cite{Bsanders,Ani,Abi},
EIT and ATS occur in two different driving regimes, respectively,
defined by the relation between the Rabi frequency $|\Omega_{\rm D}|$ and the threshold $|\gamma_{11}-\gamma_{22}|/2$.
EIT can only occur in the regime $|\Omega_{\rm D}| <|\gamma_{11}-\gamma_{22}|/2$,
which is called here the weak-driving regime,
while ATS can only occur in the regime $|\Omega_{\rm D}| >|\gamma_{11}-\gamma_{22}|/2$,
which is called here the strong-driving regime.
The two resonances $R_{\pm}^{(01)}(\delta_1)$ of $\chi_{01}(\omega)$
have different shapes in these two driving regimes.
In the weak-driving regime, ${\rm Re}(\delta_{\pm})=0$, and then
${\rm Im}[R_{+}^{(01)}(\delta_1)]$ and ${\rm Im}[R_{-}^{(01)}(\delta_1)]$
are two Lorentzians with different signs, both centered at $\delta_1=0$.
In the strong-driving regime, ${\rm Re}(\delta_{+})=-{\rm Re}(\delta_{-}) \neq 0$
and ${\rm Im}(\delta_{+})={\rm Im}(\delta_{-})$,
and then ${\rm Im}[R_{+}^{(01)}(\delta_1)]$ and ${\rm Im}[R_{-}^{(01)}(\delta_1)]$
are two positive Lorentzians centered at $\delta_1=\pm\Omega_{\rm T}$, respectively.
The bifurcation point $|\Omega_{\rm D}|=|\gamma_{11}-\gamma_{22}|/2$ is a special case
where the decomposition in Eq.~(\ref{X01D}) is invalid.
In this study, we do not consider the conditions for realizing EIT and ATS at this bifurcation point.

The conditions $|\Omega_{\rm D}| <|\gamma_{11}-\gamma_{22}|/2$
and $|\Omega_{\rm D}| >|\gamma_{11}-\gamma_{22}|/2$
can be used to distinguish EIT from ATS in the $|0\rangle\leftrightarrow|1\rangle$ frequency range,
however, these are not the sufficient conditions for the realization of EIT (ATS).
The dip in the absorption spectrum of EIT (ATS) implies that the resonant point
${\rm Im}[\chi_{01}(\omega_1)]$ must be a local minimum when EIT (ATS) is realized.
By analyzing the derivative of ${\rm Im}[\chi_{01}(\omega)]$ over $\omega$,
we find that ${\rm Im}[\chi_{01}(\omega_1)]$ is the local minimum if
$|\Omega_{\rm D}|>\gamma_{22} \sqrt{\gamma_{22}/(\gamma_{11}+2\gamma_{22})}$ .
Combining this condition with $|\Omega_{\rm D}|<|\gamma_{11}-\gamma_{22}|/2$
($|\Omega_{\rm D}| >|\gamma_{11}-\gamma_{22}|/2$),
we finally obtain the condition for realizing EIT (ATS) in the $|0\rangle\leftrightarrow|1\rangle$ frequency range:
\begin{align}
\label{EIT01}
{\rm EIT} &:
  \left\{
   \begin{aligned}
   \gamma_{11} &> 2\gamma_{22},\\
   |\Omega_{\rm D}| &> \gamma_{22}\sqrt{\frac{\gamma_{22}}{\gamma_{11}+2\gamma_{22}}} \, ,\\
   |\Omega_{\rm D}| &< \frac{|\gamma_{11}-\gamma_{22}|}{2}.\\
   \end{aligned}
  \right.\\
\label{ATS01}
{\rm ATS} &:
  \left\{
   \begin{aligned}
   |\Omega_{\rm D}| &> \frac{|\gamma_{11}-\gamma_{22}|}{2}   &\gamma_{11} > 2\gamma_{22},\\
   |\Omega_{\rm D}| &> \gamma_{22}\sqrt{\frac{\gamma_{22}}{\gamma_{11}+2\gamma_{22}}}  &\gamma_{11} \leq 2\gamma_{22}.\\
   \end{aligned}
  \right.
\end{align}

The above discussions can also be applied to the $|0\rangle\leftrightarrow|2\rangle$ frequency range.
Considering that the damping rates $\gamma_{11}$ and $\gamma_{22}$ play different roles
in the expressions for $\chi_{01}(\omega)$ and $\chi_{02}(\omega)$ in Eqs.~(\ref{X01}) and (\ref{X02}),
the conditions for realizing EIT (ATS) in the $|0\rangle\leftrightarrow|2\rangle$ frequency range are
\begin{align}
\label{EIT02}
{\rm EIT} &:
  \left\{
   \begin{aligned}
   \gamma_{22} &> 2\gamma_{11},\\
   |\Omega_{\rm D}| &> \gamma_{11}\sqrt{\frac{\gamma_{11}}{\gamma_{22}+2\gamma_{11}}}\, ,\\
   |\Omega_{\rm D}| &< \frac{|\gamma_{11}-\gamma_{22}|}{2}.\\
   \end{aligned}
  \right.\\
\label{ATS02}
{\rm ATS} &:
  \left\{
   \begin{aligned}
   |\Omega_{\rm D}| &> \frac{|\gamma_{11}-\gamma_{22}|}{2}   &\gamma_{22} > 2\gamma_{11},\\
   |\Omega_{\rm D}| &> \gamma_{11}\sqrt{\frac{\gamma_{11}}{\gamma_{22}+2\gamma_{11}}}  &\gamma_{22} \leq 2\gamma_{11}.\\
   \end{aligned}
  \right.
\end{align}

We can introduce two thresholds, $\Omega_{\rm W}$ and $\Omega_{\rm M}$,
of the Rabi frequency $|\Omega_{\rm D}|$ to better interpret the conditions in Eqs.~(\ref{EIT01})--(\ref{ATS02}).
Here, $\Omega_{\rm W}$ is just the threshold $|\gamma_{11}-\gamma_{22}|/2$ mentioned above,
which separates the weak-driving from the strong-driving regime.
While $\Omega_{\rm M}$ is used to describe the characteristic of
the resonant points ${\rm Im}[\chi_{01}(\omega_1)]$ and ${\rm Im}[\chi_{02}(\omega_2)]$.
Note that $\Omega_{\rm M}$ has different definitions for $\chi_{01}(\omega)$ and $\chi_{02}(\omega)$:
$\Omega_{\rm M}^{(01)}=\gamma_{22} \sqrt{\gamma_{22}/(\gamma_{11}+2\gamma_{22})}$ for $\chi_{01}(\omega)$,
while $\Omega_{\rm M}^{(02)}=\gamma_{11} \sqrt{\gamma_{11}/(\gamma_{22}+2\gamma_{11})}$ for $\chi_{02}(\omega)$.
Similar to the threshold $\Omega_{\rm W}$, the threshold $\Omega_{\rm M}$ also defines two regimes.
We denote $|\Omega_{\rm D}| \leq \Omega_{\rm M}^{(01)}$
($|\Omega_{\rm D}| \leq \Omega_{\rm M}^{(02)}$) as the maximum regime.
In this regime, the resonant point ${\rm Im}[\chi_{01}(\omega_1)]$ (${\rm Im}[\chi_{02}(\omega_2)]$)
is a local maximum of ${\rm Im}[\chi_{01}(\omega)]$ (${\rm Im}[\chi_{02}(\omega)]$).
By contrast, we denote $|\Omega_{\rm D}|>\Omega_{\rm M}^{(01)}$
($|\Omega_{\rm D}|>\Omega_{\rm M}^{(02)}$) as the minimum regime.
In this regime, the resonant point ${\rm Im}[\chi_{01}(\omega_1)]$ (${\rm Im}[\chi_{02}(\omega_2)]$)
is a local minimum of ${\rm Im}[\chi_{01}(\omega)]$ (${\rm Im}[\chi_{02}(\omega)]$).

The realization of EIT requires that the Rabi frequency $|\Omega_{\rm D}|$
lies in both the weak-driving regime and the minimum regime.
The former condition guarantees that the two resonances ${\rm Im}[R_{\pm}^{(01)}(\delta_1)]$
(${\rm Im}[R_{\pm}^{(02)}(\delta_2)]$) are centered at the same place,
and lead to destructive interference.
While the latter condition guarantees that the destructive interference between the two resonances
is strong enough to make a dip appear in the absorption spectrum.
In contrast to the realization of EIT, the realization of ATS requires that
the Rabi frequency $|\Omega_{\rm D}|$ lies in both the strong-driving regime and the minimum regime.
The former condition guarantees that the two resonances ${\rm Im}[R_{\pm}^{(01)}(\delta_1)]$
(${\rm Im}[R_{\pm}^{(02)}(\delta_2)]$) are centered at different places.
While the latter condition guarantees that the distance between the two resonances
is large enough to make a dip appear in the absorption spectrum.

When $\gamma_{11}>2\gamma_{22}$ ($\gamma_{22}>2\gamma_{11}$),
we find $\Omega_{\rm M}^{(01)}<\Omega_{\rm W}$ ($\Omega_{\rm M}^{(02)}<\Omega_{\rm W}$).
In this case, as shown in Fig.~\ref{fig3}(a), the weak-driving regime overlaps with the minimum regime,
while the strong-driving regime is totally included by the minimum regime.
Thus EIT can be realized in the $|0\rangle\leftrightarrow|1\rangle$ ($|0\rangle\leftrightarrow|2\rangle$) frequency range
when the Rabi frequency $|\Omega_{\rm D}|$ lies in the interval
where the weak-driving regime and the minimum regime overlaps,
i.e., $\Omega_{\rm M}^{(01)}<|\Omega_{\rm D}|<\Omega_{\rm W}$
($\Omega_{\rm M}^{(02)}<|\Omega_{\rm D}|<\Omega_{\rm W}$).
While ATS can be realized in the $|0\rangle\leftrightarrow|1\rangle$ ($|0\rangle\leftrightarrow|2\rangle$) frequency range
when the Rabi frequency $|\Omega_{\rm D}|$ lies in the strong-driving regime,
i.e., $|\Omega_{\rm D}|>\Omega_{\rm W}$.

When $\gamma_{11} \leq 2\gamma_{22}$ ($\gamma_{22} \leq 2\gamma_{11}$),
we find $\Omega_{\rm M}^{(01)} \geq \Omega_{\rm W}$ ($\Omega_{\rm M}^{(02)} \geq \Omega_{\rm W}$).
In this case, as shown in Fig.~\ref{fig3}(b), the weak-driving regime does not overlap with the minimum regime,
while the strong-driving regime includes the minimum regime.
Thus EIT cannot be realized in the $|0\rangle\leftrightarrow|1\rangle$ ($|0\rangle\leftrightarrow|2\rangle$) frequency range
for any value of the Rabi frequency $|\Omega_{\rm D}|$.
While ATS can be realized in the $|0\rangle\leftrightarrow|1\rangle$ ($|0\rangle\leftrightarrow|2\rangle$) frequency range
when the Rabi frequency $|\Omega_{\rm D}|$ lies in the minimum regime,
i.e., $|\Omega_{\rm D}|>\Omega_{\rm M}^{(01)}$ ($|\Omega_{\rm D}|>\Omega_{\rm M}^{(02)}$).

Combining the above discussions, we can straightforwardly obtain the conditions
for realizing EIT (ATS) in Eqs.~(\ref{EIT01})--(\ref{ATS02}).
We also find that EIT cannot be realized simultaneously
in both the $|0\rangle\leftrightarrow|1\rangle$ and $|0\rangle\leftrightarrow|2\rangle$ frequency ranges,
because the damping rates $\gamma_{11}$ and $\gamma_{22}$ cannot satisfy
both the conditions $\gamma_{11}>2\gamma_{22}$ and $\gamma_{22}>2\gamma_{11}$
in Eqs.~(\ref{EIT01}) and (\ref{EIT02}).
However, ATS can be realized simultaneously in both the $|0\rangle\leftrightarrow|1\rangle$ and $|0\rangle\leftrightarrow|2\rangle$ frequency ranges,
as long as the Rabi frequency $|\Omega_{\rm D}|$ is larger than
$\Omega_{\rm W}$, $\Omega_{\rm M}^{(01)}$, and $\Omega_{\rm M}^{(02)}$.

\begin{figure}
\includegraphics[bb=0 200 530 600, width=4.2 cm, clip]{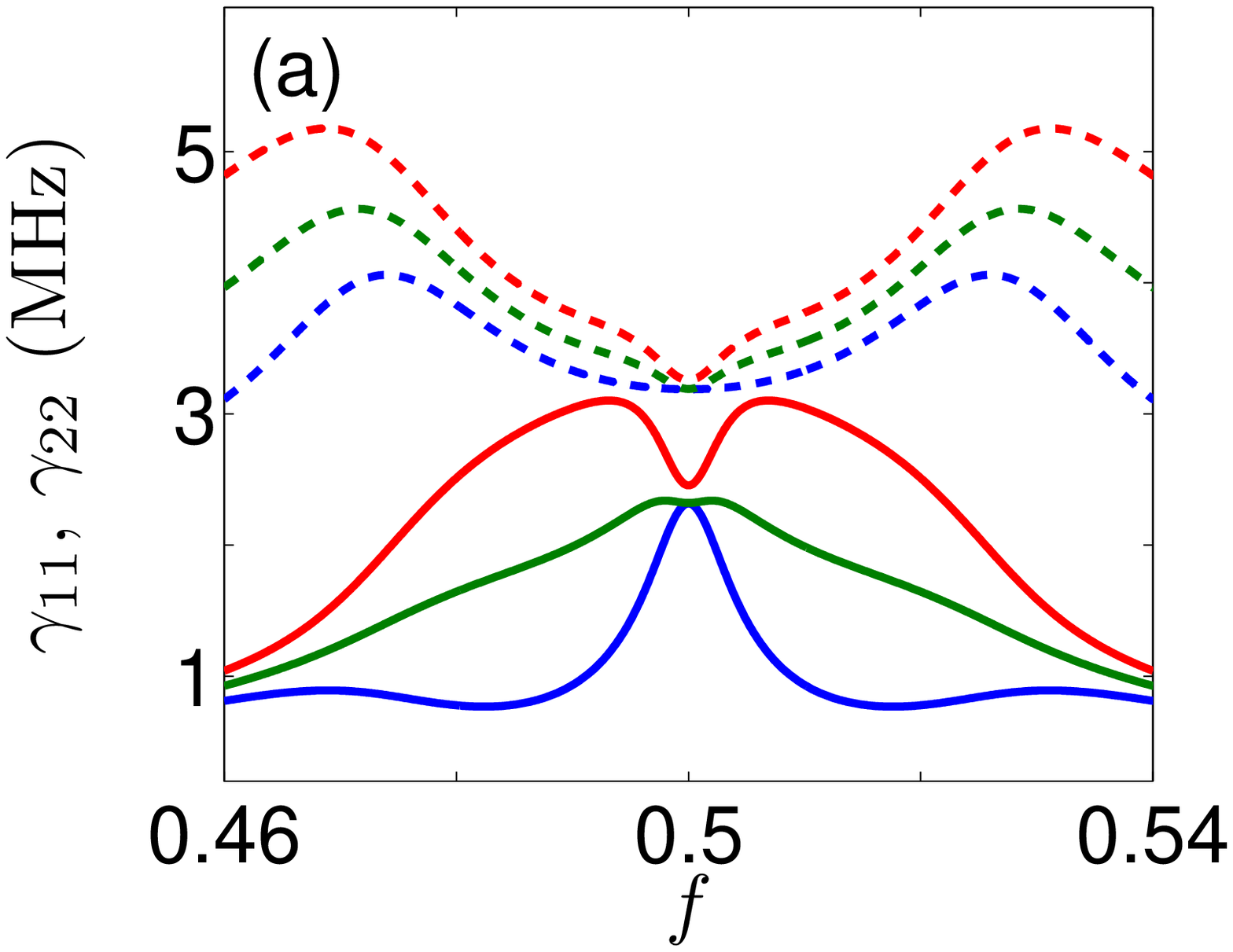}
\includegraphics[bb=0 200 530 600, width=4.2 cm, clip]{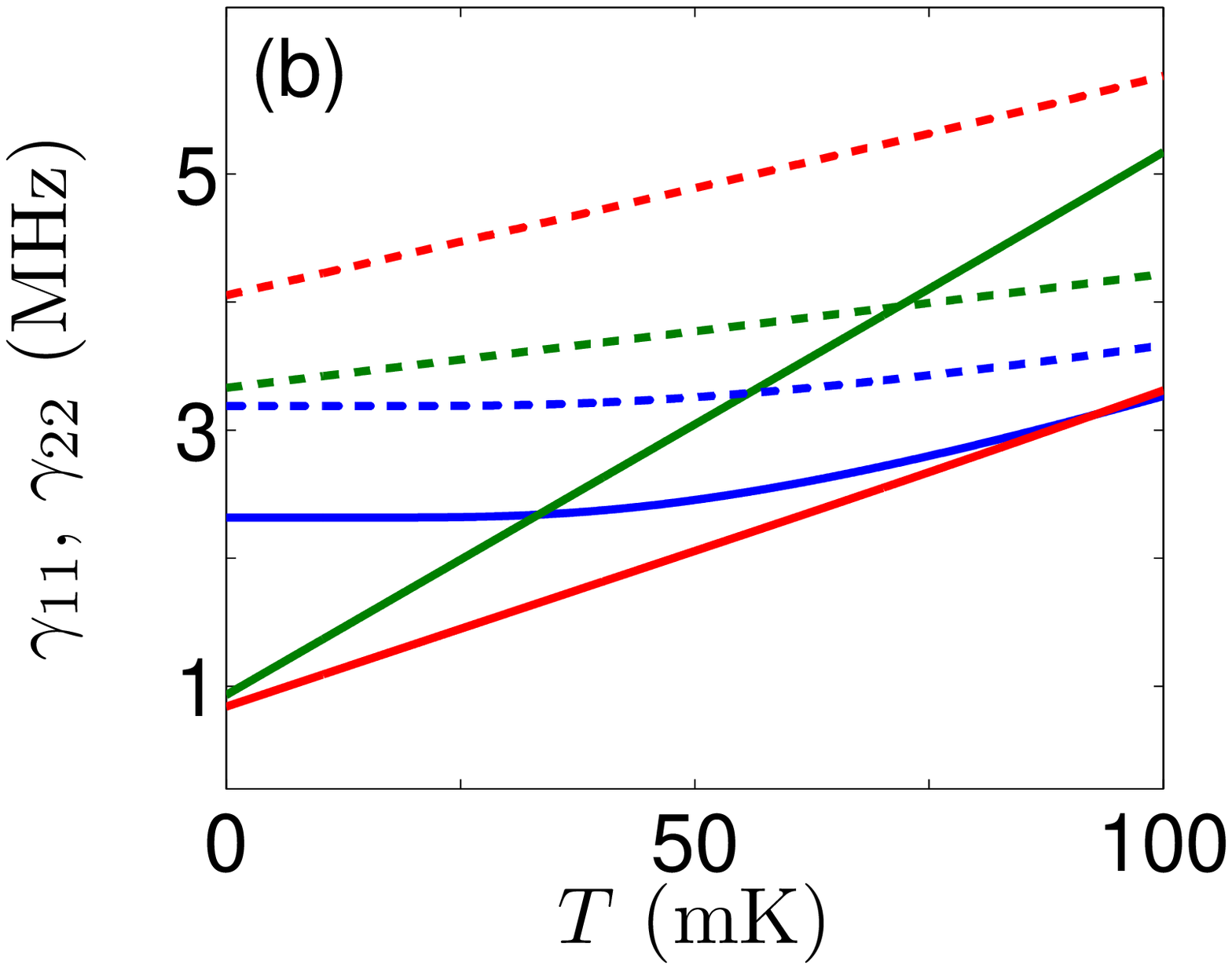}
\includegraphics[bb=0 200 530 600, width=4.2 cm, clip]{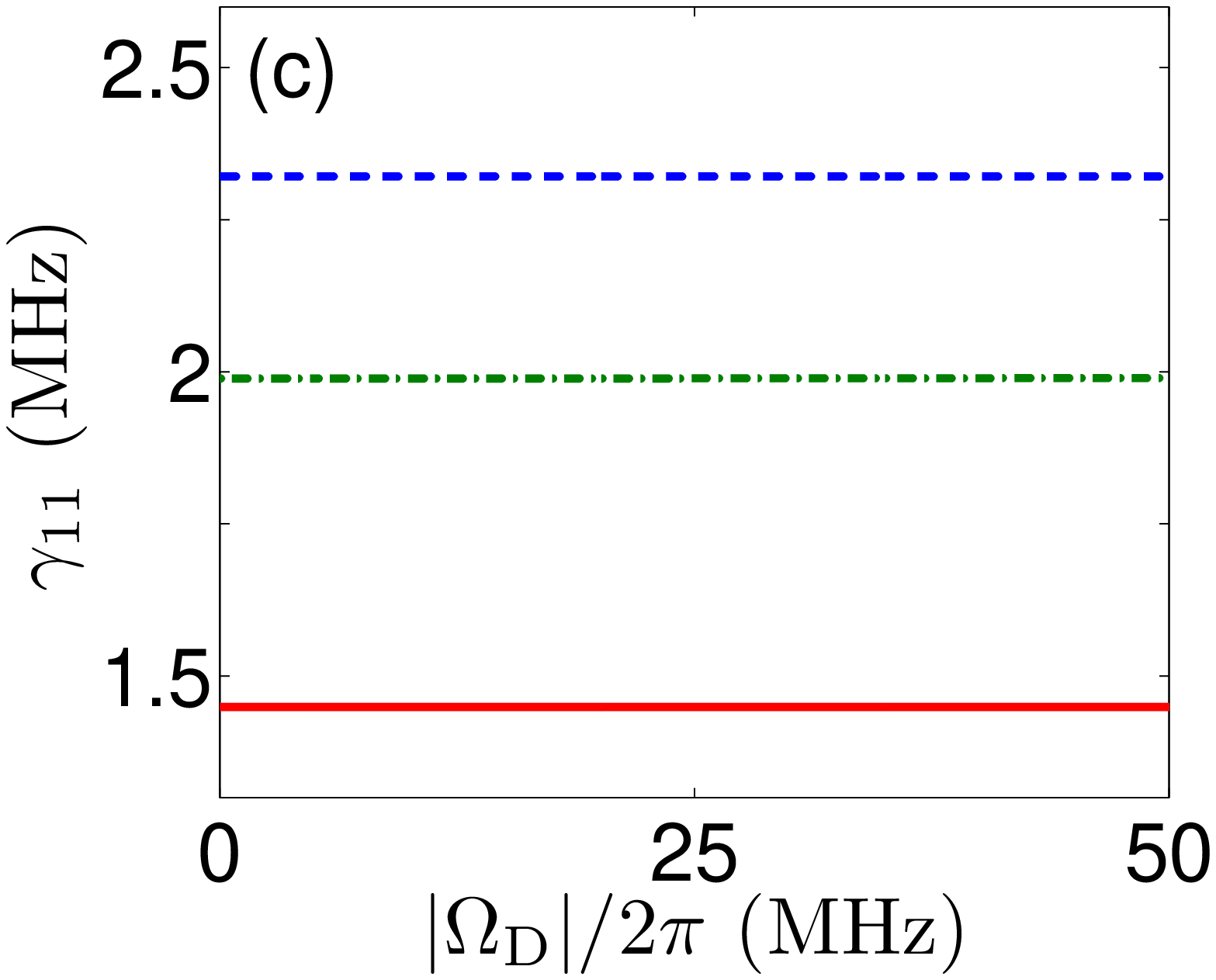}
\includegraphics[bb=0 200 530 600, width=4.2 cm, clip]{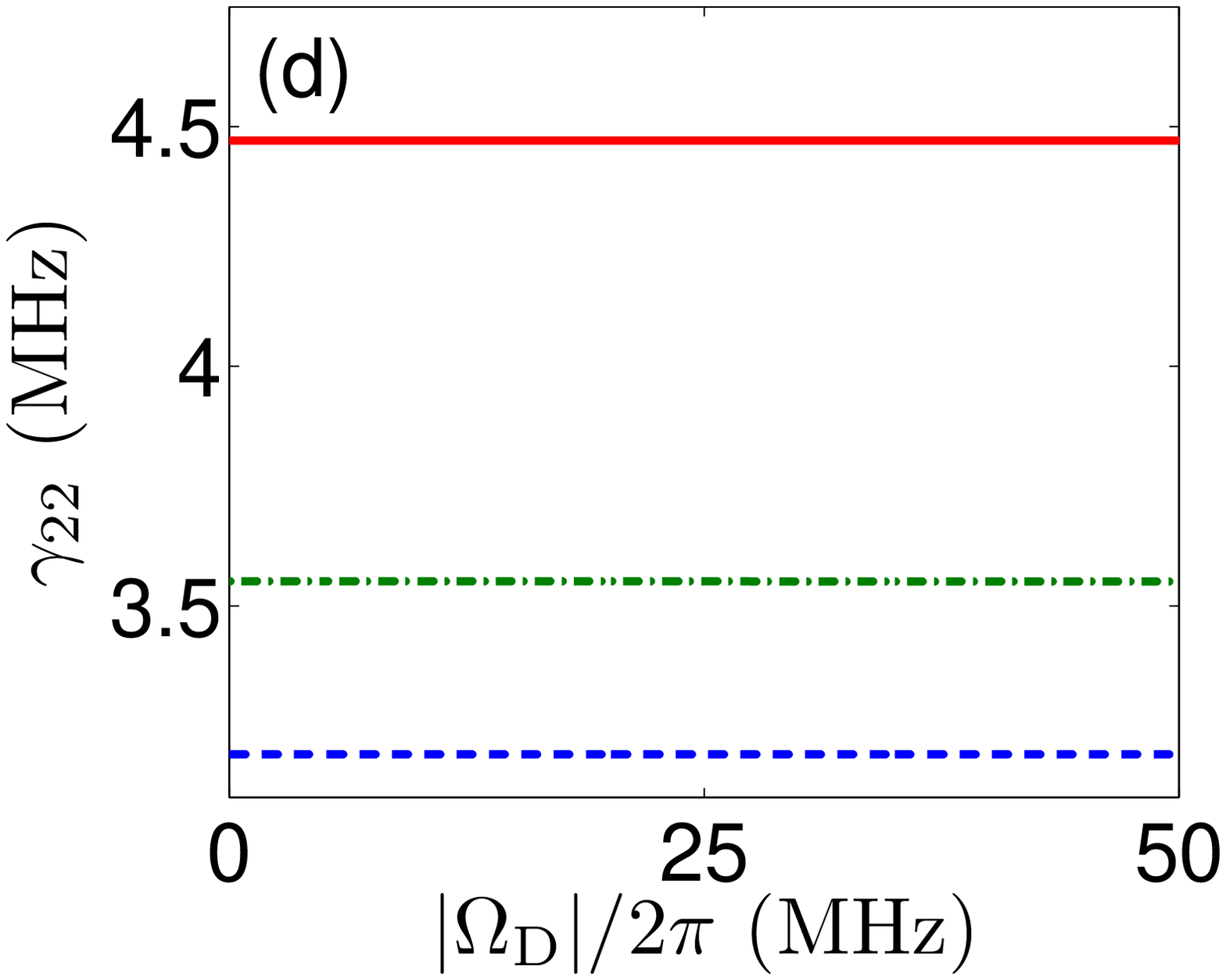}
\includegraphics[bb=0 200 530 600, width=4.2 cm, clip]{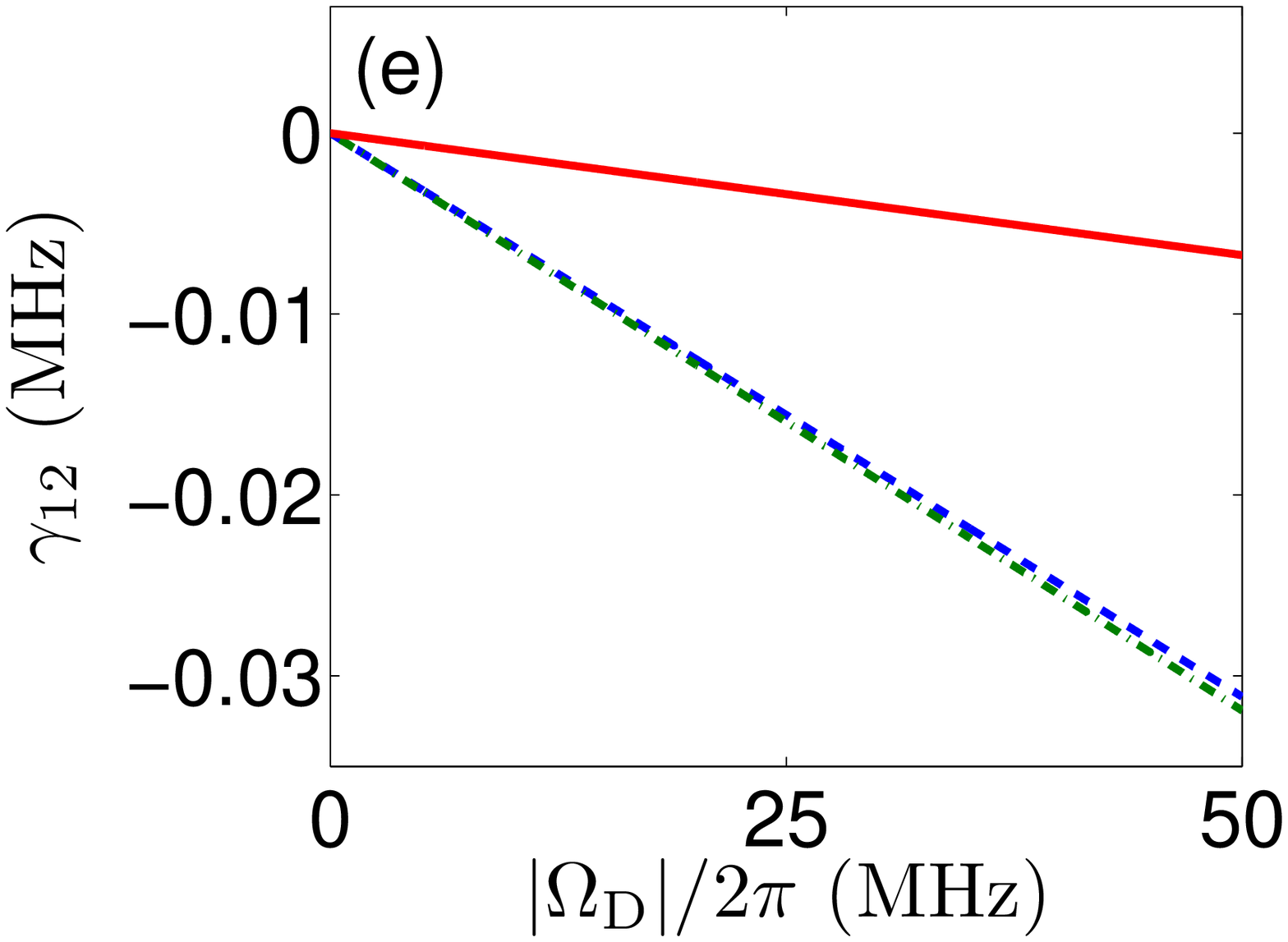}
\includegraphics[bb=0 200 530 600, width=4.2 cm, clip]{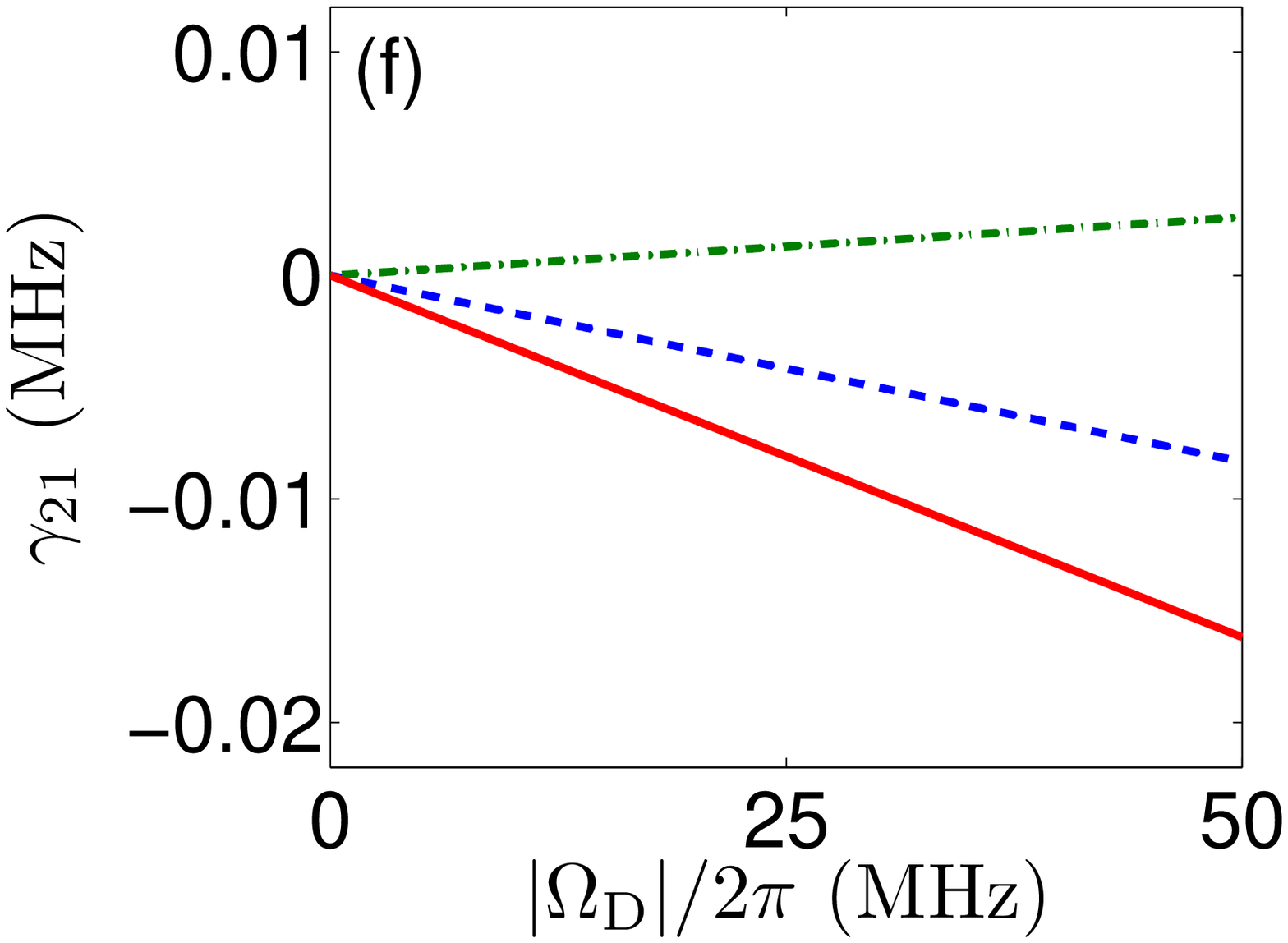}
\caption[]{(Color online)
Damping rates $\gamma_{11}$ (solid curves) and $\gamma_{22}$ (dashed curves)
in (a) as a function of the reduced magnetic flux $f$,
plotted for the temperatures $T=0$ (blue curves), $T=25$ mK (green curves), and $T=50$ mK (red curves).
Damping rates $\gamma_{11}$ (solid curves) and $\gamma_{22}$ (dashed curves)
in (b) as a function of the temperature $T$,
plotted for the reduced magnetic flux $f=0.5$ (blue curves), $f=0.51$ (green curves), and $f=0.525$ (red curves).
Damping rates $\gamma_{11}$ in (c), $\gamma_{22}$ in (d), $\gamma_{12}$ in (e), and $\gamma_{21}$ in (f)
as a function of the Rabi frequency $|\Omega_{\rm D}|$,
plotted for the reduced magnetic flux $f=0.5$ (blue dashed curve),
$f=0.51$ (green dash-dotted curve), and $f=0.525$ (red solid curve).
Here $\beta=10^{-4}$ and $\Delta=0$.
In (a) and (b), $|\Omega_{\rm D}|$ is assumed to be 0.
In (e) and (f), $T$ is assumed to be 25 mK.
Other parameters of the three-level SFQC have been given in the two items at the beginning of Sec.~\ref{NUM}.
}\label{fig4}
\end{figure}

\subsection{The effects of the driving-field detuning $\Delta$ and the damping rates $\gamma_{12}$ and $\gamma_{21}$}

We now discuss the effects of the driving-field detuning $\Delta$ and the damping rates $\gamma_{12}$ and $\gamma_{21}$ on
the linear response.
In this study, we only consider the case when the driving-field detuning $\Delta$
and the Rabi frequency $|\Omega_{\rm D}| $ satisfy the condition
$\Delta \leq |\Omega_{\rm D}| \ll\omega_i$ ($i=1,\,2,\,3$).
In addition, the condition $\gamma_{12}, \gamma_{21}\ll|\Omega_{\rm D}|$ is fulfilled
because the coupling between the three-level SFQC and its environment is weak.
In this case, the driving-field detuning $\Delta$ and the damping rates $\gamma_{12}$ and $\gamma_{21}$
only slightly affect the magnetic susceptibilities $\chi_{01}(\omega)$ and $\chi_{02}(\omega)$.
Thus we can approximately use the conditions for realizing EIT and ATS in Eqs.~(\ref{EIT01})--(\ref{ATS02})
even when the driving-field detuning $\Delta$ is nonzero
and the damping rates $\gamma_{12}$ and $\gamma_{21}$ are taken into account.
However, it is necessary to mention the special effect of the damping rates $\gamma_{12}$ and $\gamma_{21}$
on the magnetic susceptibilities $\chi_{01}(\omega)$ and $\chi_{02}(\omega)$.
That is, the damping rates $\gamma_{12}$ and $\gamma_{21}$ can lead to different heights of the two peaks
in the absorption spectrum for ATS.
For example, when the conditions in Eq.~(\ref{ATS01}) for realizing ATS in the $|0\rangle\leftrightarrow|1\rangle$ frequency range are fulfilled,
and the driving-field detuning $\Delta$ is zero,
we find that the damping rates $\gamma_{12}$ and $\gamma_{21}$ can make the complex roots $\delta_{\pm}$
have different imaginary parts, i.e., ${\rm Re}(\delta_{+})=-{\rm Re}[\delta_{-}] \neq 0$
and ${\rm Im}(\delta_{+}) \neq{\rm Im}(\delta_{-})$.
In this case, ${\rm Im}[R_{+}^{(01)}(\delta_1)]$ and ${\rm Im}[R_{-}^{(01)}(\delta_1)]$
are two positive Lorentzians with different heights,
resulting in two different heights for the two peaks in the absorption spectrum for ATS.
Similar results for ATS in the $|0\rangle\leftrightarrow|2\rangle$ frequency range can also be found.
Because the transmission coefficient $t$ in Ref.~\cite{atsEF}
and the magnetic susceptibility $\chi_{01}(\omega)$ in Eq.~(\ref{X01})
satisfy the relation $\chi_{01}(\omega) \propto i(1-t)$,
we speculate that this effect of the damping rates $\gamma_{12}$ and $\gamma_{21}$
is a possible reason for the asymmetric transmission spectrum in the ATS experiment~\cite{atsEF,Bsanders}.

\section{Numerical results}\label{NUM}

In this section, we numerically study the damping rates $\gamma_{lm}$
and the magnetic susceptibility $\chi_{q}(\omega)$.
Our numerical calculations include the following considerations:

\begin{enumerate}

\item
Same as in Fig.~\ref{fig2}, our discussions below are based on the assumption
that the ratio between the Josephson energy $E_{\rm J}$
and the charging energy $E_{\rm c}$ is chosen as $E_{\rm J}=\Emu\,E_{\rm c}$,
and the ratio $\alpha$ between the larger Josephson junction and the smaller one is $\alpha=\Ea$.
We take an experimentally accessible Josephson energy,
$E_{\rm J}/\hbar =2\pi\times \Ejv$ GHz, for example.
In this case, when the reduced magnetic flux $f$ is at the optimal point, i.e., $f=0.5$,
we can obtain the transition frequencies via Fig.~\ref{fig2}(a) as
$\omega_{1} \approx \woA\,E_{\rm J}/\hbar \approx 2\pi\times \wvA$ GHz,
and $\omega_{3} \approx \woC \,E_{\rm J}/\hbar \approx 2\pi\times \wvC$ GHz.
We can also obtain the matrix elements of the loop current operator $\hat{I}$
via Figs.~\ref{fig2}(b) and \ref{fig2}(c) as
$|I_{01}|=\ioA\,I_{0}$, $|I_{02}|=0$, $|I_{12}|=\ioC\,I_{0}$,
and $I_{00}=I_{11}=I_{22}=0$.
We will also discuss the physics when the reduced magnetic flux $f$
is not at the optimal point, e.g., $f= \fB$.
In this case, the corresponding transition frequencies are
$\omega_{1} \approx\WoA\,E_{\rm J}/\hbar \approx2\pi\times \WvA$ GHz
and $\omega_{3} \approx\WoC\,E_{\rm J}/\hbar \approx2\pi\times \WvC$ GHz , respectively.
The corresponding loop current matrix elements are
$|I_{01}|=\IoA\,I_{0}$, $|I_{02}|=\IoB\,I_{0}$, $|I_{12}|=\IoC\,I_{0}$,
$I_{00}=\IoD\,I_{0}$, $I_{11}=\IoE\,I_{0}$, and $I_{22}=\IoF\,I_{0}$.
Here, we note again $I_{0}=2\pi E_{\rm J}/\Phi_{0}$.

\item
We assume the cutoff frequency in Eq.~(\ref{Xohmic}) as $\omega_{c}=100\,\omega_{s}$,
with $\omega_{s}$ corresponding to the transition frequency $\omega_{3}=(E_{2}-E_{0})/\hbar$ when $f=0.5$.
We set a dimensionless constant $\beta=\eta I_{s}^2/(2\pi)$,
with $I_{s}\equiv|I_{01}|$ when $f=0.5$.
In this way, the spectral function $\chi^{\prime\prime}(\omega)$ in Eq.~(\ref{Xohmic})
can be expressed as
\begin{equation}
\chi^{\prime\prime}(\omega)=\frac{\beta}{I_{s}^2}\,\omega\,
{\rm exp} \left(-\frac{|\omega|}{100\omega_{s}}\right).
\end{equation}

\end{enumerate}

\begin{figure}
\includegraphics[bb=25 200 520 600, width=4.2 cm, clip]{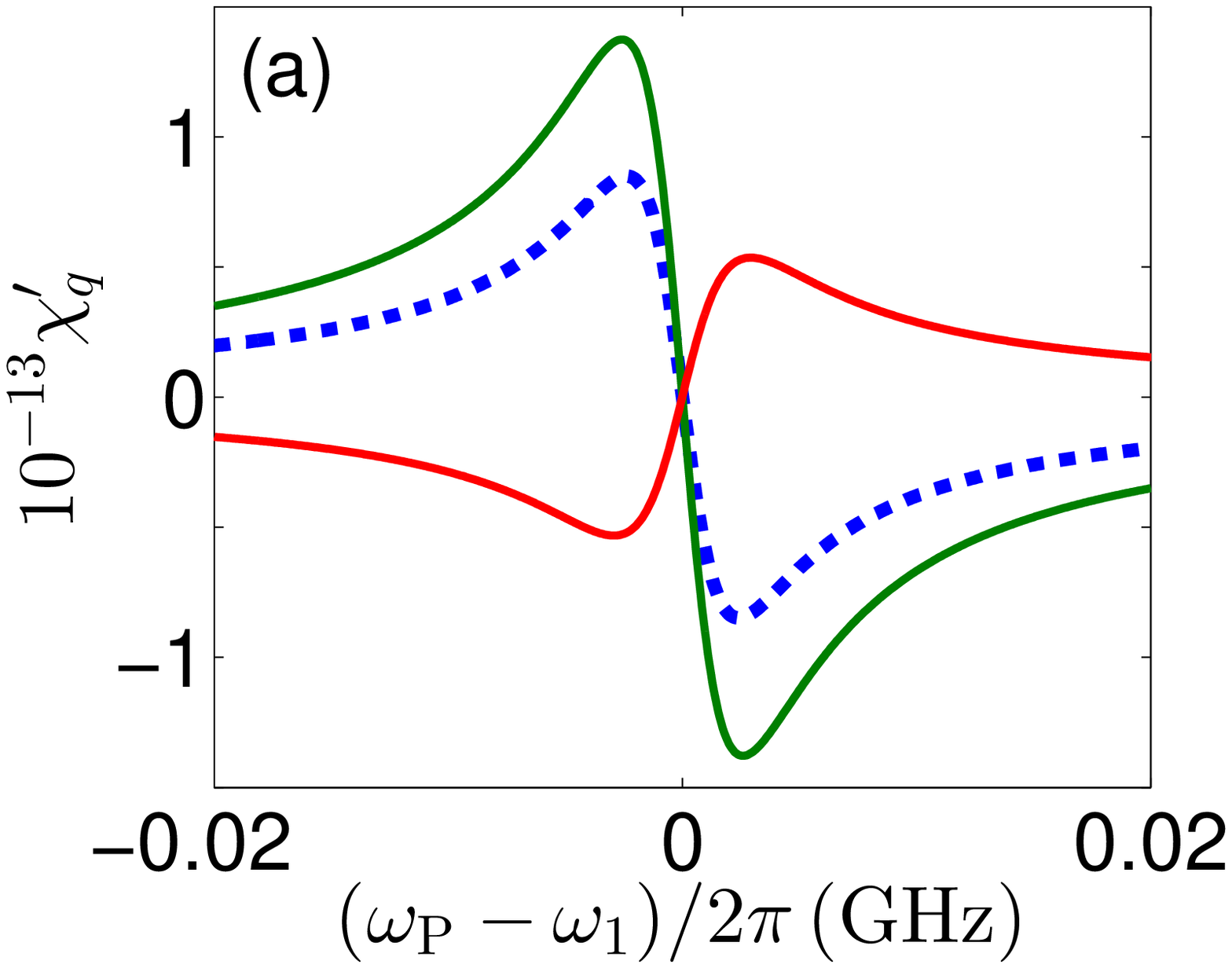}
\includegraphics[bb=25 200 520 600, width=4.2 cm, clip]{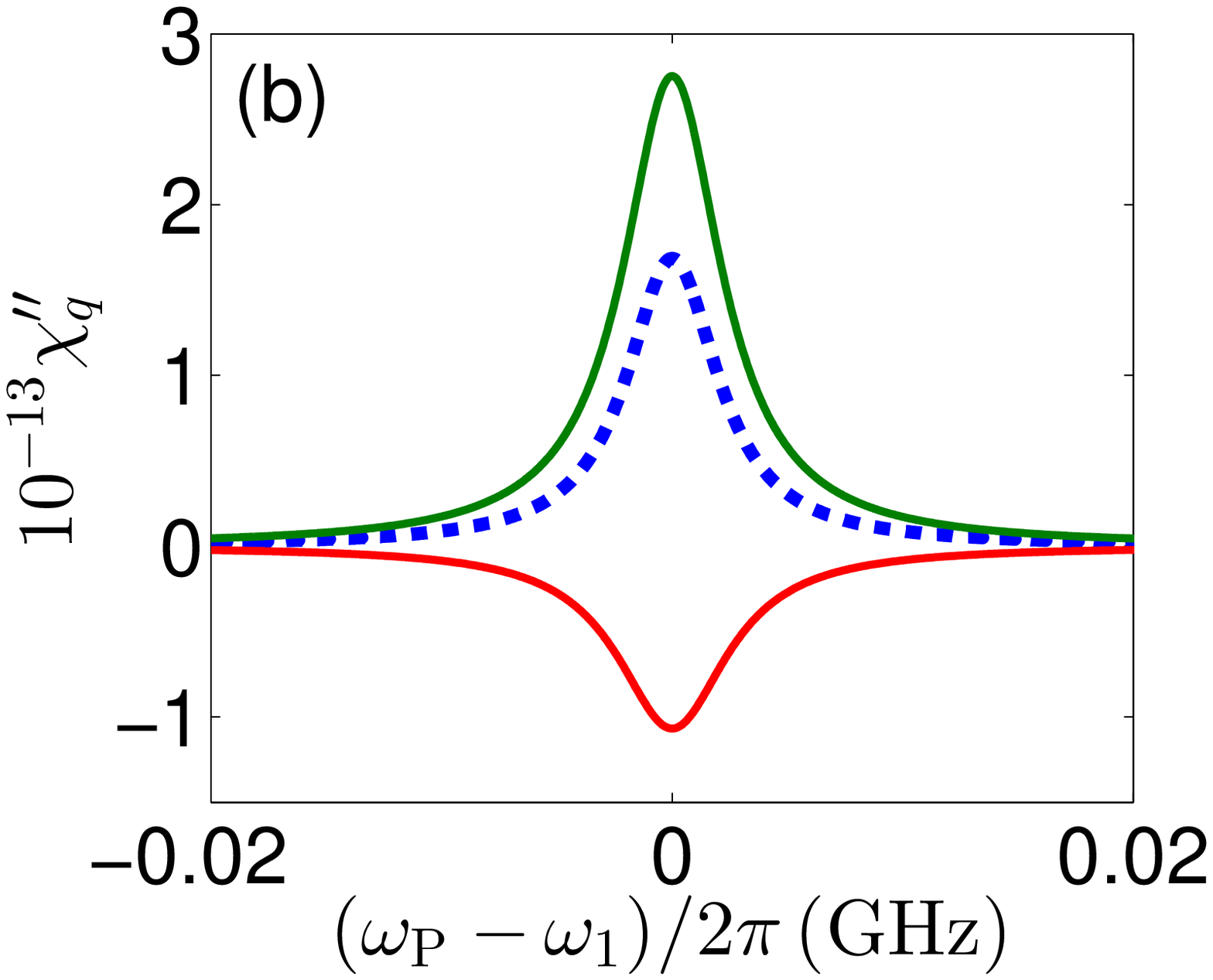}
\includegraphics[bb=25 200 520 600, width=4.2 cm, clip]{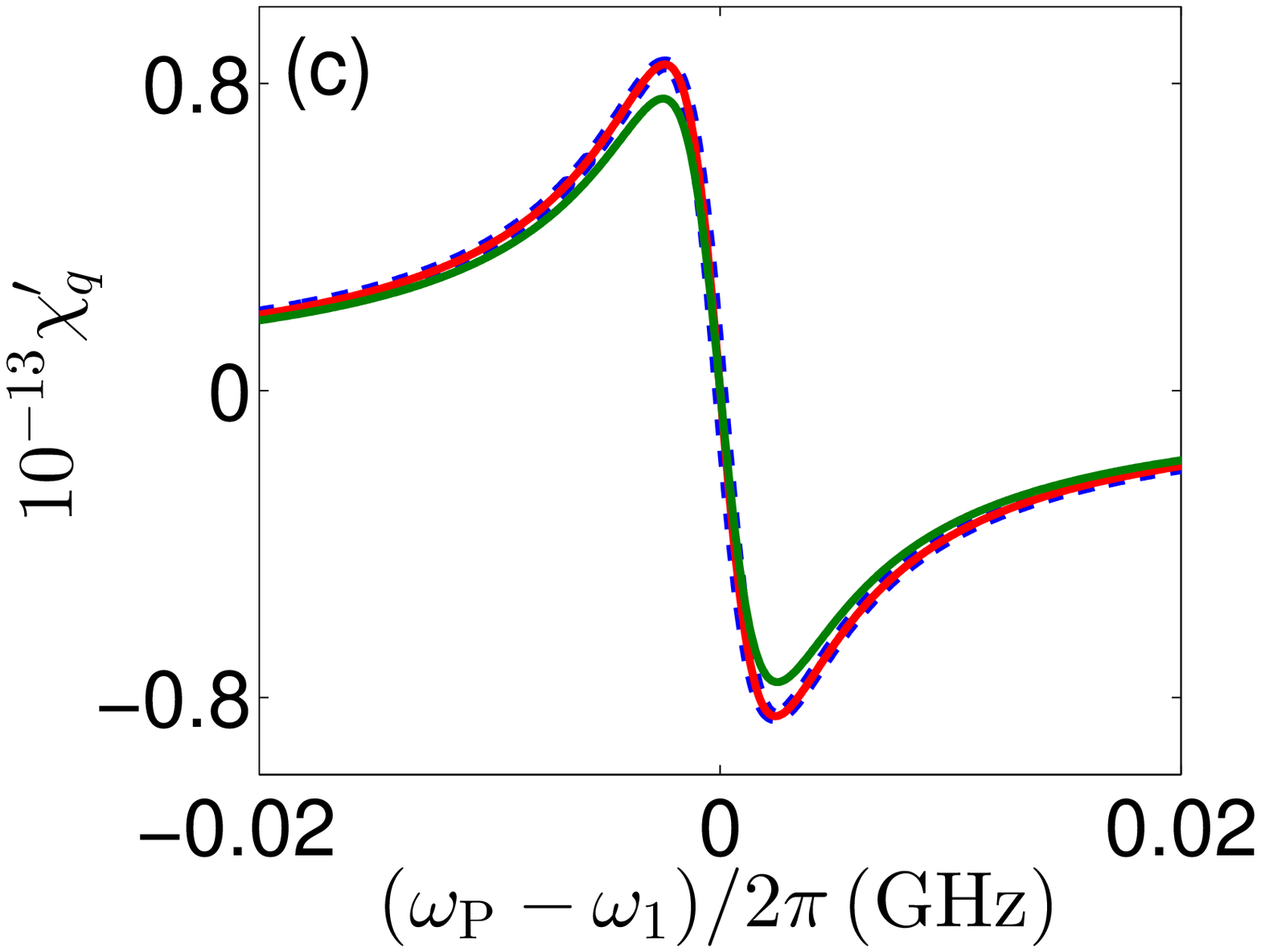}
\includegraphics[bb=25 200 520 600, width=4.2 cm, clip]{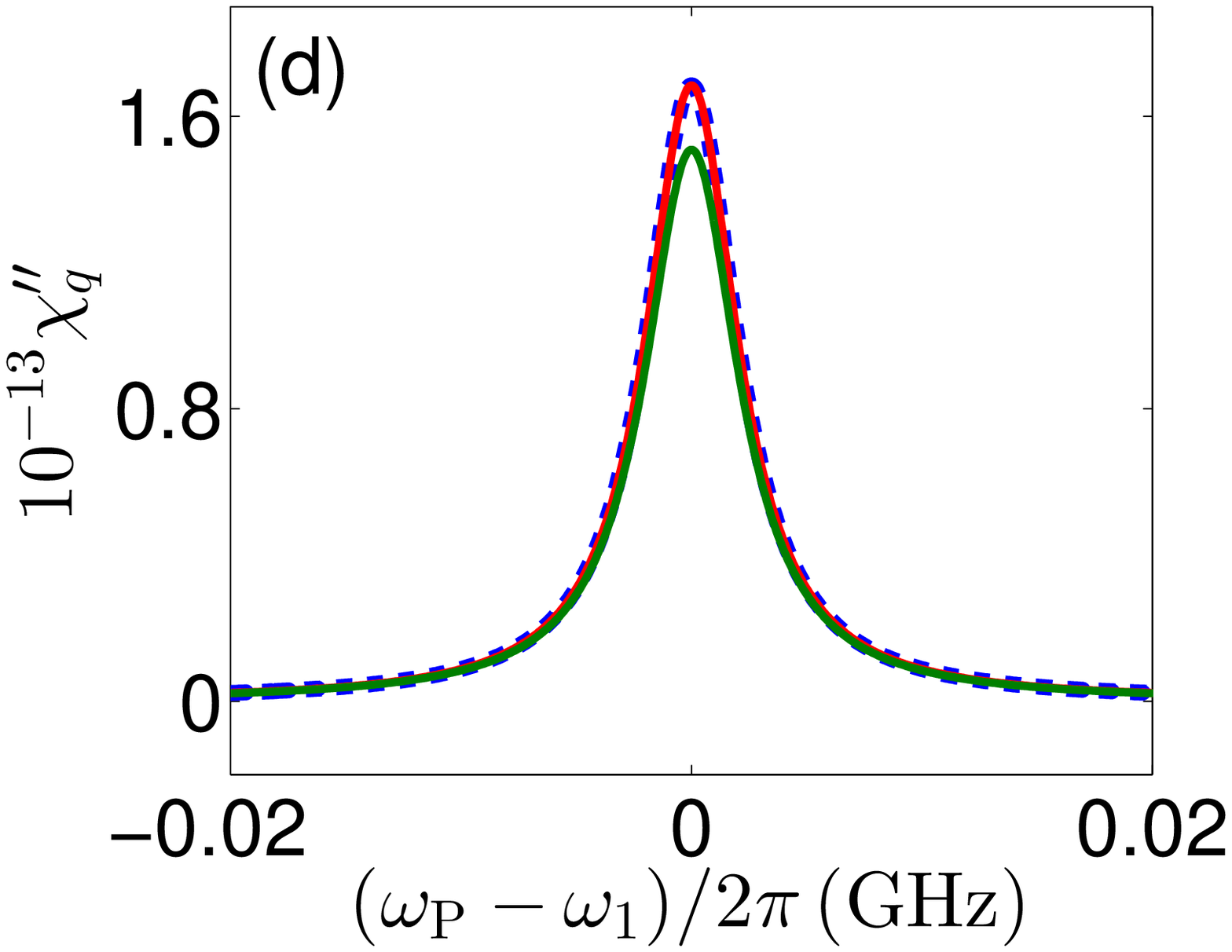}
\caption[]{(Color online)
The real and imaginary parts of $\chi_{q}(\omega)$ (blue dashed curve),
$R_{+}^{(01)}(\delta_1)$ (green solid curve), and $R_{-}^{(01)}(\delta_1)$ (red solid curve)
in (a) and (b) as a function of the detuning of the probe field $\omega_{\rm P}-\omega_1$.
These are plotted for $T=25$ mK and $\Delta=0$.
The real and imaginary parts of $\chi_{q}(\omega)$ in (c) and (d)
as a function of the detuning of the probe field $\omega_{\rm P}-\omega_1$.
These are plotted for:
(i) same parameters as in (a) and (b) (blue dashed curve);
(ii) $T=25$ mK, $\Delta/2\pi=0.37$ MHz (red solid curve);
(iii) $T=50$ mK, $\Delta=0$ (green solid curve).
Here we assume that $f=0.5$, $\beta=10^{-4}$, and $|\Omega_{\rm D}|/2\pi=0.37$ MHz.
The other parameters of the three-level SFQC are provided in the two numbered items at the beginning of Sec.~\ref{NUM}.
}\label{fig5}
\end{figure}

\begin{figure}
\includegraphics[bb=25 200 520 600, width=4.2 cm, clip]{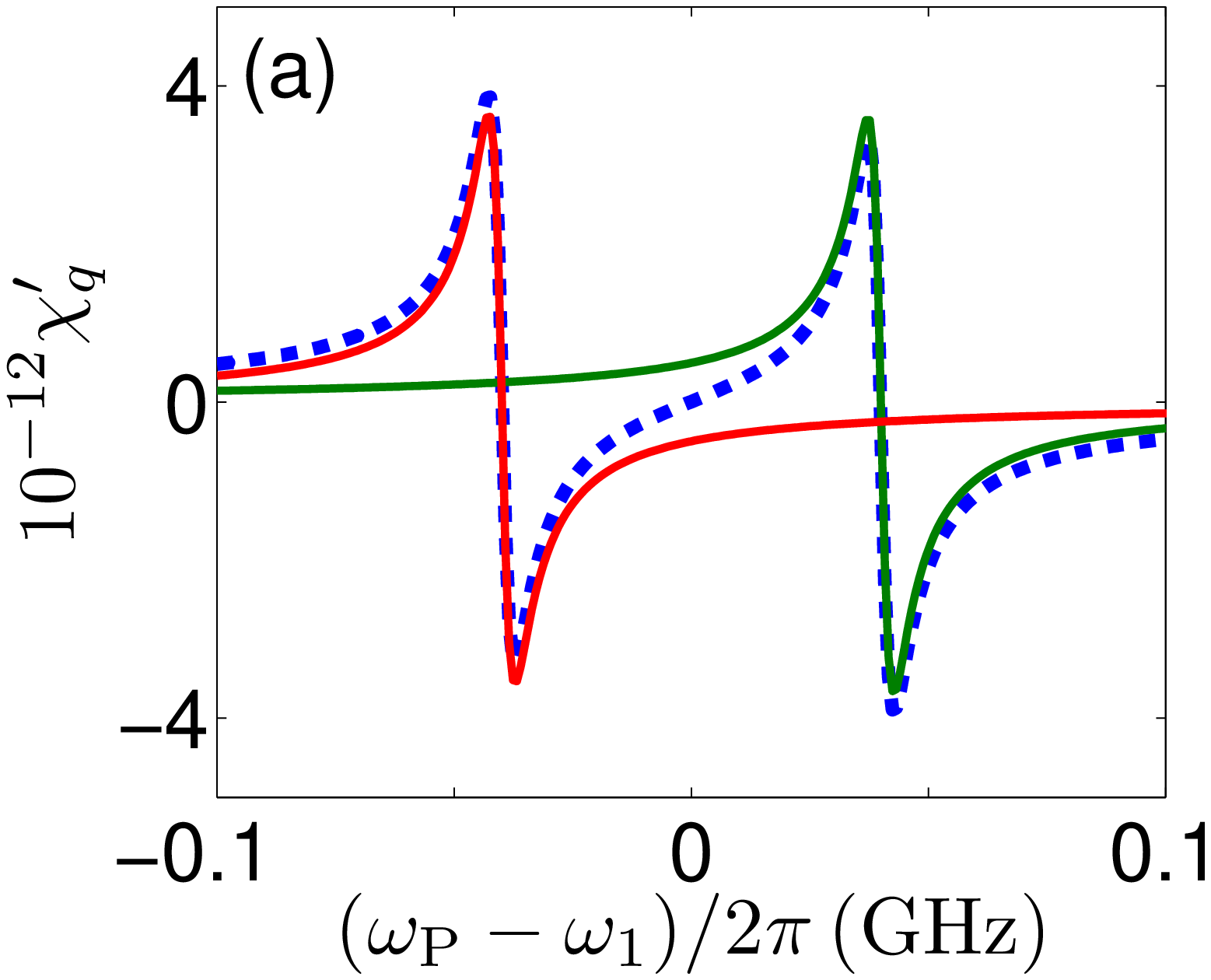}
\includegraphics[bb=25 200 520 600, width=4.2 cm, clip]{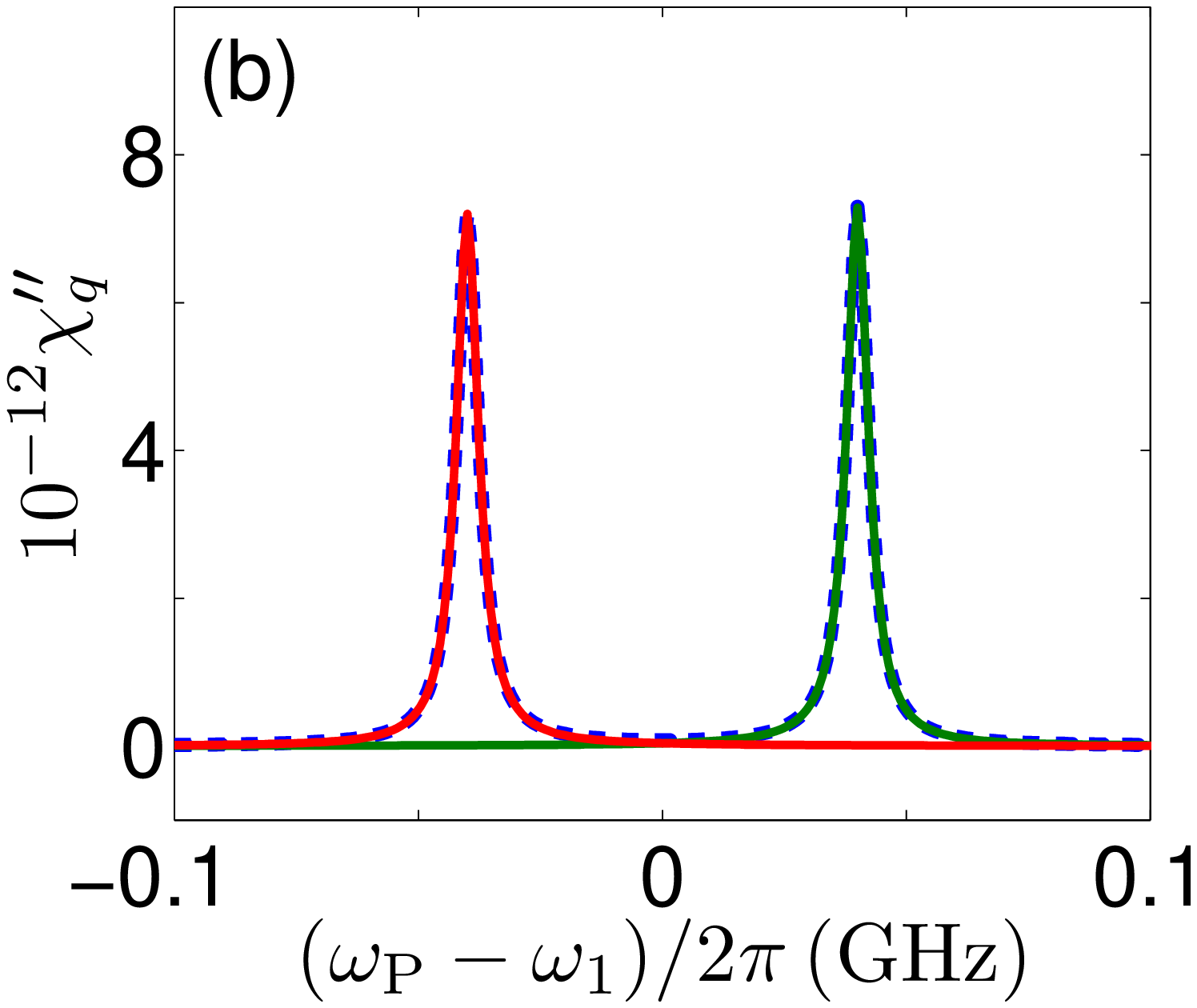}
\includegraphics[bb=25 200 520 600, width=4.2 cm, clip]{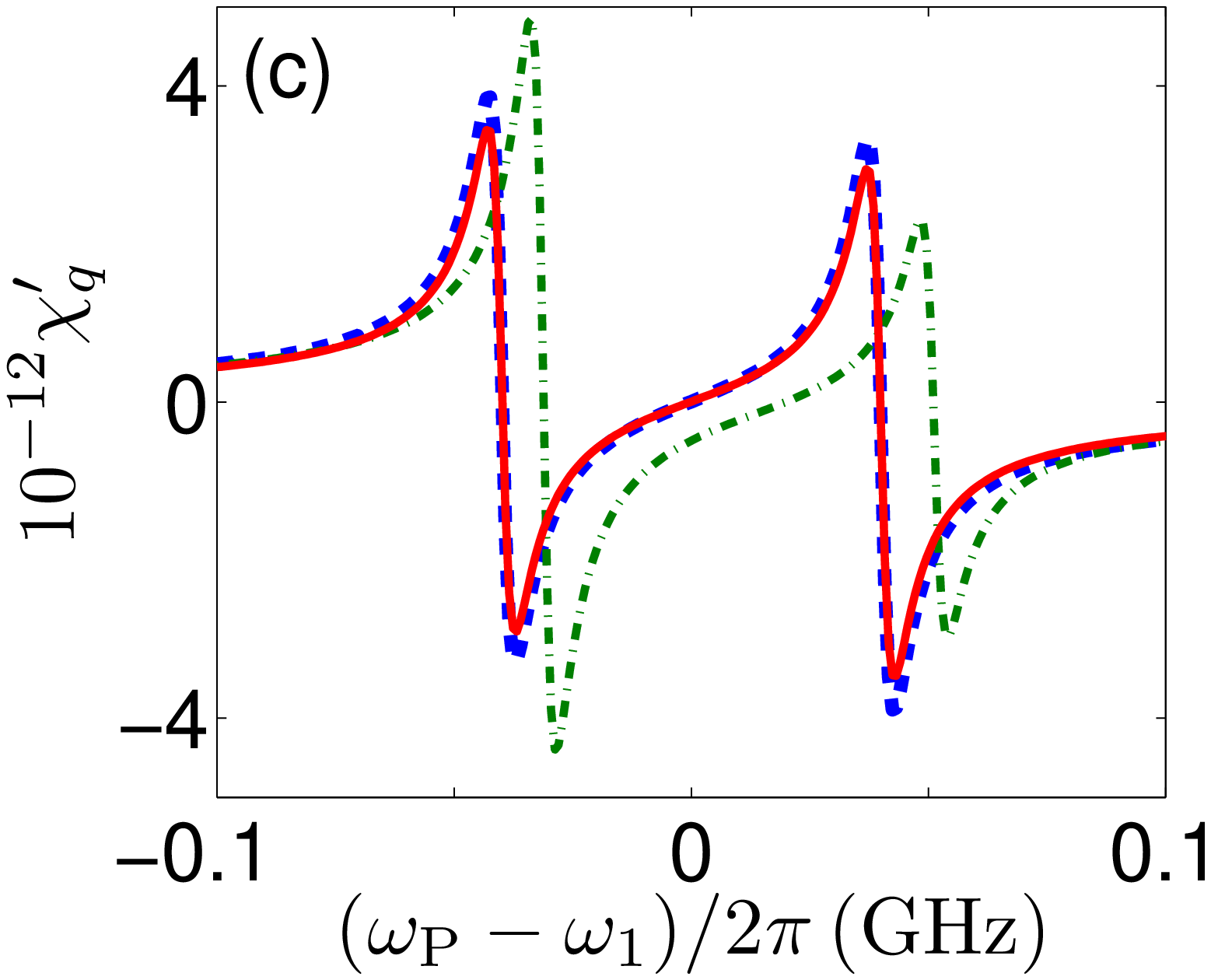}
\includegraphics[bb=25 200 520 600, width=4.2 cm, clip]{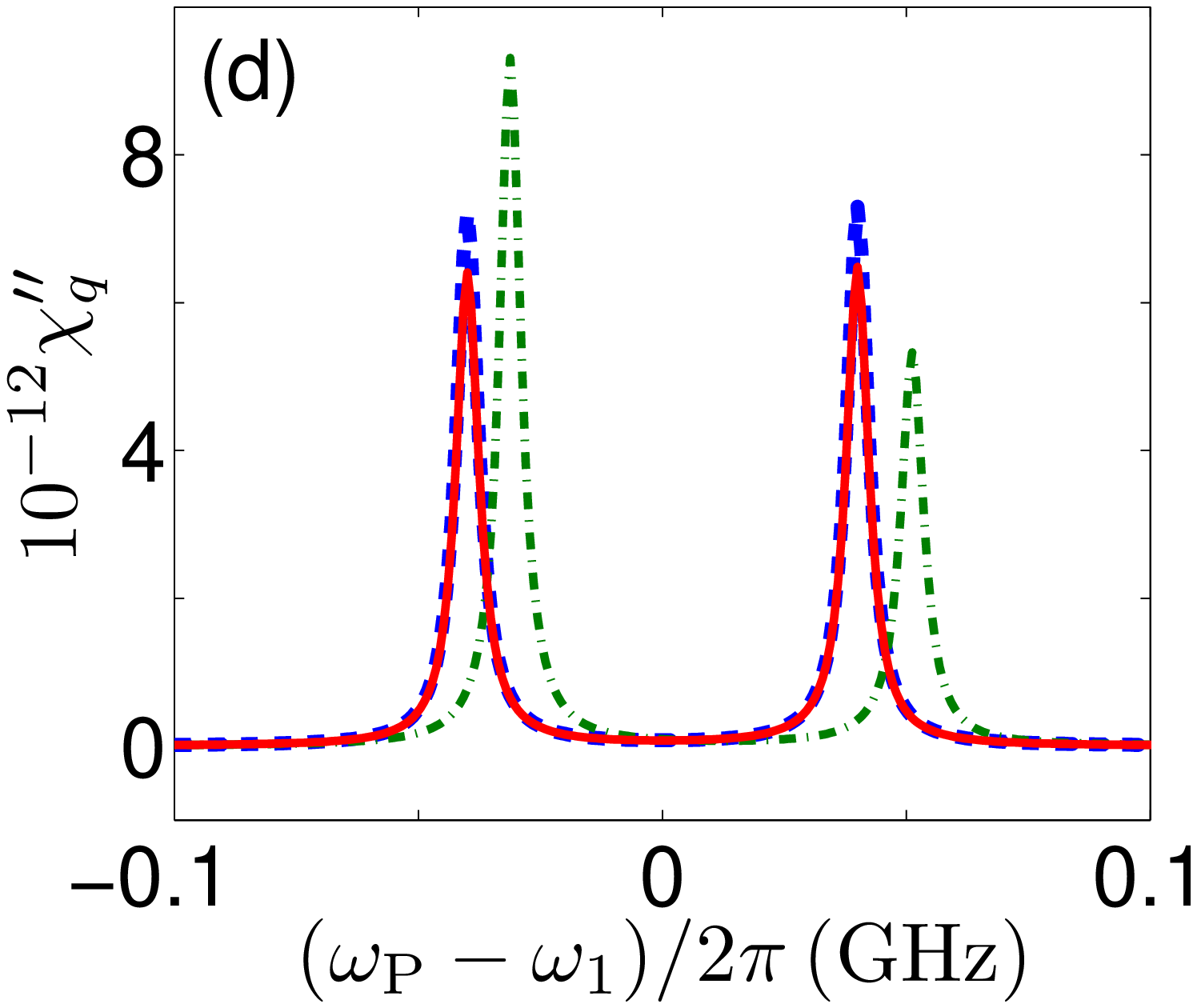}
\caption[]{(Color online)
The real and imaginary parts of $\chi_{q}(\omega)$ (blue dashed curve),
$R_{+}^{(01)}(\delta_1)$ (green solid curve), and $R_{-}^{(01)}(\delta_1)$ (red solid curve)
in (a) and (b) as a function of the detuning of the probe field $\omega_{\rm P}-\omega_1$.
These are plotted for $T=25$ mK and $\Delta=0$.
The real and imaginary parts of $\chi_{q}(\omega)$ in (c) and (d)
as a function of the detuning of the probe field $\omega_{\rm P}-\omega_1$.
These are plotted for:
(i) same parameters as in (a) and (b) (blue dashed curve);
(ii) $T=25$ mK, $\Delta/2\pi=20$ MHz (green dash-dotted curve);
(iii) $T=50$ mK, $\Delta=0$ (red solid curve).
Here we assume that $f=0.5$, $\beta=10^{-4}$, and $|\Omega_{\rm D}|/2\pi=40$ MHz.
The other parameters of the three-level SFQC are provided in the two numbered items at the beginning of Sec.~\ref{NUM}.
}\label{fig6}
\end{figure}

\begin{figure*}
\includegraphics[bb=45 260 650 620, width=8 cm, clip]{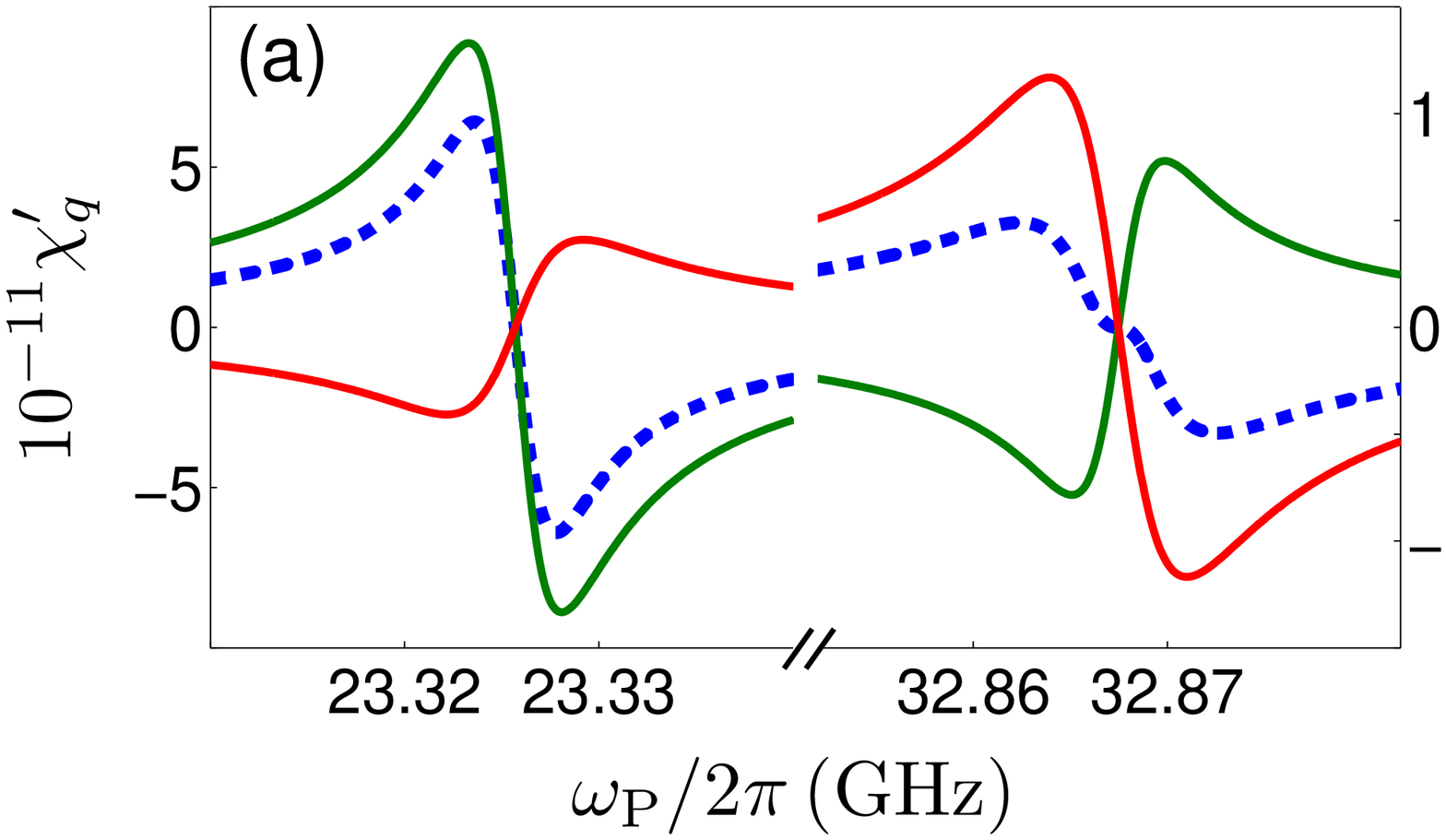}
\includegraphics[bb=45 260 650 620, width=8 cm, clip]{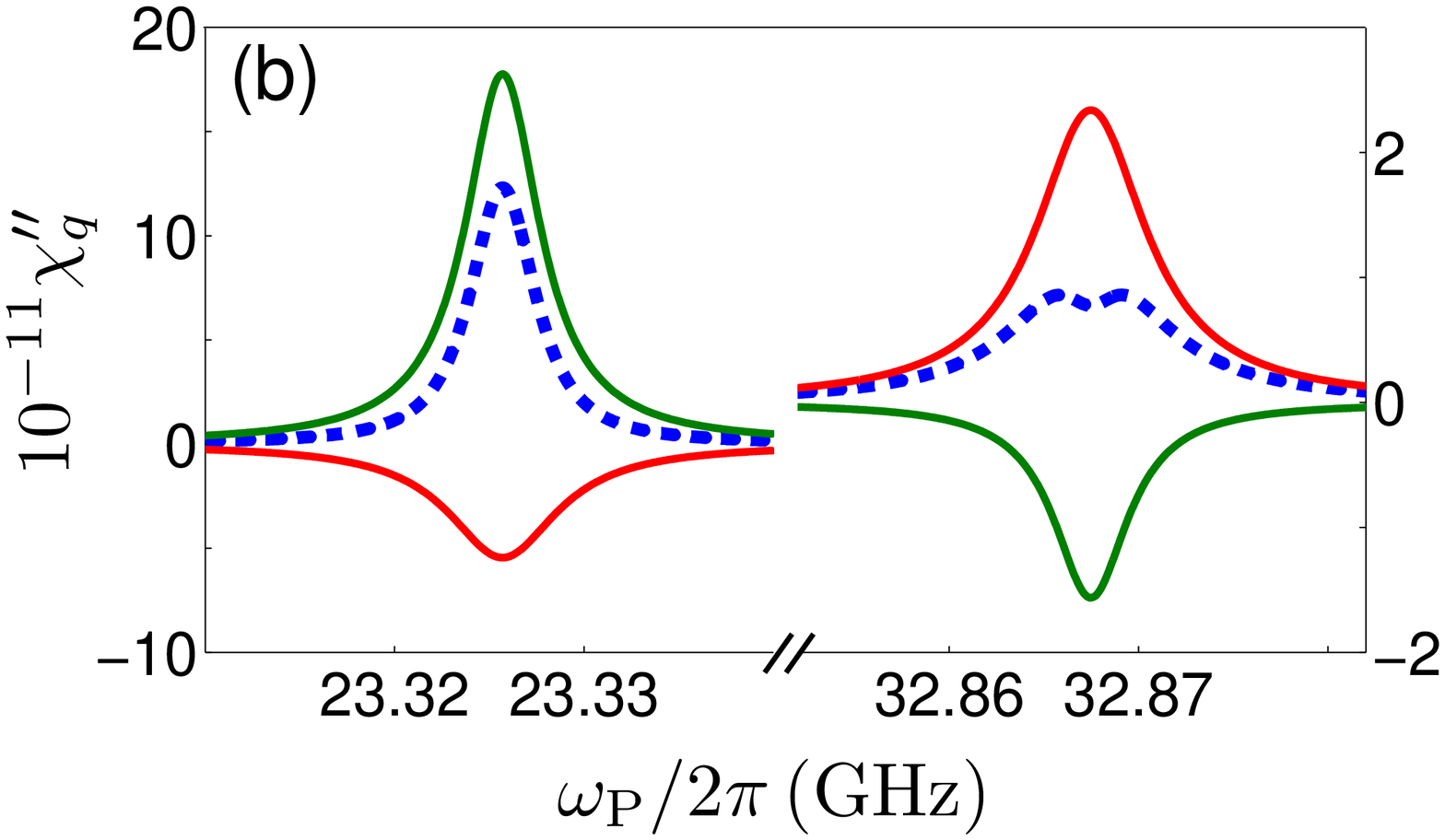}
\includegraphics[bb=45 260 650 620, width=8 cm, clip]{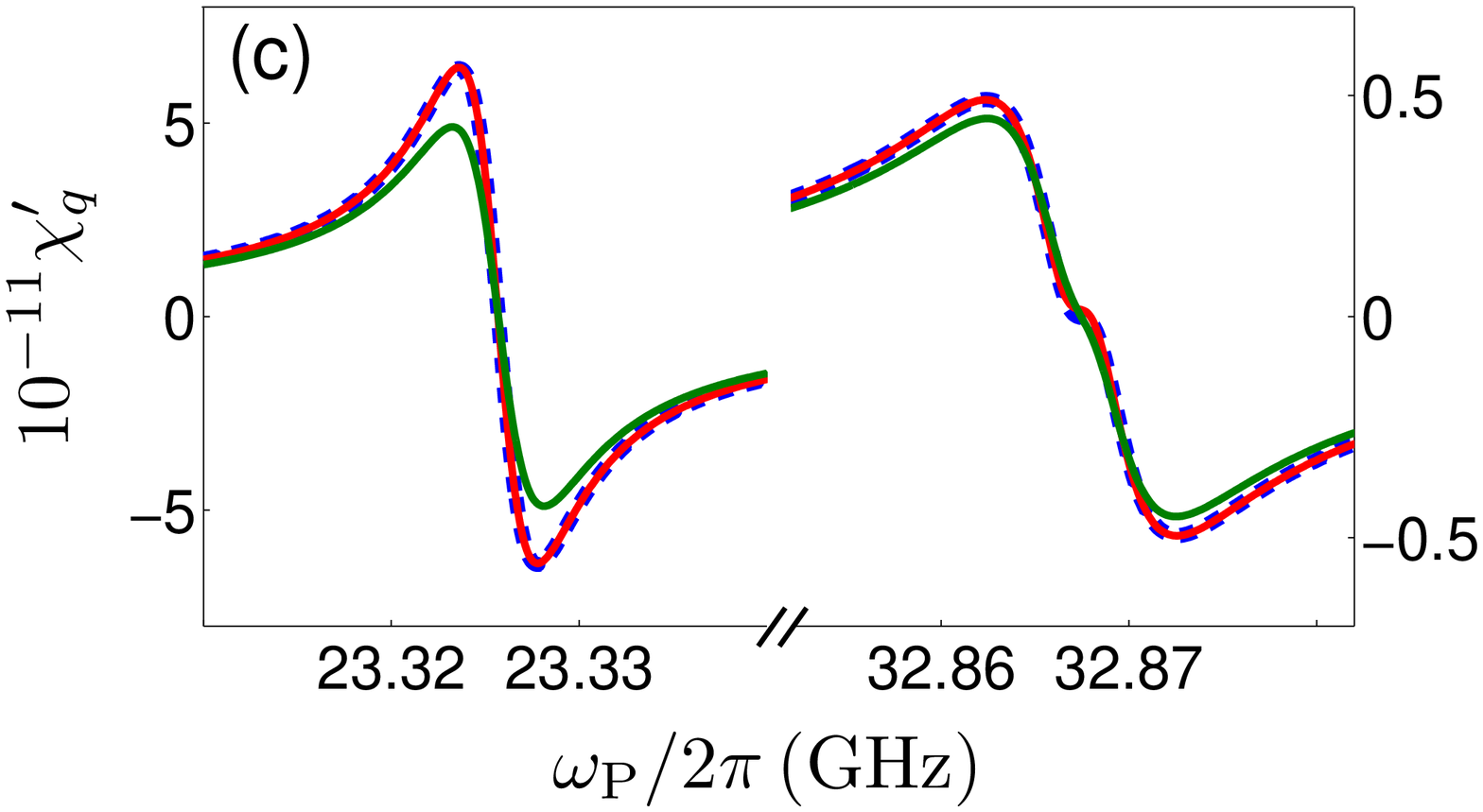}
\includegraphics[bb=45 260 650 620, width=8 cm, clip]{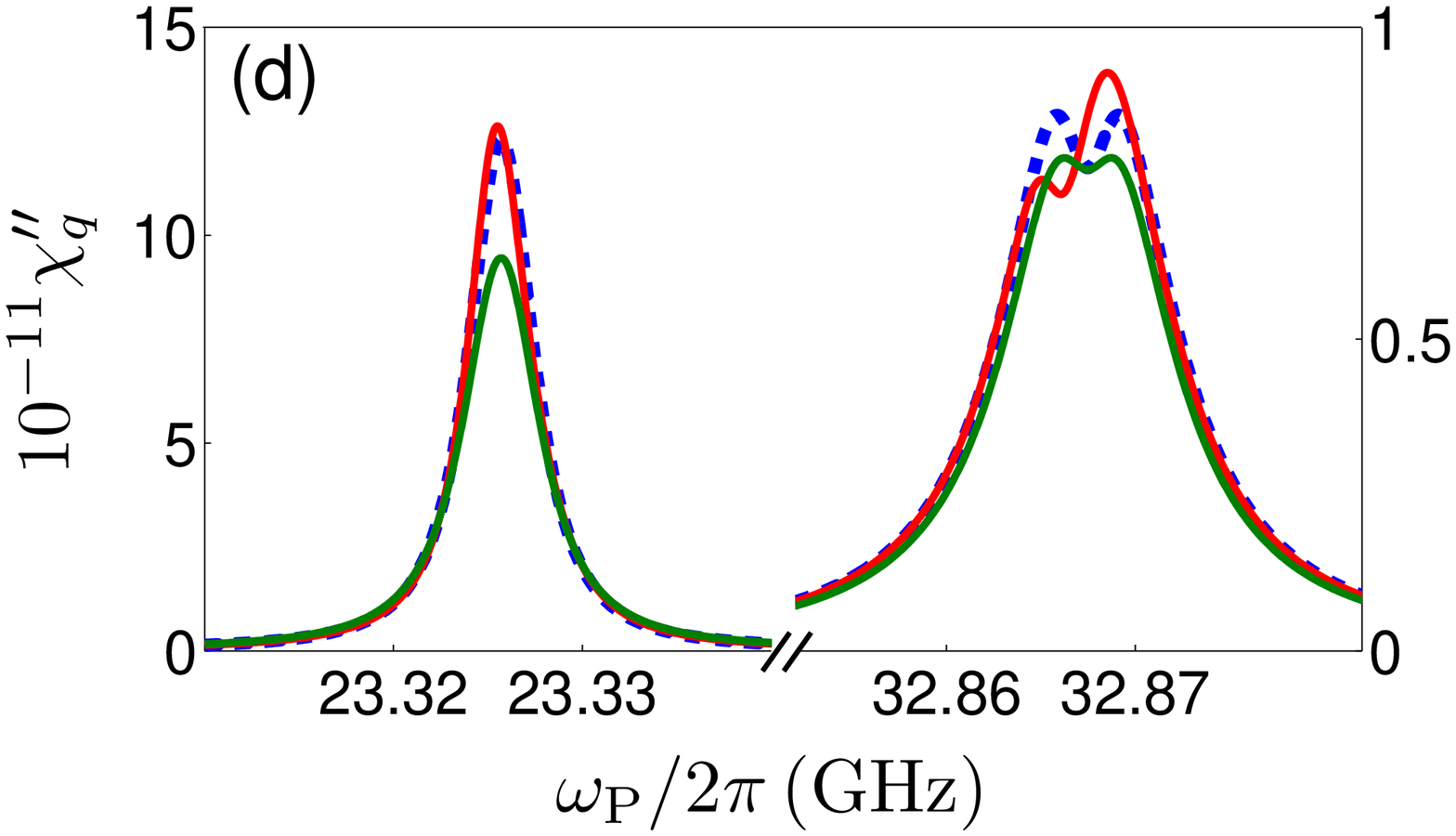}
\caption[]{(Color online)
The real and imaginary parts of $\chi_{q}(\omega)$ (blue dashed curve) in (a) and (b)
are shown as a function of the probe field frequency $\omega_{\rm P}/(2\pi)$.
In the left parts of (a) and (b), the real and imaginary parts of
$R_{+}^{(01)}(\delta_1)$ (green solid curve) and $R_{-}^{(01)}(\delta_1)$ (red solid curve)
are shown as a function of the probe field frequency $\omega_{\rm P}/(2\pi)$.
In the right parts of (a) and (b), the real and imaginary parts of
$R_{+}^{(02)}(\delta_1)$ (green solid curve) and $R_{-}^{(02)}(\delta_1)$ (red solid curve)
are shown as a function of the probe field frequency $\omega_{\rm P}/(2\pi)$.
In (a) and (b), all curves are plotted with $T=25$ mK and $\Delta=0$.
The real and imaginary parts of the susceptibility $\chi_{q}(\omega)$ in (c) and (d)
are shown as a function of the probe field frequency $\omega_{\rm P}/(2\pi)$ for:
(i) same parameters as in (a) and (b) (blue dashed curve);
(ii) $T=25$ mK, $\Delta/2\pi=1.4$ MHz (red solid curve);
(iii) $T=50$ mK, $\Delta=0$ (green solid curve).
Here we assume that $f=0.525$, $\beta=10^{-4}$, and $|\Omega_{\rm D}|/2\pi=1.4$ MHz.
The other parameters of the three-level SFQC are provided in the two numbered items at the beginning of Sec.~\ref{NUM}.
Note that in (a)--(d), the left vertical axes are for the left parts of the figures,
while the right vertical axes are for the right parts of the figures.
}\label{fig7}
\end{figure*}

\subsection{Damping rates}

Using the fluctuation-dissipation theorem in Eq.~(\ref{FDT}),
the damping rates $\gamma_{lm}={\rm Im}(\Gamma_{lm})$ ($l,\,m=1,\,2$) can be rewritten (see Appendix \ref{ApGM}).
In principle, $\gamma_{lm}$ ($l,\,m=1,\,2$) can be numerically calculated by their expressions.
The damping rates $\gamma_{lm}$ ($l,\,m=1,\,2$) are plotted in
Fig.~\ref{fig4} as a function of $f$, $T$, and $|\Omega_{\rm D}|$.

We find in Fig.~\ref{fig4}(a) that the damping rate $\gamma_{11}$ reaches its maximum at the optimal point
and decreases as $f$ deviates from 0.5 when the environmental equilibrium temperature $T=0$.
However, when $T \neq 0$, the damping rate $\gamma_{11}$ reaches its minimum at the optimal point.
As $f$ deviates from 0.5, $\gamma_{11}$ first increases,
and then decreases after reaching its maximum point.
We find in Fig.~\ref{fig4}(a) that the damping rate $\gamma_{22}$ reaches its minimum at the optimal point.
As $f$ deviates from 0.5, $\gamma_{22}$ first increases,
and then decreases after reaching its maximum point.
In Fig.~\ref{fig4}(b),
we find that the damping rates $\gamma_{11}$ and $\gamma_{22}$ increase when the environmental equilibrium temperature $T$ goes up.
In Figs.~\ref{fig4}(c) and \ref{fig4}(d),
we find that $\gamma_{11}$ and $\gamma_{22}$ are almost not affected by the Rabi frequency $|\Omega_{\rm D}|$.
Whereas in Figs.~\ref{fig4}(e) and \ref{fig4}(f),
we find that $\gamma_{12}$ and $\gamma_{12}$ are almost linearly dependent on the Rabi frequency $|\Omega_{\rm D}|$.

\begin{figure*}
\includegraphics[bb=45 260 650 620, width=8 cm, clip]{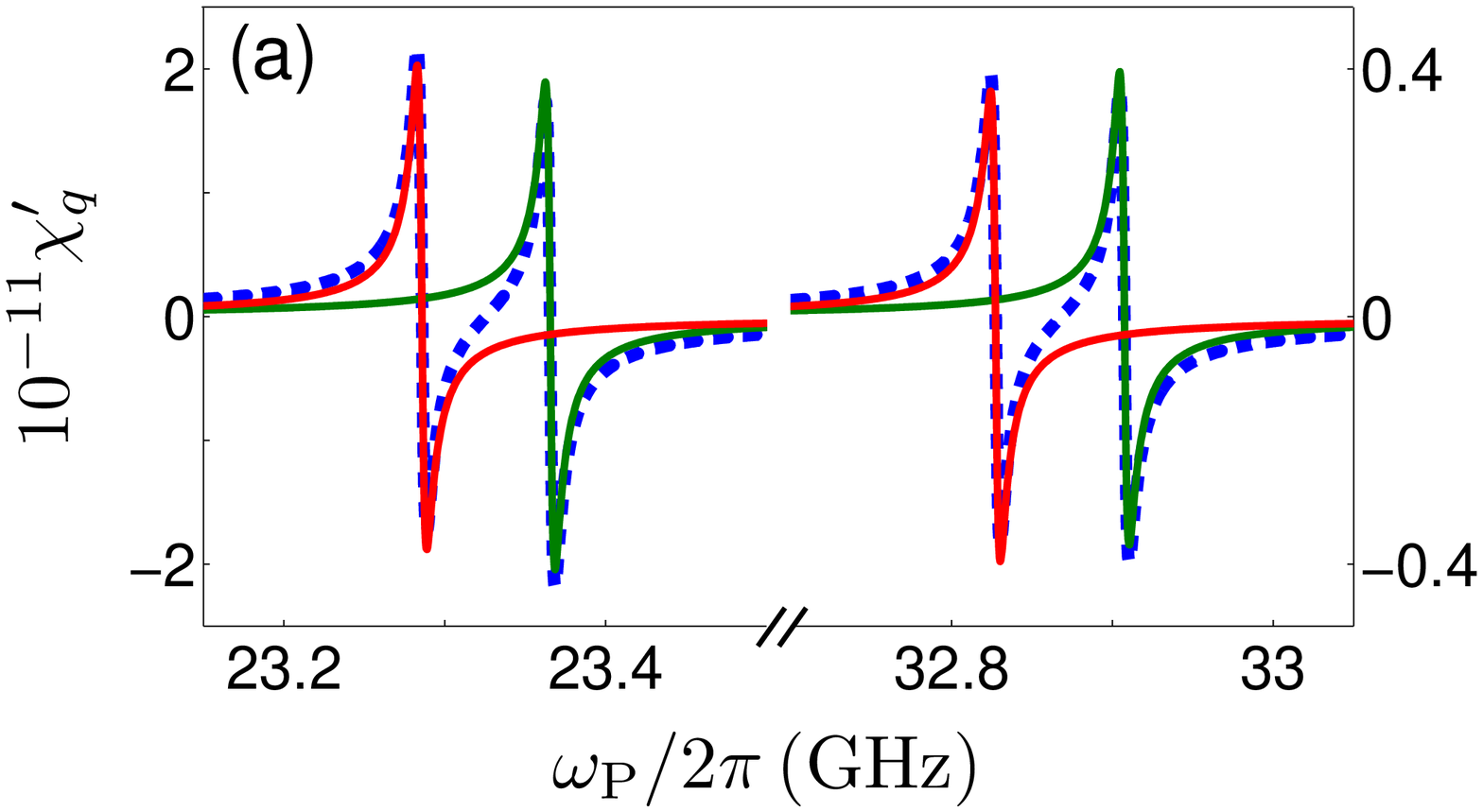}
\includegraphics[bb=45 260 650 620, width=8 cm, clip]{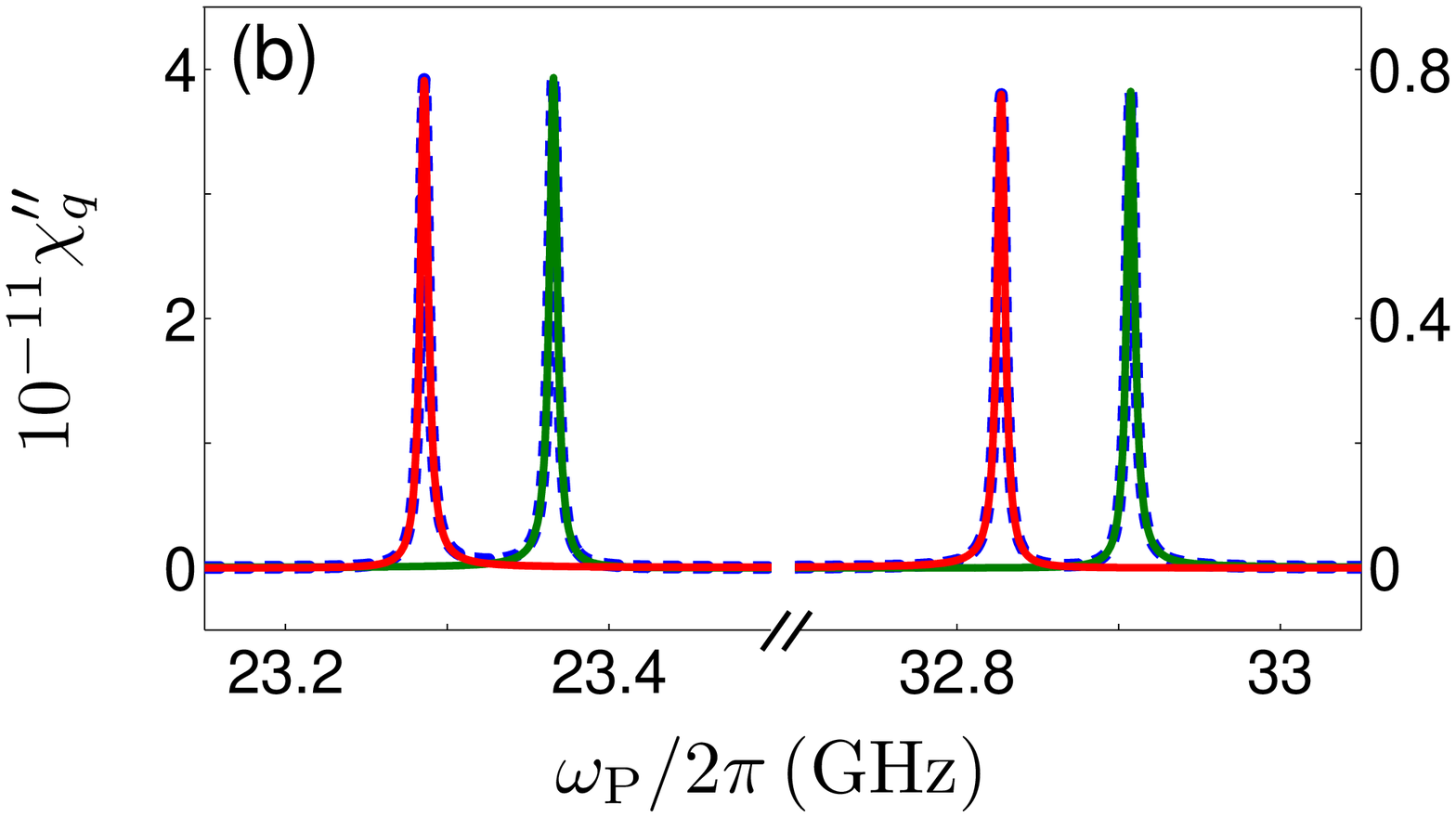}
\includegraphics[bb=45 260 650 620, width=8 cm, clip]{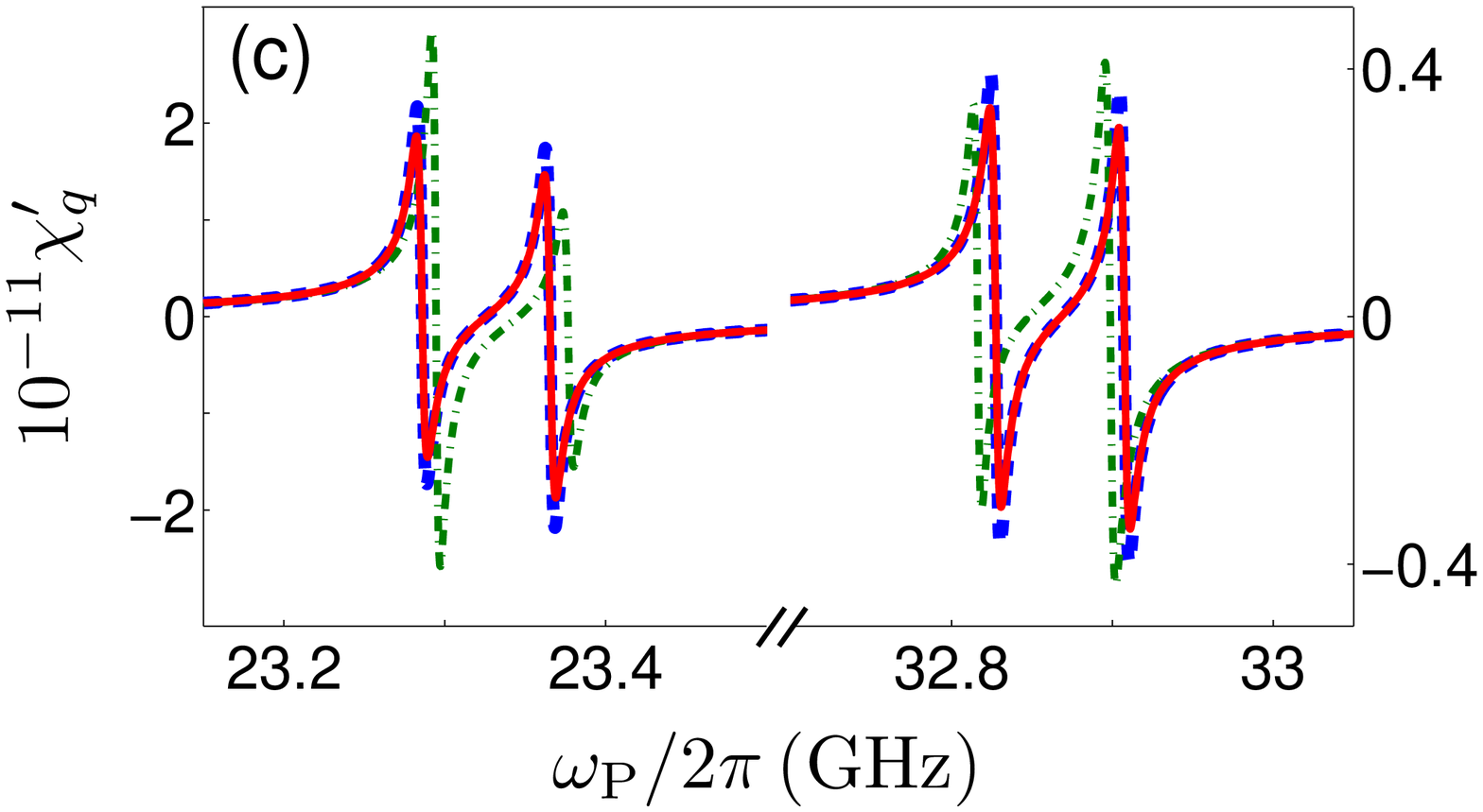}
\includegraphics[bb=45 260 650 620, width=8 cm, clip]{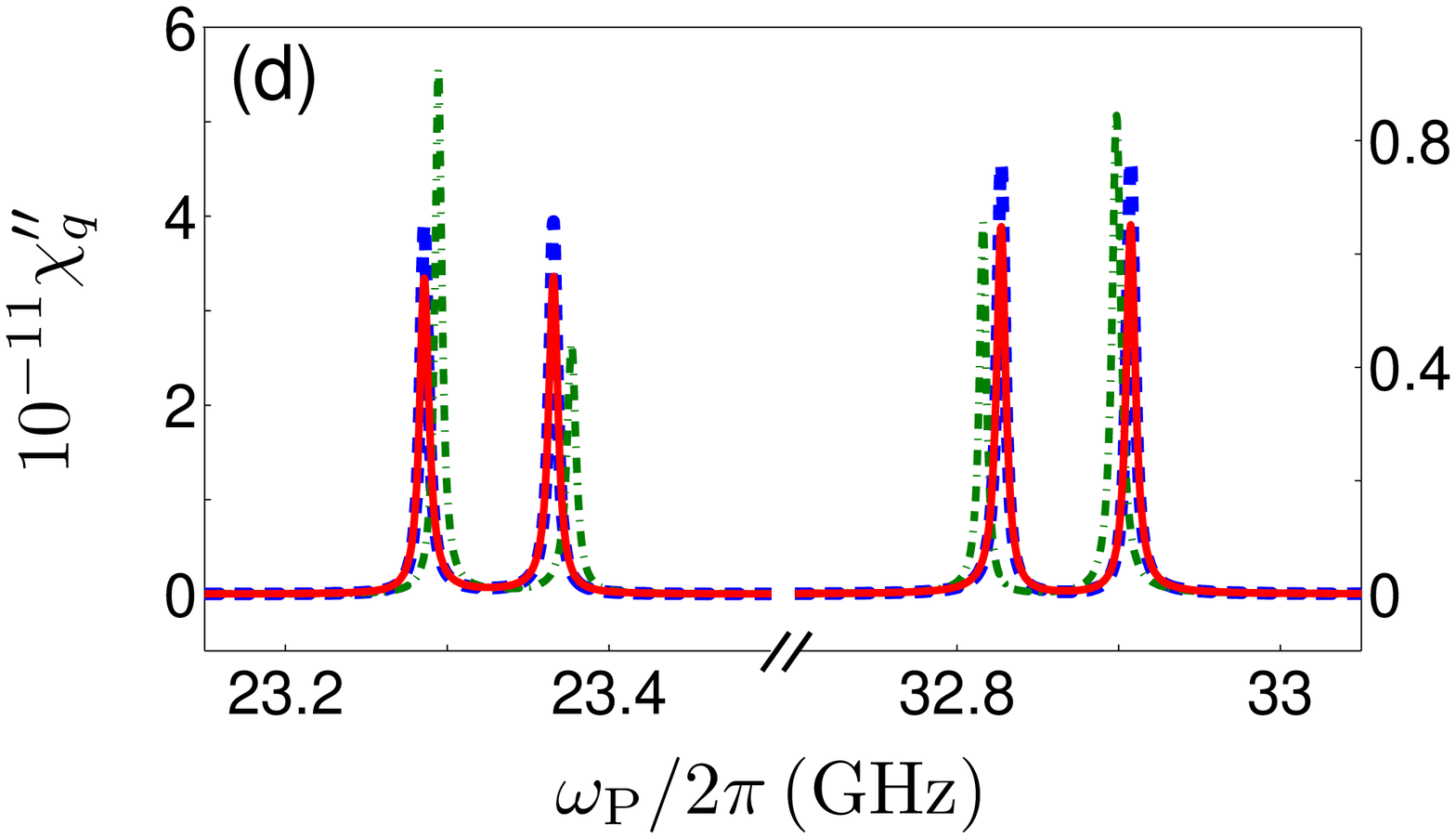}
\caption[]{(Color online)
The real and imaginary parts of $\chi_{q}(\omega)$ (blue dashed curve) in (a) and (b)
are shown as a function of the probe field frequency $\omega_{\rm P}/(2\pi)$.
In the left parts of (a) and (b), the real and imaginary parts of
$R_{+}^{(01)}(\delta_1)$ (green solid curve) and $R_{-}^{(01)}(\delta_1)$ (red solid curve)
are shown as a function of the probe field frequency $\omega_{\rm P}/(2\pi)$.
In the right parts of (a) and (b), the real and imaginary parts of
$R_{+}^{(02)}(\delta_1)$ (green solid curve) and $R_{-}^{(02)}(\delta_1)$ (red solid curve)
are shown as a function of the probe field frequency $\omega_{\rm P}/(2\pi)$.
In (a) and (b), all curves are plotted with $T=25$ mK and $\Delta=0$.
The real and imaginary parts of the susceptibility $\chi_{q}(\omega)$ in (c) and (d)
are shown as a function of the probe field frequency $\omega_{\rm P}/(2\pi)$
for:
(i) same parameters as in (a) and (b) (blue dashed curve);
(ii) $T=25$ mK, $\Delta/2\pi=20$ MHz (green dash-dotted curve);
(iii) $T=50$ mK, $\Delta=0$ (red solid curve).
Here we assume that $f=0.525$, $\beta=10^{-4}$, and $|\Omega_{\rm D}|/2\pi=40$ MHz.
The other parameters of the three-level SFQC are provided in the two numbered items at the beginning of Sec.~\ref{NUM}.
Note that in (a)--(d), the left vertical axes are for the left parts of the figures,
while the right vertical axes are for the right parts of the figures.
}\label{fig8}
\end{figure*}

\subsection{Susceptibilities of the three-level SFQC to the probe field when $f=0.5$}

Let us first study the linear response of the three-level SFQC at the optimal point,
i.e., the reduced magnetic flux $f=0.5$,
where the three-level SFQC has a ladder-type transition which can also occur in three-level natural atoms.
In this case, the susceptibility $\chi_{q}(\omega_{\rm P})$ can be given as
$\chi_{q}(\omega_{\rm P})=\chi_{01}(\omega_{\rm P})$,
and EIT (ATS) may occur in the $|0\rangle\leftrightarrow|1\rangle$ frequency range.

First, we study the susceptibility $\chi_{q}(\omega_{\rm P})$ in the weak-driving regime.
In Figs.~\ref{fig5}(a) and \ref{fig5}(b), the real and imaginary parts of the susceptibility
$\chi_{q}(\omega_{\rm P})$ and the two resonances $R_{\pm}^{(01)}(\omega_{\rm P}-\omega_1)$ are plotted.
In Figs.~\ref{fig5}(c) and \ref{fig5}(d), the real and imaginary parts of the susceptibility
$\chi_{q}(\omega_{\rm P})$ are plotted for different parameters.
All the parameters in Figs.~\ref{fig5}(a)--\ref{fig5}(d) are appropriately chosen to guarantee
that the Rabi frequency $|\Omega_{\rm D}|$ lies in the weak-driving regime.

In Fig.~\ref{fig5}(b), we find that ${\rm Im}[R_{\pm}^{(01)}(\omega_{\rm P}-\omega_1)]$
are centered at $\omega_{\rm P}=\omega_1$ with different signals;
however their summation ${\rm Im}[\chi_{q}(\omega_{\rm P})]$ is an absorption peak
without a dip at $\omega_{\rm P}=\omega_1$ and EIT does not occur.
Actually, the necessary condition $\gamma_{11}>2\gamma_{22}$
for realizing EIT in the $|0\rangle\leftrightarrow|1\rangle$ frequency range
cannot be satisfied at the optimal point when $0$ mK $<T<100$ mK (see blue curves in Fig.~\ref{fig4}(b)),
thus EIT cannot occur for any value of the Rabi frequency $|\Omega_{\rm D}|$
in this temperature range with the damping rates given in Fig.~\ref{fig4}.
In Fig.~\ref{fig5}(d), we find that the height of the absorption peak decreases
when the environmental equilibrium temperature $T$ goes up.
This is because the damping rates $\gamma_{11}$ and $\gamma_{22}$ increase
when increasing $T$ as shown in Fig.~\ref{fig4}(b).
We also find that $\chi_{q}(\omega_{\rm P})$ is not apparently affected
by the nonzero driving-field detuning $\Delta$.

Second, we study the susceptibility $\chi_{q}(\omega_{\rm P})$ in the strong-driving regime.
In Figs.~\ref{fig6}(a) and \ref{fig6}(b), the real and imaginary parts of the susceptibility
$\chi_{q}(\omega_{\rm P})$ and the two resonances $R_{\pm}^{(01)}(\omega_{\rm P}-\omega_1)$ are plotted.
In Figs.~\ref{fig6}(c) and \ref{fig6}(d), the real and imaginary parts of the susceptibility
$\chi_{q}(\omega_{\rm P})$ are plotted for different parameters.
All the parameters in Figs.~\ref{fig6}(a)--\ref{fig6}(d) are appropriately chosen to guarantee
that the Rabi frequency $|\Omega_{\rm D}|$ lies in the strong-driving regime.

In Fig.~\ref{fig6}(b), the condition for realizing ATS in the $|0\rangle\leftrightarrow|1\rangle$ frequency range is fulfilled,
and ATS can be realized when $\omega_{\rm P}$ is
near resonant to the the $|0\rangle\leftrightarrow|1\rangle$ transition.
The imaginary part of the susceptibility $\chi_{q}(\omega_{\rm P})$ is made up with two absorption peaks,
corresponding to the two resonances ${\rm Im}[R_{+}^{(01)}(\omega_{\rm P}-\omega_1)]$
and ${\rm Im}[R_{-}^{(01)}(\omega_{\rm P}-\omega_1)]$.
In Fig.~\ref{fig6}(d), we find that the heights of the absorption peaks decrease
when the environmental equilibrium temperature $T$ goes up.
Same as in the weak-driving regime,
this is because the damping rates $\gamma_{11}$ and $\gamma_{22}$ increase when increasing $T$.
We also find that the nonzero driving-field detuning $\Delta$
makes the two absorption peaks have different heights and asymmetric positions.
In addition, we find that two absorption peaks have different heights
(see blue dashed curves in Figs.~\ref{fig6}(b) and \ref{fig6}(d))
when the three-level SFQC is resonantly driven, i.e., $\Delta=0$.
This is because the damping rates $\gamma_{12}$ and $\gamma_{21}$
make the imaginary parts of the two resonances ${\rm Im}[R_{+}^{(01)}(\omega_{\rm P}-\omega_1)]$
and ${\rm Im}[R_{-}^{(01)}(\omega_{\rm P}-\omega_1)]$ have different heights,
as discussed in Subsec.~C of Sec.~\ref{MS}.

\subsection{Susceptibilities of the three-level SFQC to the probe field when $f\neq 0.5$}

We now study the linear response of the three-level SFQC
when the bias magnetic flux is not at the optimal point, e.g., $f= \fB$,
where the three-level SFQC has a cyclic transition which cannot occur in three-level natural atoms.
In this case, the susceptibility $\chi_{q}(\omega_{\rm P})$ can be given as
$\chi_{q}(\omega_{\rm P})=\chi_{01}(\omega_{\rm P}) +\chi_{02}(\omega_{\rm P})$,
and then EIT (ATS) may occur in the $|0\rangle\leftrightarrow|1\rangle$ frequency range and the $|0\rangle\leftrightarrow|2\rangle$ frequency range.

First, we study the susceptibility $\chi_{q}(\omega_{\rm P})$ in the weak-driving regime.
In Figs.~\ref{fig7}(a) and \ref{fig7}(b),
the real and imaginary parts of the susceptibility $\chi_{q}(\omega_{\rm P})$ are plotted.
The real and imaginary parts of the two resonances
$R_{\pm}^{(01)}(\omega_{\rm P}-\omega^{\prime})$ ($R_{\pm}^{(02)}(\omega_{\rm P}-\omega^{\prime})$),
corresponding to $\chi_{01}(\omega_{\rm P})$ ($\chi_{02}(\omega_{\rm P})$),
are also plotted in the left (right) parts of Figs.~\ref{fig7}(a) and \ref{fig7}(b)
with the same parameters as for $\chi_{q}(\omega_{\rm P})$.
In Figs.~\ref{fig7}(c) and \ref{fig7}(d),
the real and imaginary parts of the susceptibility $\chi_{q}(\omega_{\rm P})$ are plotted for different parameters.
All the parameters in Figs.~\ref{fig7}(a)--\ref{fig7}(d) are appropriately chosen to guarantee
that the Rabi frequency $|\Omega_{\rm D}|$ lies in the weak-driving regime.

In the right part of Fig.~\ref{fig7}(b),
the condition for realizing EIT in the $|0\rangle\leftrightarrow|2\rangle$ frequency range is fulfilled,
and EIT can be realized when $\omega_{\rm P}$ is near resonant to the $|0\rangle\leftrightarrow|2\rangle$ transition.
Note that ${\rm Im}[R_{\pm}^{(01)}(\omega_{\rm P}-\omega^{\prime})]$
are centered at $\omega_{\rm P}=\omega^{\prime}$ with different signals.
The summation of the two resonances makes ${\rm Im}[\chi_{q}(\omega_{\rm P})]$
exhibit a dip at $\omega_{\rm P}=\omega^{\prime}$.
However EIT does not occur in the $|0\rangle\leftrightarrow|1\rangle$ frequency range (see the left part of Fig.~\ref{fig7}(b)),
where the curve of ${\rm Im}[\chi_{q}(\omega_{\rm P})]$ is similar to that in Fig.~\ref{fig5}(b).
This is in accordance with our conclusion in Subsec.~B of Sec.~\ref{MS},
i.e., EIT cannot be realized simultaneously in both the $|0\rangle\leftrightarrow|1\rangle$ and $|0\rangle\leftrightarrow|2\rangle$ frequency ranges.
In the right part of Fig.~\ref{fig7}(d), we find that the dip of ${\rm Im}[\chi_{q}(\omega_{\rm P})]$
becomes shallower when the environmental equilibrium temperature $T$ goes up.
This is because the increase of $T$ makes the damping rates $\gamma_{11}$ and $\gamma_{22}$ increase
and further changes the value of ${\rm Im}[\chi_{q}(\omega_{\rm P})]$
at the minimum point $\omega_{\rm P}=\omega^{\prime}$.
We also find that the nonzero driving-field detuning $\Delta$
makes ${\rm Im}[\chi_{q}(\omega_{\rm P})]$ become asymmetric,
and changes the position of the minimum point of the dip.

Second, we study the susceptibility $\chi_{q}(\omega_{\rm P})$ in the strong-driving regime.
In Figs.~\ref{fig8}(a) and \ref{fig8}(b),
the real and imaginary parts of the susceptibility $\chi_{q}(\omega_{\rm P})$ are plotted.
The real and imaginary parts of the two resonances
$R_{\pm}^{(01)}(\omega_{\rm P}-\omega^{\prime})$ ($R_{\pm}^{(02)}(\omega_{\rm P}-\omega^{\prime})$),
corresponding to $\chi_{01}(\omega_{\rm P})$ ($\chi_{02}(\omega_{\rm P})$),
are also plotted in the left (right) parts of Figs.~\ref{fig8}(a) and \ref{fig8}(b)
with the same parameters as for $\chi_{q}(\omega_{\rm P})$.
In Figs.~\ref{fig8}(c) and \ref{fig8}(d),
the real and imaginary parts of the susceptibility
$\chi_{q}(\omega_{\rm P})$ are plotted for different parameters.
All the parameters in Figs.~\ref{fig8}(a)--\ref{fig8}(d) are appropriately chosen to guarantee
that the Rabi frequency $|\Omega_{\rm D}|$ lies in the strong-driving regime.

As discussed in Subsec.~B of Sec.~\ref{MS},
when the Rabi frequency of the driving field is large enough,
ATS can be realized not only in the $|0\rangle\leftrightarrow|1\rangle$ frequency range (see the left parts of Figs.~\ref{fig8}(a) and \ref{fig8}(b)),
but also in the $|0\rangle\leftrightarrow|2\rangle$ frequency range (see the right parts of Figs.~\ref{fig8}(a) and \ref{fig8}(b)).
In this case, there are four absorption peaks and two transparency windows.
However, in three-level natural atoms, there are only two absorption peaks and one transparency window for the ATS spectrum.
In Fig.~\ref{fig8}(d), we find that the effects of the environmental equilibrium temperature $T$
and the nonzero driving-field detuning $\Delta$ on the ATS spectrum are similar to those in Fig.~\ref{fig6}(d).

\section{Conclusions}\label{CS}

In conclusion, we have studied the linear response of a three-level SFQC,
with a driving field applied to the two upper energy levels.
We include the environmental effect on the three-level SQFC within the Born-Markov approximation.
In particular, we study electromagnetically induced transparency (EIT) and  Autler-Townes splitting (ATS).
We find that when the bias magnetic flux is at the optimal point,
the three-level SFQC can respond to the probe field in the frequency range corresponding to
the transition between the ground and the first excited energy levels of the three-level SFQC,
like natural atoms with ladder-type transitions.
However, when the bias magnetic flux deviates from the optimal point,
the three-level SFQC can respond to the probe field in two different frequency ranges,
of which one frequency range corresponds to
the transition between the ground and the first excited energy levels of the three-level SFQC,
while the other frequency range corresponds to
the transition between the ground and the second excited energy levels of the three-level SFQC.
In this case, the three-level SFQC acts like a combination of three-level natural atoms with ladder-type transitions
and three-level natural atoms with $\Lambda$-type transitions.

We derive the conditions for realizing EIT and ATS in three-level SFQCs.
We find that the realization of EIT in a three-level SFQC requires that
the damping rates of the three-level SFQC fulfill certain conditions
and that the Rabi frequency of the driving field lie in a certain frequency interval defined by the damping rates of the three-level SFQC. While the realization of ATS in a three-level SFQC only requires that
the Rabi frequency of the driving field is large enough.
We emphasize that these conditions for realizing EIT and ATS can apply not only to three-level SFQCs,
but also to other similar three-level quantum systems.

Using the conditions for realizing EIT and ATS, we analyze the linear response of the driven three-level SFQC.
When the bias magnetic flux is at the optimal point,
we find that EIT cannot be realized in the three-level SFQC
for the parameters chosen by us, however, ATS can be realized.
We note that EIT may be realized in three-level SFQCs like natural atoms with ladder-type transitions~\cite{eitEA}
when the parameters of the three-level SFQCs are further optimized.
When the bias magnetic flux is not at the optimal point,
EIT (ATS) can be realized in two different frequency ranges.
In addition, EIT cannot be realized simultaneously in these two frequency ranges
due to the restrictions laid by the conditions for realizing EIT.
However ATS can be realized simultaneously in these two frequency ranges
as long as the Rabi frequency of the driving field is large enough.

We also find that the damping rates modified by the driving field
can result in two different heights for the two peaks in the ATS absorption spectrum even in the resonant driving.
This phenomenon is a possible reason for the asymmetric transmission spectrum in the ATS experiment~\cite{atsEF,Bsanders}.

\section{Acknowledgement}

We thank Dr. Anatoly Smirnov for enlightening discussions.
Y.X.L. is supported by the NSFC Grants No.~61025022 and No.~60836001.
J.Q.Y. is supported by the NSFC Grant No.~91121015,
the National Basic Research Program of China Grant No.~2014CB921401,
and the NSAF Grant No.~U1330201.
E.I. is supported by the European Community¡¯s Seventh Framework Programme (FP7/2007-2013)
under Grant No.~270843 (iQIT).
F.N. is partially supported by
the ARO,
RIKEN iTHES Project,
MURI Center for Dynamic Magneto-Optics,
JSPS-RFBR Contract No.~12-02-92100,
Grant-in-Aid for Scientific Research (S),
MEXT Kakenhi on Quantum Cybernetics,
and the JSPS via its FIRST program.

\appendix

\section{Calculations of commutators and anti-commutators}\label{ApCA}

With the assumptions that the coupling between the three-level SFQC
and its environment is weak, and that the environmental correlation time $\tau_{c}$ is very small,
now we present a concrete method to calculate
the commutators and anti-commutators in Eqs.~(\ref{cmtor}) and (\ref{acmtor}).

In the rotating reference frame  with the unitary operator shown in Eq.~(\ref{eq:15}),
and neglecting the interaction between the three-level SFQC and its environment,
the time evolution of the three-level SFQC operators is governed by an effective Hamiltonian
\begin{align}\label{eq:28}
H_{\rm eff}
& =\hbar\Delta\sigma_{22} +\hbar\Omega_{\rm D}\sigma_{12} +\hbar\Omega_{\rm D}^{*}\sigma_{21}
\nonumber\\
& \equiv\varepsilon_{1}|\tilde{1}\rangle\langle\tilde{1}| +\varepsilon_{2}|\tilde{2}\rangle\langle\tilde{2}|.
\end{align}
The eigenvalues of the Hamiltonian in Eq.~(\ref{eq:28})
$\varepsilon_{1}=\hbar(\Delta-\Omega)/2$
and $\varepsilon_{2}=\hbar(\Delta+\Omega)/2$ correspond to the eigenstates
\begin{align}
|\tilde{1}\rangle
&= \cos(\theta) |1\rangle-\nu^{*}\sin(\theta) |2\rangle, \label{eq:29}
\\
|\tilde{2}\rangle
&= \nu\sin(\theta) |1\rangle+\cos(\theta) |2\rangle, \label{eq:30}
\end{align}
with the parameter $\nu=\Omega_{\rm D}/|\Omega_{\rm D}|$.
The parameter $\theta$, which characterizes the mixing
between the states $|1\rangle$ and $|2\rangle$, is given by
\begin{equation}
\tan(\theta) =\sqrt{\frac{\Omega-\Delta}{\Omega+\Delta}} \, ,
\end{equation}
with
\begin{equation}
\Omega=\sqrt{\Delta^2+4|\Omega_{\rm D}|^2} \, .
\end{equation}
Note that the parameter $\theta=0$ if $|\Omega_{\rm D}|=0$.

With the effective Hamiltonian in Eq.~(\ref{eq:28}),
the operator $\sigma_{lm}(t^{\prime})$ at the moment $t^{\prime}$ evolves to the operator
\begin{equation}\label{eq:33}
\sigma_{lm}(t)=U^{\dagger}_{\rm eff}(\tau) \sigma_{lm}(t^{\prime}) U_{\rm eff}(\tau),
\end{equation}
with the time interval $\tau=t-t^{\prime}$.
The time-evolution operator $U_{\rm eff}(\tau)$ in Eq.~(\ref{eq:33}) is
\begin{equation}\label{eq:34}
U_{\rm eff}(\tau)=|0\rangle\langle0|
+\sum_{l=1}^{2}\exp\left({-i\frac{\varepsilon_{l}}{\hbar}\tau}\right)|\tilde{l}\rangle\langle
\tilde{l}|,
\end{equation}
which is given via the effective Hamiltonian in Eq.~(\ref{eq:28}).
Here, the completeness relation
$|0\rangle\langle0| +|\tilde{1}\rangle \langle\tilde{1}| +|\tilde{2}\rangle \langle\tilde{2}|=1$
has been used in the derivation of Eq.~(\ref{eq:34})
and the states $|\tilde{l}\rangle$ ($l=1,\,2$) are given by Eqs.~(\ref{eq:29}) and (\ref{eq:30}).
Therefore, all operators $\sigma_{lm}(t^{\prime})$ at the moment $t^{\prime}$
can be given by the operators at the moment $t$ via Eqs.~(\ref{eq:33}) and (\ref{eq:34}).
The explicit expressions for $\sigma_{lm}(t^{\prime})$ are listed as follows:
\begin{align}
\label{cm00}
\sigma_{00}(t^{\prime}) &= \sigma_{00}(t),
\\
\label{cm11}
\sigma_{11}(t^{\prime}) &= |A(\tau)|^2 \sigma_{11}(t)+|B(\tau)|^2 \sigma_{22}(t)
\nonumber\\
&+ iA(\tau)B(\tau)\sigma_{12}(t) -iA^{*}(\tau)B^{*}(\tau)\sigma_{21}(t),
\\
\label{cm22}
\sigma_{22}(t^{\prime}) &= |B(\tau)|^2 \sigma_{11}(t)+|A(\tau)|^2 \sigma_{22}(t)
\nonumber\\
&- iA(\tau)B(\tau)\sigma_{12}(t) +iA^{*}(\tau)B^{*}(\tau)\sigma_{21}(t),
\\
\label{cm01}
\sigma_{01}(t^{\prime}) &= e^{i\frac{\Delta}{2}\tau} A^{*}(\tau) \sigma_{01}(t)
+ie^{i\frac{\Delta}{2}\tau} B(\tau) \sigma_{02}(t),
\\
\label{cm02}
\sigma_{02}(t^{\prime}) &= ie^{i\frac{\Delta}{2}\tau} B^{*}(\tau) \sigma_{01}(t)
+e^{i\frac{\Delta}{2}\tau} A(\tau) \sigma_{02}(t),
\\
\label{cm12}
\sigma_{12}(t^{\prime}) &= iA(\tau)B^{*}(\tau) \sigma_{11}(t) -iA(\tau)B^{*}(\tau)\sigma_{22}(t)
\nonumber\\
&+ A^{2}(\tau) \sigma_{12}(t)+B^{*2}(\tau) \sigma_{21}(t).
\end{align}
The expressions for the operators $\sigma_{10}(t^{\prime})$, $\sigma_{20}(t^{\prime})$,
and $\sigma_{21}(t^{\prime})$ can be obtained by taking the conjugates of Eqs.~(\ref{cm01})--(\ref{cm12}).

\section{Expressions for $\Gamma_{lm}$}\label{ApGM}

The explicit expressions for the complex coefficients $\Gamma_{lm}$ ($l,\,m=1,\,2$)
in Eqs.~(\ref{SL01}) and (\ref{SL02}) are given by
\begin{align}
\label{Gm11}
\Gamma_{11}
&= |I_{01}|^2 A_{11} +|I_{02}|^2 A_{12} +|I_{12}|^2 A_{13}
\nonumber\\
& +(I_{00}-I_{11})I_{00} A_{14} +(I_{00}-I_{11})I_{11} A_{15}
\nonumber\\
& +(I_{00}-I_{11})I_{22} A_{16},
\\
\label{Gm12}
\Gamma_{12}
&= |I_{01}|^2 A_{21} +|I_{12}|^2 A_{22}
\nonumber\\
& +(I_{00}-I_{11})I_{11} A_{23} +(I_{00}-I_{11})I_{22} A_{24},
\\
\label{Gm21}
\Gamma_{21}
&= |I_{02}|^2 B_{11} +|I_{12}|^2 B_{12}
\nonumber\\
& +(I_{00}-I_{22})I_{11} B_{13} +(I_{00}-I_{22})I_{22} B_{14},
\\
\label{Gm22}
\Gamma_{22}
&= |I_{01}|^2 B_{21} +|I_{02}|^2 B_{22} +|I_{12}|^2 B_{23}
\nonumber\\
& +(I_{00}-I_{22})I_{00} B_{24} +(I_{00}-I_{22})I_{11} B_{25}
\nonumber\\
& +(I_{00}-I_{22})I_{22} B_{26},
\end{align}
with the parameters
\begin{align}
A_{11}
&= 2i\sin^2(\theta) \widetilde{S}\left(\omega_1^{(+)}\right)
+2i\cos^2(\theta) \widetilde{S}\left(\omega_1^{(-)}\right),
\end{align}
\begin{align}
A_{12}
&= -\frac{\cos^2(\theta)}{2}
\left[ \chi \left(\omega^{\prime}_{(+)}\right)
-2i\widetilde{S} \left(\omega^{\prime}_{(+)}\right) \right]
\nonumber\\
& -\frac{\sin^2(\theta)}{2}
\left[ \chi \left(\omega^{\prime}_{(-)}\right)
-2i\widetilde{S} \left(\omega^{\prime}_{(-)}\right) \right],
\end{align}
\begin{align}
A_{13}
&= \frac{\cos^4(\theta)}{2}
\left[ \chi \left(-\omega_{0}^{(+)}\right)
+2i\widetilde{S} \left(-\omega_{0}^{(+)}\right) \right]
\nonumber\\
&+ \frac{\sin^4(\theta)}{2}
\left[ \chi \left(-\omega_{0}^{(-)}\right)
+2i\widetilde{S} \left(-\omega_{0}^{(-)}\right) \right]
\nonumber\\
&+ \frac{\sin^2(2\theta)}{4}
\left[ \chi \left(-\omega_{0}\right) +2i\widetilde{S} \left(-\omega_{0}\right) \right],
\end{align}
\begin{align}
A_{14}
&= -\frac{1}{2}
\left[ \chi \left(0\right) -2i\widetilde{S} \left(0\right) \right],
\end{align}
\begin{align}
A_{15}
&= -\frac{\sin^2(2\theta)}{8}
\left[ \chi \left(\Omega\right) +2i\widetilde{S} \left(\Omega\right) \right]
\nonumber\\
& -\frac{\sin^2(2\theta)}{8}
\left[ \chi \left(-\Omega\right) +2i\widetilde{S} \left(-\Omega\right) \right]
\nonumber\\
& -\frac{1+\cos^2(2\theta)}{4}
\left[ \chi \left(0\right) +2i\widetilde{S} \left(0\right) \right],
\end{align}
\begin{align}
A_{16}
&= \frac{\sin^2(2\theta)}{8}
\left[ \chi \left(\Omega\right) +2i\widetilde{S} \left(\Omega\right) \right]
\nonumber\\
& +\frac{\sin^2(2\theta)}{8}
\left[ \chi \left(-\Omega\right) +2i\widetilde{S} \left(-\Omega\right) \right]
\nonumber\\
& -\frac{\sin^2(2\theta)}{4}
\left[ \chi \left(0\right) +2i\widetilde{S} \left(0\right) \right],
\end{align}
\begin{align}
A_{21}
&= \frac{\nu \sin(2\theta)}{4}
\left[ \chi \left(\omega_1^{(+)}\right)
+2i\widetilde{S} \left(\omega_1^{(+)}\right) \right]
\nonumber\\
&- \frac{\nu \sin(2\theta)}{4}
\left[ \chi \left(\omega_1^{(-)}\right)
+2i\widetilde{S} \left(\omega_1^{(-)}\right) \right],
\end{align}
\begin{align}
A_{22}
&= -\frac{\nu \sin(2\theta)\cos^2(\theta)}{4} \left[ \chi \left(-\omega_{0}^{(+)}\right)
+2i\widetilde{S} \left(-\omega_{0}^{(+)}\right) \right]
\nonumber\\
& +\frac{\nu \sin(2\theta)\sin^2(\theta)}{4} \left[ \chi \left(-\omega_{0}^{(-)}\right)
+2i\widetilde{S} \left(-\omega_{0}^{(-)}\right) \right]
\nonumber\\
& +\frac{\nu \sin(2\theta)\cos(2\theta)}{4}
\left[ \chi \left(-\omega_0\right) +2i\widetilde{S} \left(-\omega_0\right) \right],
\end{align}
\begin{align}
A_{23}
&= -\frac{\nu \sin(2\theta)\cos^2(\theta)}{4} \left[ \chi \left(\Omega\right)
+2i\widetilde{S} \left(\Omega\right) \right]
\nonumber\\
& +\frac{\nu \sin(2\theta)\sin^2(\theta)}{4} \left[ \chi \left(-\Omega\right)
+2i\widetilde{S} \left(-\Omega\right) \right]
\nonumber\\
& +\frac{\nu \sin(2\theta)\cos(2\theta)}{4}
\left[ \chi \left(0\right) +2i\widetilde{S} \left(0\right) \right],
\end{align}
\begin{align}
A_{24}
&= \frac{\nu \sin(2\theta)\cos^2(\theta)}{4} \left[ \chi \left(\Omega\right)
+2i\widetilde{S} \left(\Omega\right) \right]
\nonumber\\
& -\frac{\nu \sin(2\theta)\sin^2(\theta)}{4} \left[ \chi \left(-\Omega\right)
+2i\widetilde{S} \left(-\Omega\right) \right]
\nonumber\\
& -\frac{\nu \sin(2\theta)\cos(2\theta)}{4}
\left[ \chi \left(0\right) +2i\widetilde{S} \left(0\right) \right],
\end{align}
\begin{align}
B_{11}
&= \frac{\nu^* \sin(2\theta)}{4}
\left[ \chi \left(\omega^{\prime}_{(+)}\right)
+2i\widetilde{S} \left(\omega^{\prime}_{(+)}\right) \right]
\nonumber\\
&- \frac{\nu^* \sin(2\theta)}{4}
\left[ \chi \left(\omega^{\prime}_{(-)}\right)
+2i\widetilde{S} \left(\omega^{\prime}_{(-)}\right) \right],
\end{align}
\begin{align}
B_{12}
&= \frac{\nu^* \sin(2\theta)\cos^2(\theta)}{4} \left[ \chi \left(\omega_{0}^{(+)}\right)
+2i\widetilde{S} \left(\omega_{0}^{(+)}\right) \right]
\nonumber\\
&- \frac{\nu^* \sin(2\theta)\sin^2(\theta)}{4} \left[ \chi \left(\omega_{0}^{(-)}\right)
+2i\widetilde{S} \left(\omega_{0}^{(-)}\right) \right]
\nonumber\\
&- \frac{\nu^* \sin(2\theta)\cos(2\theta)}{4}
\left[ \chi \left(\omega_0\right) +2i\widetilde{S} \left(\omega_0\right) \right],
\end{align}
\begin{align}
B_{13}
&= \frac{\nu^* \sin(2\theta)\sin^2(\theta)}{4} \left[ \chi \left(\Omega\right)
+2i\widetilde{S} \left(\Omega\right) \right]
\nonumber\\
& -\frac{\nu^* \sin(2\theta)\cos^2(\theta)}{4} \left[ \chi \left(-\Omega\right)
+2i\widetilde{S} \left(-\Omega\right) \right]
\nonumber\\
& +\frac{\nu^* \sin(2\theta)\cos(2\theta)}{4}
\left[ \chi \left(0\right) +2i\widetilde{S} \left(0\right) \right],
\end{align}
\begin{align}
B_{14}
&= -\frac{\nu^* \sin(2\theta)\sin^2(\theta)}{4} \left[ \chi \left(\Omega\right)
+2i\widetilde{S} \left(\Omega\right) \right]
\nonumber\\
& +\frac{\nu^* \sin(2\theta)\cos^2(\theta)}{4} \left[ \chi \left(-\Omega\right)
+2i\widetilde{S} \left(-\Omega\right) \right]
\nonumber\\
& -\frac{\nu^* \sin(2\theta)\cos(2\theta)}{4}
\left[ \chi \left(0\right) +2i\widetilde{S} \left(0\right) \right],
\end{align}
\begin{align}
B_{21}
&= -\frac{\sin^2(\theta)}{2}
\left[ \chi \left(\omega_1^{(+)}\right)
-2i\widetilde{S} \left(\omega_1^{(+)}\right) \right]
\nonumber\\
& -\frac{\cos^2(\theta)}{2}
\left[ \chi \left(\omega_1^{(-)}\right)
-2i\widetilde{S} \left(\omega_1^{(-)}\right) \right],
\end{align}
\begin{align}
B_{22}
&= 2i\cos^2(\theta) \widetilde{S}\left(\omega^{\prime}_{(+)}\right)
+2i\sin^2(\theta) \widetilde{S}\left(\omega^{\prime}_{(-)}\right),
\end{align}
\begin{align}
B_{23}
&= \frac{\cos^4(\theta)}{2}
\left[ \chi \left(\omega_{0}^{(+)}\right) +2i\widetilde{S} \left(\omega_{0}^{(+)}\right) \right]
\nonumber\\
&+ \frac{\sin^4(\theta)}{2}
\left[ \chi \left(\omega_{0}^{(-)}\right) +2i\widetilde{S} \left(\omega_{0}^{(-)}\right) \right]
\nonumber\\
&+ \frac{\sin^2(2\theta)}{4}
\left[ \chi \left(\omega_{0}\right) +2i\widetilde{S} \left(\omega_{0}\right) \right],
\end{align}
\begin{align}
B_{24}
&= -\frac{1}{2}
\left[ \chi \left(0\right) -2i\widetilde{S} \left(0\right) \right],
\end{align}
\begin{align}
B_{25}
&= \frac{\sin^2(2\theta)}{8}
\left[ \chi \left(\Omega\right) +2i\widetilde{S} \left(\Omega\right) \right]
\nonumber\\
& +\frac{\sin^2(2\theta)}{8}
\left[ \chi \left(-\Omega\right) +2i\widetilde{S} \left(-\Omega\right) \right]
\nonumber\\
& -\frac{\sin^2(2\theta)}{4}
\left[ \chi \left(0\right) +2i\widetilde{S} \left(0\right) \right],
\end{align}
\begin{align}
B_{26}
&= -\frac{\sin^2(2\theta)}{8}
\left[ \chi \left(\Omega\right) +2i\widetilde{S} \left(\Omega\right) \right]
\nonumber\\
& -\frac{\sin^2(2\theta)}{8}
\left[ \chi \left(-\Omega\right) +2i\widetilde{S} \left(-\Omega\right) \right]
\nonumber\\
& -\frac{1+\cos^2(2\theta)}{4}
\left[ \chi \left(0\right) +2i\widetilde{S} \left(0\right) \right].
\end{align}
Here $\omega_{0}^{(\pm)}$, $\omega_1^{(\pm)}$, and $\omega^{\prime}_{(\pm)}$
are defined as
\begin{align*}
\omega_{0}^{(\pm)}&= \omega_{0} \pm\Omega,
\\
\omega_1^{(\pm)}&= \omega_1 +\frac{\Delta \pm\Omega}{2},
\\
\omega^{\prime}_{(\pm)}&= \omega^{\prime} +\frac{\Delta \pm\Omega}{2}.
\end{align*}

According to the fluctuation-dissipation theorem in Eq.~(\ref{FDT}),
the damping rates $\gamma_{lm}={\rm Im}(\Gamma_{lm})$ ($l,\,m=1,\,2$)
can be given from Eqs.~(\ref{Gm11})--(\ref{Gm22}) as
\begin{align}
\label{gm11}
\gamma_{11}
&= |I_{01}|^2 a_{11} +|I_{02}|^2 a_{12} +|I_{12}|^2 a_{13}
\nonumber\\
& +(I_{00}-I_{11})I_{00}a_{14} +(I_{00}-I_{11})I_{11} a_{15}
\nonumber\\
& +(I_{00}-I_{11})I_{22} a_{16},
\\
\label{gm12}
\gamma_{12}
&= |I_{01}|^2 a_{21} +|I_{12}|^2 a_{22}
\nonumber\\
& +(I_{00}-I_{11})I_{11} a_{23} +(I_{00}-I_{11})I_{22} a_{24},
\\
\label{gm21}
\gamma_{21}
&= |I_{02}|^2 b_{11} +|I_{12}|^2 b_{12}
\nonumber\\
& +(I_{00}-I_{22})I_{11} b_{13} +(I_{00}-I_{22})I_{22} b_{14},
\\
\label{gm22}
\gamma_{22}
&= |I_{01}|^2 b_{21} +|I_{02}|^2 b_{22} +|I_{12}|^2 b_{23}
\nonumber\\
& +(I_{00}-I_{22})I_{00} b_{24} +(I_{00}-I_{22})I_{11} b_{25}
\nonumber\\
& +(I_{00}-I_{22})I_{22} b_{26},
\end{align}
with the parameters
\begin{align}
a_{11}
&= \sin^2(\theta) S\left(\omega_1^{(+)}\right) +\cos^2(\theta) S\left(\omega_1^{(-)}\right),
\end{align}
\begin{align}
a_{12}
&= \frac{\cos^2(\theta)}{2} \Re \left(-\omega^{\prime}_{(+)}\right)
+\frac{\sin^2(\theta)}{2} \Re \left(-\omega^{\prime}_{(-)}\right),
\end{align}
\begin{align}
a_{13}
&=  \frac{\cos^4(\theta)}{2} \Re \left(-\omega_{0}^{(+)}\right)
+\frac{\sin^4(\theta)}{2} \Re \left(-\omega_{0}^{(-)}\right)
\nonumber\\
& +\frac{\sin^2(2\theta)}{4} \Re \left(-\omega_{0}\right),
\end{align}
\begin{align}
a_{14}
&= \frac{1}{2} \Re \left(0\right),
\end{align}
\begin{align}
a_{15}
&= -\frac{\sin^2(2\theta)}{8} \Re \left(\Omega\right)
-\frac{\sin^2(2\theta)}{8} \Re \left(-\Omega\right)
\nonumber\\
& -\frac{1+\cos^2(2\theta)}{4} \Re \left(0\right),
\end{align}
\begin{align}
a_{16}
&= \frac{\sin^2(2\theta)}{8} \Re \left(\Omega\right)
+\frac{\sin^2(2\theta)}{8} \Re \left(-\Omega\right)
\nonumber\\
& -\frac{\sin^2(2\theta)}{4} \Re \left(0\right),
\end{align}
\begin{align}
a_{21}
&= \frac{\nu \sin(2\theta)}{4} \Re \left(\omega_1^{(+)}\right)
  -\frac{\nu \sin(2\theta)}{4} \Re \left(\omega_1^{(-)}\right),
\end{align}
\begin{align}
a_{22}
&= -\frac{\nu \sin(2\theta)\cos^2(\theta)}{4} \Re \left(-\omega_{0}^{(+)}\right)
\nonumber\\
& +\frac{\nu \sin(2\theta)\sin^2(\theta)}{4} \Re \left(-\omega_{0}^{(-)}\right)
\nonumber\\
& +\frac{\nu \sin(2\theta)\cos(2\theta)}{4} \Re \left(-\omega_0\right),
\end{align}
\begin{align}
a_{23}
&= -\frac{\nu \sin(2\theta)\cos^2(\theta)}{4} \Re \left(\Omega\right)
\nonumber\\
& +\frac{\nu \sin(2\theta)\sin^2(\theta)}{4} \Re \left(-\Omega\right)
\nonumber\\
& +\frac{\nu \sin(2\theta)\cos(2\theta)}{4} \Re \left(0\right),
\end{align}
\begin{align}
a_{24}
&= \frac{\nu \sin(2\theta)\cos^2(\theta)}{4} \Re \left(\Omega\right)
\nonumber\\
& -\frac{\nu \sin(2\theta)\sin^2(\theta)}{4} \Re \left(-\Omega\right)
\nonumber\\
& -\frac{\nu \sin(2\theta)\cos(2\theta)}{4} \Re \left(0\right),
\end{align}
\begin{align}
b_{11}
&= \frac{\nu^* \sin(2\theta)}{4} \Re \left(\omega^{\prime}_{(+)}\right)
  -\frac{\nu^* \sin(2\theta)}{4} \Re \left(\omega^{\prime}_{(-)}\right),
\end{align}
\begin{align}
b_{12}
&= \frac{\nu^* \sin(2\theta)\cos^2(\theta)}{4} \Re \left(\omega_{0}^{(+)}\right)
\nonumber\\
& -\frac{\nu^* \sin(2\theta)\sin^2(\theta)}{4} \Re \left(\omega_{0}^{(-)}\right)
\nonumber\\
& -\frac{\nu^* \sin(2\theta)\cos(2\theta)}{4} \Re \left(\omega_0\right),
\end{align}
\begin{align}
b_{13}
&= \frac{\nu^* \sin(2\theta)\sin^2(\theta)}{4} \Re \left(\Omega\right)
\nonumber\\
& -\frac{\nu^* \sin(2\theta)\cos^2(\theta)}{4} \Re \left(-\Omega\right)
\nonumber\\
& +\frac{\nu^* \sin(2\theta)\cos(2\theta)}{4} \Re \left(0\right),
\end{align}
\begin{align}
b_{14}
&= -\frac{\nu^* \sin(2\theta)\sin^2(\theta)}{4} \Re \left(\Omega\right)
\nonumber\\
& +\frac{\nu^* \sin(2\theta)\cos^2(\theta)}{4} \Re \left(-\Omega\right)
\nonumber\\
& -\frac{\nu^* \sin(2\theta)\cos(2\theta)}{4} \Re \left(0\right),
\end{align}
\begin{align}
b_{21}
&= \frac{\sin^2(\theta)}{2} \Re \left(-\omega_1^{(+)}\right)
+\frac{\cos^2(\theta)}{2} \Re \left(-\omega_1^{(-)}\right),
\end{align}
\begin{align}
b_{22}
&= \cos^2(\theta) S\left(\omega^{\prime}_{(+)}\right)
+\sin^2(\theta) S\left(\omega^{\prime}_{(-)}\right),
\end{align}
\begin{align}
b_{23}
&= \frac{\cos^4(\theta)}{2} \Re \left(\omega_{0}^{(+)}\right)
+\frac{\sin^4(\theta)}{2} \Re \left(\omega_{0}^{(-)}\right)
\nonumber\\
& +\frac{\sin^2(2\theta)}{4} \Re \left(\omega_{0}\right),
\end{align}
\begin{align}
b_{24}
&= \frac{1}{2} \Re \left(0\right),
\end{align}
\begin{align}
b_{25}
&= \frac{\sin^2(2\theta)}{8} \Re \left(\Omega\right)
+\frac{\sin^2(2\theta)}{8} \Re \left(-\Omega\right)
\nonumber\\
& -\frac{\sin^2(2\theta)}{4} \Re \left(0\right),
\end{align}
\begin{align}
b_{26}
&= -\frac{\sin^2(2\theta)}{8} \Re \left(\Omega\right)
-\frac{\sin^2(2\theta)}{8} \Re \left(-\Omega\right)
\nonumber\\
& -\frac{1+\cos^2(2\theta)}{4} \Re \left(0\right).
\end{align}
Here, the function $\Re(\omega)$ is defined as
\begin{equation}
\Re(\omega) =\chi^{\prime\prime}(\omega) \left[ 1 +\coth\left(\frac{\hbar\omega}{2k_{\rm B}T}\right) \right].
\end{equation}

\end{document}